\documentclass[preprintnumbers,article,amsmath,amssymb,floatfix,10pt,prd,superscriptaddress,nofootinbib,twocolumn]{revtex4}

\usepackage{doi}
\usepackage{hyperref}
\hypersetup{
  colorlinks=true,        
  linkcolor=magenta,         
  citecolor=red,         
}

\providecommand{\dif}{\mathrm{d}} \def\d{\dif}
\usepackage{graphicx}
\usepackage{dcolumn}
\usepackage{bm}
\usepackage{color}
\usepackage{enumitem}
\usepackage{amsmath}
\usepackage{amssymb}

\def\EE{{\cal E}}
\def\LL{{\cal L}}
\newcommand{\beq}{\begin{equation}}
\newcommand{\eeq}{\end{equation}}
\newcommand{\bea}{\begin{eqnarray}}
\newcommand{\eea}{\end{eqnarray}}

\usepackage{bbm}
\usepackage{amsfonts}
\usepackage{mathrsfs}
\usepackage{latexsym}
\usepackage{epsfig}
\usepackage{epstopdf}
\usepackage{epstopdf}
\usepackage{graphicx}
\usepackage{amssymb}
\usepackage{amsmath}
\usepackage{dcolumn}
\usepackage{bm}
\usepackage{color}
\usepackage{comment}
\usepackage{xcolor}

\begin{document}
\title{Dynamics, Ringdown, and Accretion-Driven Multiple Quasi-Periodic Oscillations of Kerr–Bertotti–Robinson Black Holes}

\author{G.Mustafa}
\email{gmustafa3828@gmail.com}
\affiliation{Department of Physics, Zhejiang Normal University, Jinhua 321004, People’s Republic of China,}

\author{Orhan Donmez}
\email{orhan.donmez@aum.edu.kw}
\affiliation{College of Engineering and Technology, American University of the Middle East, Egaila 54200, Kuwait}

\author{Dhruba Jyoti Gogoi}
\email{moloydhruba@yahoo.in}
\affiliation{Department of Physics, Madhabdev University, Narayanpur, Lakhimpur 784164, Assam, India}
\affiliation{Research Center of Astrophysics and Cosmology, Khazar University, 41 Mehseti Street, AZ1096 Baku, Azerbaijan}

\author{Sushant G. Ghosh }
\email{sghosh2@jmi.ac.in}
\affiliation{Centre for Theoretical Physics, Jamia Millia Islamia, New Delhi 110025, India}
\affiliation{Astrophysics and Cosmology Research Unit, School of Mathematics, Statistics and Computer Science, University of KwaZulu-Natal, Private Bag 54001, Durban 4000, South Africa}

\author{Ibrar Hussain}
\email{ ibrar.hussain@seecs.nust.edu.pk}
\affiliation{School of Electrical Engineering and Computer Science, National University of Sciences and Technology, H-12, Islamabad, Pakistan}
\affiliation{Research Center of Astrophysics and Cosmology, Khazar University, 41 Mehseti Street, AZ1096 Baku, Azerbaijan}

\author{Chengxun Yuan}
\email{yuancx@hit.edu.cn}
\affiliation{School of Physics, Harbin Institute of Technology, Harbin 150001, People's Republic of China}

\begin{abstract}
We study the motion of test particles around the Kerr--Bertotti--Robinson (KBR) black hole (BH) and explore how the three defining parameters, the mass $M$, rotation parameter $a$, and magnetic parameter $B$ influence their dynamics. We derive analytical expressions for the energy and angular momentum of stable equatorial circular orbits, along with the corresponding radial and latitudinal oscillation frequencies, as functions of $M$, $a$, and $B$.
We also examine the key features of the quasi-periodic oscillations (QPOs) of test particles near stable circular orbits, including the precession effects such as periastron precession and the Lense-Thirring effect. We compare our results with those corresponding to the Kerr BH. We find that the particle motion is strongly shaped by the BH parameters. Using a WKB approach, we also study scalar quasinormal modes of rotating KBR BH in an external magnetic field and show that the magnetic field increases damping, while rotation and angular momentum mainly set the oscillation frequencies.  Alternatively, general relativistic modeling of Bondi-Hoyle-Lyttleton (BHL) accretion onto rapidly rotating KBR BH shows that two distinct physical structures emerge and cyclically transform into one another over time. These processes produce either a strongly oscillating flip–flop shock cone or a nearly stationary toroidal structure, with their formation governed by the BH spin and magnetic curvature. Power spectral analysis shows that these configurations give rise to low- and high-frequency QPO, providing a unified theoretical framework to understand how multiple QPO–like features can arise in rapidly spinning accreting systems.
\\
\textbf{Keywords}: Circular orbits; Particle dynamics; Bondi-Hoyle-Lyttleton; Flip-flop instability; Toroidal oscillation
\end{abstract}

\maketitle

\date{\today}


\section{Introduction}
In 1915, Einstein gave his theory of General Relativity (GR), which fundamentally establishes the relationship between mass, space, and time. Despite some initial objections, GR has proven to be a mathematically consistent framework and has successfully passed numerous experimental and observational tests \cite{1a,2a}. In recent years, the detection of gravitational waves from merging BHs \cite{3a}, along with direct imaging of BH shadows at the centers of galaxies such as the Milky Way and M87 \cite{4a,5a}, has provided further confirmation of Einstein’s theory. Nevertheless, although GR is one of the most successful and widely accepted theories of modern physics, several phenomena in the universe remain beyond its full explanation. Two of the most compelling challenges are the accelerated expansion of the universe \cite{6a} and the long-standing problem of the quantization of the gravitational field \cite{7a}. To overcome such limitations, several alternative and modified theories of gravity have been proposed \cite{8a,9a,10a}, extending or generalizing GR to offer a more complete description of the universe.

One of the key predictions of GR is the existence of BHs, whose strong gravitational fields make them powerful laboratories for testing the theory. Both static and rotating black-hole solutions have been widely studied within GR \cite{21n,22n}.
In this context, a wide range of related phenomena have been studied, including the dynamics of null and time-like geodesics, gravitational lensing \cite{23n,24n,25n,26n}, BH shadows \cite{27n,28n,29n,30n}, QPOs \cite{31n}, and tidal forces \cite{32n}. In recent years, BHs have also played a central role in the study of modified and alternative theories of gravity, where they are used to probe and constrain the additional spacetime parameters introduced in such models \cite{Fatima:2026prk}, often through comparison with astrophysical and observational data, for instance, released by the Event Horizon Telescope Collaboration \cite{33n,34n,35n,36n,37n}.

Rotating BHs are of particular astrophysical interest, as observations suggest that the BHs at the centers of M87 and our own Galaxy possess significant spin \cite{spin1, spin2}, which can influence estimates of their masses. The first rotating solution of Einstein’s field equations, discovered by Roy Kerr, is therefore regarded as a realistic model for astrophysical BHs \cite{Roy} and has been studied extensively from both theoretical and observational perspectives \cite{Ap1, Ap2, Ap3}. The Kerr spacetime has since been generalized to include additional parameters such as electric charge, the NUT parameter, and dark energy, giving rise to a variety of related models explored in different contexts \cite{Nu1, Nut2, Nut3}.   However, the Kerr solution does not incorporate external magnetic fields, which are expected to be present in most astrophysical environments. To address this limitation, Podolský and Ovcharenko introduced the KBR BH \cite{KBR}, describing a spacetime permeated by a uniform electromagnetic field and representing one of the simplest exact solutions of the Einstein--Maxwell equations. Owing to its high degree of symmetry and conformal flatness, this geometry provides a convenient setting for studying spacetime curvature, field interactions, and particle dynamics. Notably, it arises as the near-horizon geometry of the extremal Reissner--Nordström and Kerr--Newman BHs and shares close connections with the Near-Horizon Extreme Kerr (NHEK) geometry \cite{NHEK}. As a result, the KBR spacetime offers a valuable framework for probing the interplay between strong gravitational and electromagnetic fields and has found applications in black-hole thermodynamics, quantum gravity, and holographic approaches such as AdS/CFT and Kerr/CFT \cite{ADS}.


BHs, when subjected to small perturbations, exhibit a characteristic oscillatory response governed by their quasinormal modes (QNMs), i.e., complex frequency spectra that encode how spacetime relaxes back to equilibrium after a disturbance \cite{Vishveshwara:1970cc, Chandrasekhar:1985kt0, Lambiase:2024lvo}. These modes represent the intrinsic “ringing” of the BH, determined solely by its physical parameters, such as mass, charge, and angular momentum, as well as by the type of perturbation, whether scalar, electromagnetic, or gravitational. The defining feature of QNMs is their independence from the initial perturbation, rendering them powerful diagnostic tools for probing the geometry and stability of BHs across a variety of gravitational theories \cite{Gogoi:2024scc, Gogoi:2024epx, Gogoi:2024eyw}. Originating from the foundational work of Regge and Wheeler \cite{Regge:1957td} and subsequently refined through the seminal analyses of Vishveshwara, Zerilli, Moncrief, Teukolsky, and Chandrasekhar \cite{Vishveshwara:1970cc,Zerilli:1970se,Zerilli:1970wzz,Zerilli:1974ai,Moncrief:1975sb,Teukolsky:1972my,Chandrasekhar:1985kt0}, the study of QNMs has evolved into a central framework for testing gravitational dynamics in both classical and modified theories of gravity. In contemporary research, QNMs provide the theoretical foundation for understanding the ringdown phase of binary BH mergers, during which the remnant emits gravitational waves that carry imprints of its internal structure. The detection of such signals by observatories such as LIGO and Virgo \cite{3a} has transformed QNMs from purely theoretical constructs into observational tools, enabling precision tests of GR and the potential discrimination between competing models of gravity, matter coupling, and higher-dimensional effects.

The interplay among strong gravitational fields, BH rotation, and external electromagnetic fields is crucial for accretion dynamics around BHs. In this context, Bondi–Hoyle–Lyttleton (BHL) accretion provides a powerful framework for exploring how relativistic spacetime geometry governs large-scale flow morphologies and gives rise to time-dependent variability, particularly in wind-fed X-ray binary systems \cite{Bondi:1952ni,Font:1998sc,Armitage:2020owb}. Although Kerr BHs produce a relatively stable downstream shock-cone structure, recent studies have demonstrated that spacetime metrics containing deviations from the Kerr geometry can trigger qualitatively new hydrodynamical behaviors. These include flip-flop instabilities and strong non-axisymmetric oscillations \cite{Foglizzo:2005in,Blondin:2009bp,Donmez2012MNRAS}. Motivated by these findings, and extending earlier studies that examined the impact of the KBR magnetic curvature parameter on particle dynamics and epicyclic frequencies, we perform fully general relativistic hydrodynamical simulations to investigate accretion dynamics and associated instabilities. By numerically evolving the accretion flow in KBR spacetime, we analyze how the combined effects of the BH spin and the magnetic curvature parameter influence shock cone stability, flow morphology, and the emergence of distinct accretion structures.

Beyond morphological features, the temporal variability of the accretion flow provides a useful phenomenological bridge between theoretical models and high-energy astrophysical observations. In particular, fluctuations in the mass accretion rate, together with the formation of spiral shock waves and toroidal configurations, imprint characteristic signatures on the power spectrum of the accretion flow \cite{Zanotti:2002it,Schnittman:2005ac}. Accordingly, we analyze the radial and azimuthal variability of the accreting matter and carry out detailed power spectral analyses for different values of the BH spin and magnetic curvature parameters. These analyses allow us to identify distinct low- and high-frequency oscillatory modes associated with flip-flop shock dynamics and toroidal oscillations, respectively. We then place the frequencies extracted from our numerical simulations in the context of the QPO phenomenology discussed for X-ray binary systems, with the aim of assessing whether the characteristic frequency ranges and mode families produced by the model are astrophysically plausible. By comparing the frequencies extracted from our numerical simulations with characteristic frequency ranges discussed in the context of QPO phenomenology in X-ray binary systems, we propose a unified physical interpretation in which transitions between shock-dominated and torus-dominated accretion states naturally account for the coexistence and recurrence of multiple QPO frequencies \cite{Remillard:2006fc,Belloni:2012sv,Ingram:2019mna}. Within this framework, variations in the magnetic configuration of the spacetime act as a control parameter governing accretion-state transitions and their observational manifestations.

The present work goes beyond a straightforward application of known techniques by presenting a unified analysis of particle dynamics, QNMs, and fully relativistic accretion flows in the KBR spacetime. While the individual methods employed are well established, their combined application reveals qualitatively new effects absent in the Kerr case. In particular, the magnetic curvature parameter intrinsic to the KBR spacetime simultaneously modifies epicyclic frequencies, photon-sphere properties, and accretion morphology, thereby linking orbital dynamics to large-scale flow instabilities. Physically, the KBR spacetime represents the near-horizon geometry of extremal rotating BHs immersed in a uniform electromagnetic field, and our results show that magnetic curvature acts as a control parameter for accretion-state transitions and dynamical variability.

This paper is organized as follows. In Section~\ref{Sec2}, we introduce the rotating KBR spacetime and analyze the motion of test particles, derive expressions for the energy, angular momentum, and effective potential for equatorial circular orbits. Section~\ref{Sec3} investigates small harmonic oscillations around stable orbits and the corresponding epicyclic frequencies measured by a distant observer. In Section~\ref{Sec4}, we compute the QNMs of a massless scalar field in the KBR background using the WKB method, highlighting the influence of the magnetic field and black-hole spin on the oscillation and damping rates. Section~~\ref{Sec5} presents general relativistic hydrodynamic simulations of BHL accretion onto a KBR BH, examining the formation of shock cones, flip-flop instabilities, and transitions to toroidal structures. The radial and azimuthal variability of the accretion flow is analyzed, including mass accretion rates and spiral density patterns. We also carry out a power spectral analysis of the accretion variability, identifying distinct families of QPOs associated with different accretion morphologies. In Section~~\ref{Sec8}, we place our numerical QPO frequencies in the context of observed QPO phenomenology in X-ray binary systems, discuss possible qualitative correspondences, and provide a physical interpretation of the model transitions together with their limitations.

\section{Rotating Kerr-Bertotti-Robinson BH}\label{Sec2}
The KBR spacetime corresponds to a solution of the Einstein–Maxwell equations in which the geometry is supported by a constant electromagnetic field. Despite its nontrivial origin, the resulting spacetime is relatively simple and, for this reason, it has often been used as a convenient background for examining the interaction between gravitation and electromagnetism in GR. After its derivation, this solution has been considered in different contexts \cite{Zeng25,Kumar25,Ali26}. The line element describing the geometry of the rotating KBR BH is given by \cite{KBR,Ovch25}
\bea\label{BH}
\d s^2 &=& \frac{1}{\chi} \left[-(dt - a \sin^2\theta d\phi)^2 \frac{Q}{\rho^2} + \frac{\rho^2}{Q}dr^2 + \frac{\rho^2}{P} d\theta^2\right. \nonumber \\ &+&\left. (a dt - (a^2 + r^2) d \phi)^2 \frac{P}{\rho^2} \sin^2\theta 
\right], 
\eea
where the metric functions take the form
\begin{eqnarray*}
\Delta &=& a^2+r^2 \left(1-\frac{B^2 I_2 M^2 }{I_1^2} \right)-\frac{2 I_{2} M r}{ I_{1} },\\
\rho^2 &=& r^2 + a^2\, \cos^2\theta,\\\
P &=& B^2 \cos^2\theta \left(\frac{I_2 M^2}{I_1^2}-a^2\right) + 1.\\
Q &=& ( 1 + B^2 r^2 ) \Delta\\
\chi &=& \left(B^2 r^2+1\right)-B^2 \Delta \cos^2\theta\\
I_1 &=& 1-\frac{a^2 B^2}{2}\\
I_2 &=& 1 - a^2 B^2.
\end{eqnarray*}
Here, $M$ is the mass of the BH, $a$ is the rotation parameter of the BH, and $B$ represents the magnetic field parameter. For $B=0$, the spacetime (\ref{BH}) reduces to the Kerr spacetime. The horizon of the BH can be found by solving $\Delta=0$. 

Next, we clarify the physical significance of the magnetic parameter $B$ in the KBR spacetime. This parameter $B$ does not correspond to an intrinsic magnetic charge of the BH. Instead, it quantifies the magnitude of an external magnetic field, where the rotating BH is immersed.
Magnetic fields play a fundamental role in contemporary astrophysics, as astrophysically realistic BHs are not isolated systems but are typically immersed in external electromagnetic environments. These environments regulate a broad spectrum of high-energy phenomena, most notably those associated with active galactic nuclei, quasars, and related relativistic outflows. The inclusion of magnetic fields significantly modifies BH properties such as event horizon structure, geodesic motion of test particles, and thermodynamic characteristics rendering them indispensable for the formulation of physically accurate models. The solution describes a BH placed in a large-scale electromagnetic background, obtained consistently from the Einstein--Maxwell equations. In realistic astrophysical situations, BHs are not isolated objects. They are usually surrounded by accreting plasma, and such plasma naturally carries magnetic fields, e.g., the supermassive BHs in AGN and stellar-mass BHs in X-ray binaries are expected to be immersed in magnetic fields generated by their accretion disks. Near the event horizon, if the magnetic field varies slowly over distances larger than the gravitational radius, $B$ can be approximated as locally uniform. In this sense, the present model should be understood as an idealized but physically motivated description of a rotating BH in an external magnetic environment. Throughout this work, the parameter $B$ therefore controls how strongly the background magnetic field influences the spacetime geometry and the associated physical observables. 

\subsection{Circular orbits around rotating KBR BH} 

\begin{figure*}
\centering 
\includegraphics[width=85mm]{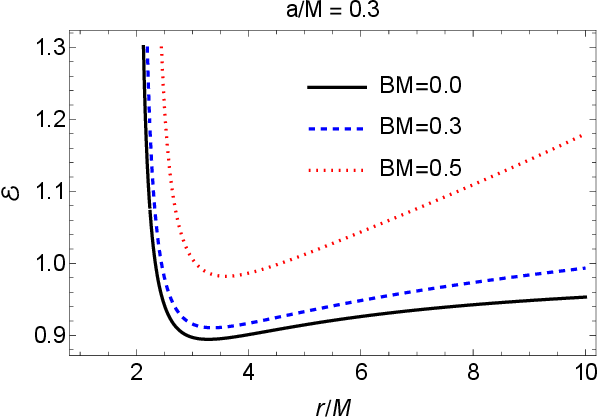}\;\;\;
\includegraphics[width=85mm]{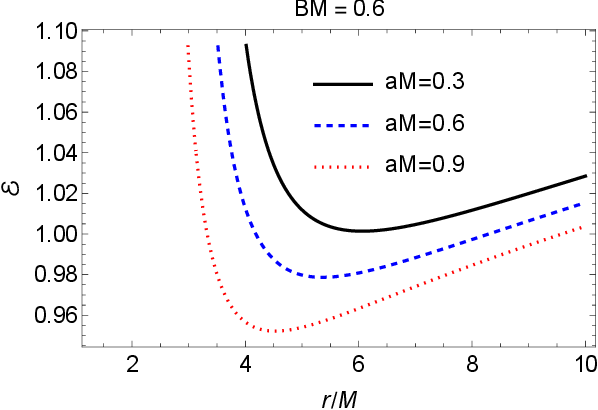}
\caption{Energy of test particles in equatorial circular orbits around a rotating KBR BH, shown as a function of radial distance $r/M$. The left panel illustrates the influence of the magnetic parameter $B$ for a fixed spin $a/M = 0.3$, and the right panel shows the effect of the spin parameter $a$ for a fixed magnetic parameter $B^2 M = 0.6$.}\label{fig_energy}
\end{figure*}

\begin{figure*}
\centering 
\includegraphics[width=85mm]{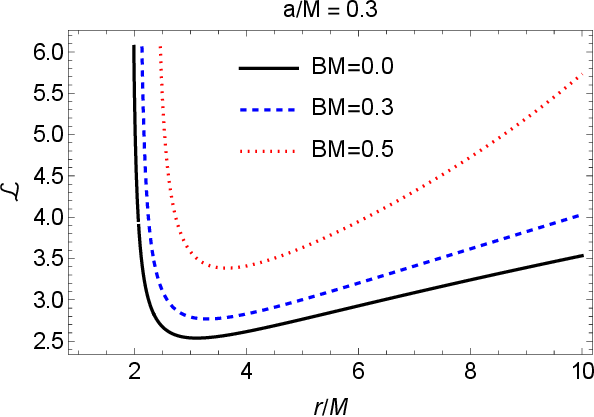}\;\;\;
\includegraphics[width=85mm]{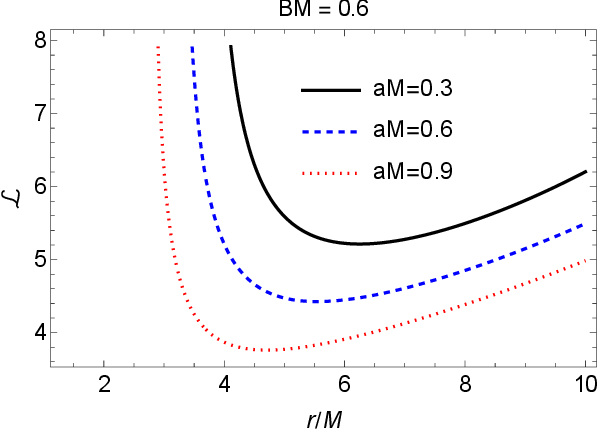}
\caption{Specific angular momentum $\mathcal{L}/M$ of test particles in equatorial circular orbits around a rotating KBR BH. The left panel displays the dependence on the magnetic parameter $B$ for fixed $a/M = 0.3$, and the right panel shows the variation with spin $a$ for fixed $B^2 M = 0.3$.
}\label{fig_ang}
\end{figure*}

The motion of a neutral particle can be discussed by the Hamiltonian, which is given as:
\beq\label{Ham}
H=\frac{1}{2}g^{\alpha \beta} p_{\alpha} p_{\beta} + \frac{1}{2}m^2,
\eeq
where $m$ is the particle's mass, $p^{\gamma}=m u^{\gamma}$ represents the four-momentum, $u^{\gamma}= d x^{\gamma}/d\tau$ denotes the four-velocity, and $\tau$ is the appropriate time of the test particle. The Hamilton equations of motion can be expressed as
\beq
\frac{dx^{\gamma}}{d\zeta}\equiv m u^{\gamma}=\frac{\partial H}{\partial p_{\gamma}}, \quad
\frac{d p_{\gamma}}{d\zeta} = -\frac{\partial H}{\partial x^{\gamma}},
\eeq
with $\zeta=\tau/m$ representing the affine parameter. The symmetries of the BH geometry lead to the emergence of two constants of motion: specific energy $E$ and specific angular momentum $L$ given by
\bea\label{EE}
\frac{p_{t}}{m}&=&g_{tt}u^{t}+g_{t\phi}u^{\phi}=-\mathcal{E}, \\\label{LL}
\frac{p_{\phi}}{m}&=&g_{\phi\phi}u^{\phi}+g_{t\phi}u^{t}=\mathcal{L},
\eea
where $\mathcal{E}=E/m$, $\mathcal{L}=L/m$. Figure \ref{fig_energy} describes the behavior of the energy of circular equatorial orbits around a rotating  KBR BH. In the first column, the influence of the magnetic parameter $B$ is shown, while in the second column, the  rotation parameter $a$ of BH is presented. The energy of circular orbits increases with increasing magnetic parameter $B$. However, an opposite behavior can be seen for the case of BH rotation $a$. Circular orbits have lower energy when the rotation parameter $a$ is small. We also compare the particle energy around a Kerr BH and a rotating  KBR BH. Particles around Kerr BH have smaller energy compared to those near a rotating KBR BH.

Figure \ref{fig_ang} depicts the behavior of the angular momentum of particles orbiting a rotating  KBR BH. In the first column, the influence of the magnetic parameter $B$ is shown, while in the second column, the behavior of the BH rotation parameter $a$ on the particle angular momentum is illustrated. Circular orbits have a greater angular momentum as the magnetic parameter $B$ increases. However, the opposite behavior can be observed for the case of BH rotation $a$. Circular orbits have a small angular momentum when the rotation parameter $a$ is large. However, the angular momentum increases as the radial distance \(r\) increases. Particles orbiting Kerr BHs have smaller angular momentum than those orbiting rotating KBR BHs.

\begin{figure*}
\centering 
\includegraphics[width=85mm]{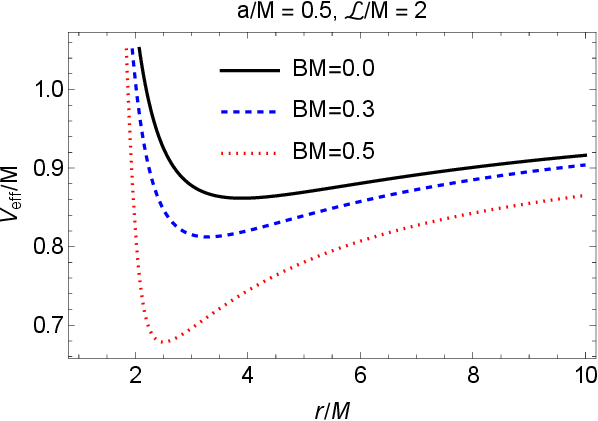}\;\;\;
\includegraphics[width=85mm]{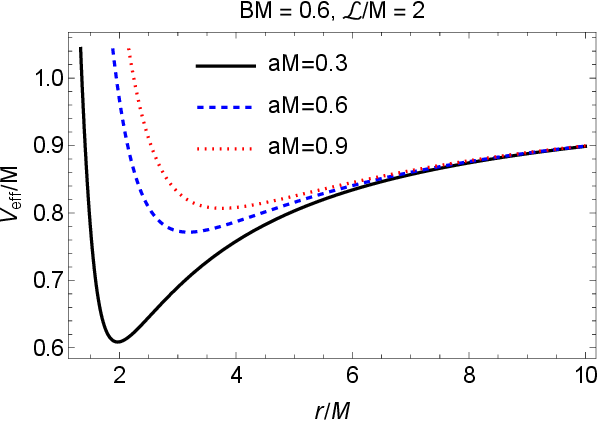}
\caption{Effective potential $V_{\text{eff}}/M$ for test particles moving in the equatorial plane of a rotating KBR BH. The left panel shows the effect of the magnetic parameter $B$ for fixed $a/M = 0.5$ and $\mathcal{L}/M = 2$, and the right panel shows the effect of the spin parameter $a$ for fixed $B^2 M = 0.6$ and $\mathcal{L}/M = 2$.}\label{fig_eff}
\end{figure*}

\subsection{Effective potential}

Using the normalization condition $g_{\nu \sigma} u^\nu u^\sigma = -1$, one can write
\beq
V_{eff}(r, \theta) = g_{rr}\, \Dot{r}^{2} + g_{\theta \theta}\, \Dot{\theta}^{2},
\eeq
where $\Dot{r} = dr/d\tau$, $\Dot{\theta} = d\theta/d\tau$, and $V_{eff}$ denotes the effective potential defined by the following relation 
\beq
V_{eff}(r, \theta) = \frac{\EE^2 g_{\phi \phi} + 2 \EE \LL g_{t \phi} + \LL^2 g_{tt}}{g_{t\phi}^{2} - g_{tt} g_{\phi\phi}} - 1.
\eeq

To illustrate the motion of test particles, the effective potential $V_{\text{eff}}(r, \theta)$ plays a crucial role. It describes particle motion without directly utilizing the equations of motion. Circular orbits in the equatorial plane $\theta=\pi/2$ can be derived under the following conditions as
\beq
V_{\rm eff}(r) = 0, \quad \frac{\d V_{\rm eff} (r)}{\d r} = 0.\label{Veff-1}
\eeq
The extrema of the effective potential specify the possible circular orbits. The minima correspond to stable orbits, while maxima indicate unstable ones. In Newtonian theory, for any given angular momentum, the effective potential always has a minimum, implying the existence of a stable circular orbit, commonly identified with the ISCO. When additional effects are present, such as particle rotation or background fields, the shape of the effective potential changes, and the locations of these orbits can shift. In GR, even for a fixed angular momentum, the effective potential near a Schwarzschild BH admits two extrema, associated with stable and unstable circular motion. In our analysis, the angular momentum is therefore crucial for determining the ISCO position.

Figure~\ref{fig_eff} shows the effective potential $V_{\mathrm{eff}}$ as a function of the radial coordinate $r$ for different BH parameters. The left panel illustrates the effect of the magnetic parameter $B$, while the right panel shows the role of the rotation parameter $a$. The two parameters influence the potential in opposite ways. An increase in $B$ shifts the minimum of $V_{\mathrm{eff}}$ to lower values, whereas an increase in $a$ has the opposite effect. For the same rotation parameter, the minimum of $V_{\mathrm{eff}}$ for the Kerr BH lies above that obtained for the rotating KBR spacetime.

\begin{figure*}
\centering 
\includegraphics[width=\hsize]{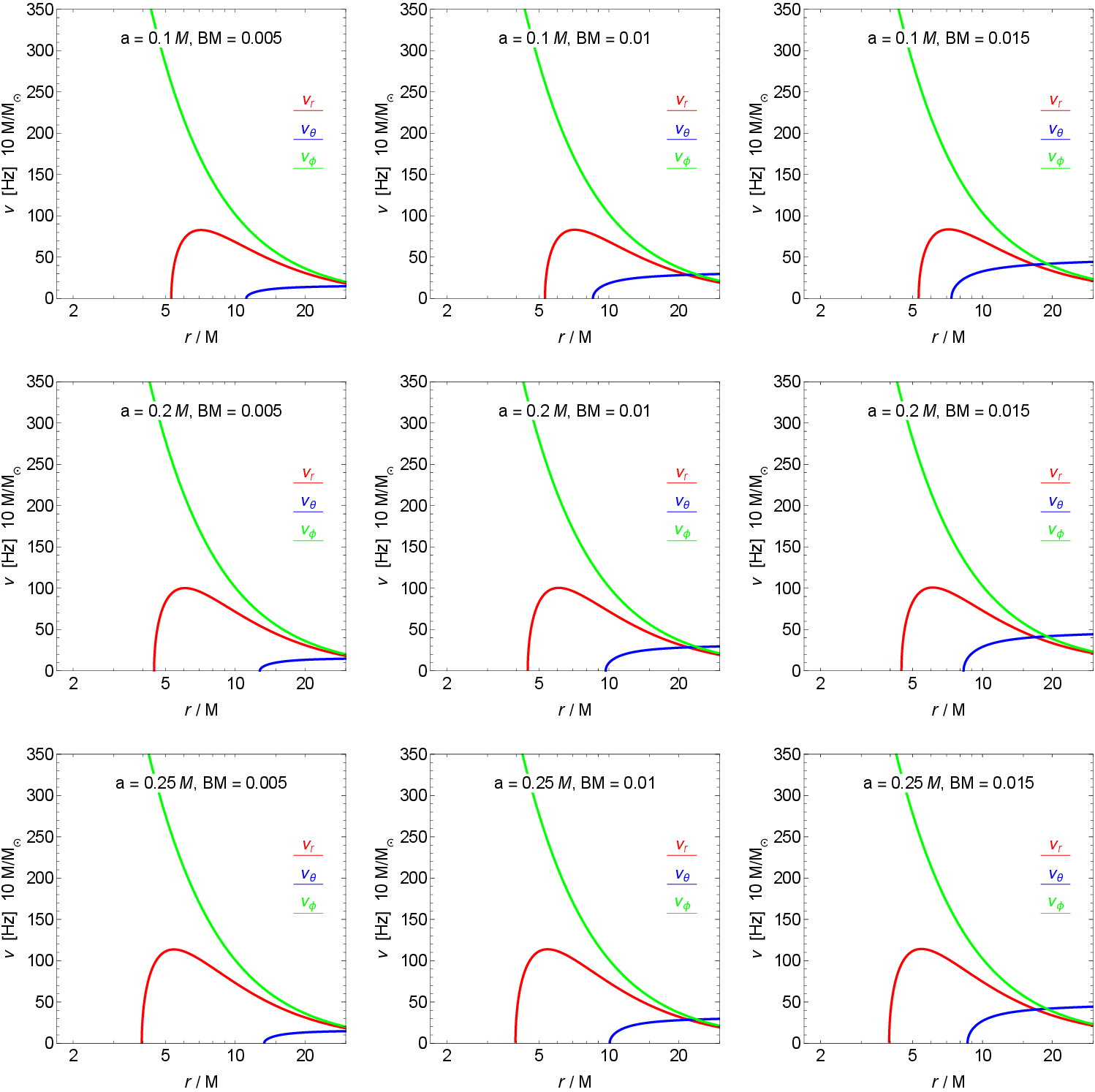}
\caption{Fundamental frequencies of small harmonic oscillations for neutral particles around a rotating KBR BH, measured by a distant observer. The plots show the radial $\nu_r$, latitudinal $\nu_\theta$, and azimuthal $\nu_\phi$ frequencies as functions of radial coordinate $r/M$, for different values of the spin parameter $a$ and magnetic parameter $B$.
}\label{figFRQ}
\end{figure*}

\section{Harmonic oscillations as perturbation of circular orbits} \label{Sec3}

To investigate the oscillatory motion of neutral particles, we perturb the equations of motion near stable circular orbits. When a test particle is slightly displaced from its equilibrium point in a stable circular orbit on the equatorial plane, it undergoes epicyclic motion. The frequencies of these harmonic oscillations, as measured by a local observer, are given by 
\bea\label{Freq-2}
\omega_{r}^{2} &=&  \frac{-1}{2\, g_{rr}} \frac{\partial^{2} V_{\rm eff} (r, \theta)}{\partial r^{2}},\\\label{Freq-3}
\omega_{\theta}^{2} &=& \frac{-1}{2\, g_{\theta \theta}} \frac{\partial^{2} V_{\rm eff}(r, \theta)}{\partial \theta^{2}},\\\label{Freq-4}
\omega_\phi &=& \frac{\d \phi}{\d \tau}.
\eea
The fundamental frequencies $\omega_r$, $\omega_\theta$, and $\omega_\phi$ characterize epicyclic motion around stable circular orbits. In Newtonian gravity, these frequencies are equal and bound orbits are closed. In the Schwarzschild spacetime this degeneracy is removed, and one finds $\omega_r < \omega_\theta = \omega_\phi$. The mismatch between the radial and angular frequencies leads to a shift in the periapsis after each orbit. The effect increases as the orbital radius decreases.

\subsection{Frequencies measured by distant observer} 

The locally measured angular frequencies \(\omega_\alpha\) are provided in Eqs.~(\ref{Freq-2})–(\ref{Freq-4}). However, the angular frequencies measured by a distant static observer, \(\Omega\), are given by
\beq\label{frequencies}
\Omega_{\alpha} = \omega_{\alpha} \frac{\d \tau}{\d t},
\eeq
where $\d \tau/\d t$ is the redshift coefficient. Using Eqs. (\ref{EE}) and (\ref{LL}), we found
\begin{equation}
    \frac{\d t}{\d \tau} = - \frac{E g_{\phi \phi} + L g_{t \phi}}{g_{tt} g_{\phi \phi} - g_{t \phi}^2}.
\end{equation}
If the small amplitude oscillation frequencies measured by a distant observer are expressed in physical units, the corresponding dimensionless frequencies are obtained by scaling with the factor $c^3/(GM)$, where $G$ is the gravitational constant, $c$ is the speed of light, and $M$ is the BH mass.
 Accordingly, the frequencies of neutral particles, as measured by distant observers, are expressed as
\beq\label{nu_rel}
\nu_{j}=\frac{1}{2\pi}\frac{c^{3}}{GM} \, \Omega_{j}[{\rm Hz}].
\eeq
Here, $j \in \{r,\theta,\phi\}$, $\Omega_r$, $\Omega_\theta$, and $\Omega_\phi$ denote the dimensionless angular frequencies measured by a distant observer in the radial, latitudinal, and axial directions, respectively. Figure~\ref{figFRQ} shows the radial dependence of the frequencies $\nu_j$ for small harmonic oscillations of neutral particles around a rotating KBR BH, for different values of the spin parameter $a$ and the magnetic parameter $B$. As $B$ increases, the frequency profiles shift toward smaller radii, approaching the event horizon.
However, the opposite behavior has been observed for the case of BH rotation $a$. The radial profiles move away from the event horizon when BH rotates rapidly.

\section{Quasinormal Modes using WKB method} \label{Sec4}

The analysis of small perturbations around stable circular orbits, discussed in the previous section, describes oscillatory motion of test particles in the vicinity of the BH and provides insight into the dynamical properties of the spacetime. A closely related phenomenon arises when the spacetime itself is perturbed. In this case, the BH responds through damped oscillations known as QNMs, which characterize the relaxation of the geometry toward equilibrium. These modes are described by complex frequencies $\omega=\omega_R-i\omega_I$, where the real part $\omega_R$ determines the oscillation frequency and the imaginary part $\omega_I$ determines the damping rate of the perturbation. Since QNMs dominate the ringdown phase of gravitational waves emitted after BH mergers, they provide a direct probe of the underlying spacetime geometry in the strong-field regime. In the following, we investigate the QNMs of the rotating KBR BH and examine how the presence of the magnetic parameter $B$ modifies the corresponding ringdown spectrum. For more details related to the WKB method used in our analysis, one may refer to \cite{Iyer:1986np, Dias:2022oqm, Konoplya:2019hlu, Konoplya:2011qq, Konoplya:2017wot, Konoplya:2003ii, Yang:2012he}.

The perturbation of a massless scalar field $\Psi$ satisfies the Klein–Gordon equation. 
\begin{equation}
    \Box \Phi = \frac{1}{\sqrt{-g}} \partial_\mu\Big(\sqrt{-g} g^{\mu\nu} \partial_\nu \Phi\Big) = 0.
\end{equation}

Exploiting the stationarity and axial symmetry of the rotating KBR spacetime, the massless Klein–Gordon equation admits the separation of variables. We therefore decompose the scalar field as $\Phi=e^{-i\omega t}e^{i m_l\phi}u_r(r)u_\theta(\theta)$ \cite{Teukolsky:1972my, Luna:2022rql, Yang:2012he, Yang:2021zqy}. 
This allows us to obtain the Schrödinger-like form (see appendix for more details):

\begin{equation}
\frac{d^2 u_r}{dr_*^2} + V^r(r,\omega)\, u_r = 0.
\end{equation}

The effective potential is given explicitly by

\begin{widetext}
    \begin{equation}
V^r(r,\omega) = \frac{(a m_l - \omega \Sigma(r))^2}{\Sigma(r)^2} - \frac{(1+B^2 r^2)\,\Delta(r)}{(r^2+a^2)^2}\Big[\mathcal{A}_{lm_l} + 2 a m_l \omega - a^2\omega^2\Big],
\end{equation}
\end{widetext}

with
$ \Delta(r) = a^2 + r^2 \Big(1 - \frac{B^2 I_2 M^2}{I_1^2}\Big) - \frac{2 I_2 M r}{I_1},
$
and
$
I_1 = 1 - \frac{a^2 B^2}{2}, $  $ I_2 = 1 - a^2 B^2.
$

Thus, the magnetic field parameter $B$ modifies the potential barrier both via the multiplicative factor $(1+B^2 r^2)$ and through the modified horizon function $\Delta(r)$.
Due to the presence of the conformal factor and the magnetic deformation, the Klein–Gordon equation in the rotating KBR background does not admit a compact closed-form exact radial potential. However, in the weak-field regime and within the eikonal approximation, the dynamics is governed by an effective radial potential of Schrödinger type, which captures all leading magnetic corrections as shown above.

QNMs correspond to solutions localized near the peak of the effective potential. Denoting the peak position by $r_0$, the oscillation frequency satisfies

$$
V^r(r_0,\omega_R) = 0, \qquad \left.\frac{\partial V^r}{\partial r}\right|_{r_0,\omega_R} = 0.
$$

From these conditions one can have
\begin{align}\label{v}
\Omega_R &= \frac{ \mu a}{r_0^2+a^2} \pm \frac{\sqrt{\Delta(r_0)}}{r_0^2+a^2}\beta(a \Omega_R) \,, \\
0 &=\frac{\partial}{\partial r}\left[\frac{\Omega_R (r^2+a^2)- \mu a}{\sqrt{\Delta(r)}} \right]_{r=r_0} \,, \label{dvdrG}
\end{align}
where $\mu = \frac{m_l}{l+1/2}$, $\Omega_R = (l+1/2) \omega_R$ and $\beta(a \Omega_R) = \sqrt{\frac{1}{2} a \Omega _R \left(a \left(\mu ^2+1\right) \Omega _R-4 \mu \right)+1}$.
From condition (\ref{dvdrG}), one gets the oscillation frequency $\omega_R$.

The damping rate is obtained from the curvature of the potential at its maximum, by using the following relation:

\begin{equation}
\omega_I = -\Big(n+\tfrac{1}{2}\Big)\, \frac{\sqrt{2\big(d^2 V^r/dr_*^2\big)_{r_0,\omega_R}}}{\big(\partial V^r/\partial \omega\big)_{r_0,\omega_R}}, \quad n=0,1,2,\ldots.
\end{equation}

\begin{table}[h]
\centering
\caption{QNM frequencies $\omega$ and corresponding percentage deviations from Kerr BH for different multipole numbers $l$ with model parameters $M=1$, $n=0$, $m_l = 1$, $B=0.005$ and $a=0.1$.}
\label{tab:qnm_values}
\begin{tabular}{ccc}
\hline
$l$ & $\omega = \omega_R - i \; \omega_I$ & Deviation from Kerr BH \\ 
\hline
1 & $0.296459 - 0.101388\, i$ & 0.00347758 \% \\
2 & $0.488953 - 0.0991498\, i$ & 0.00364738 \% \\
3 & $0.68149 - 0.0982327\, i$ & 0.00368803 \% \\
4 & $0.874041 - 0.0977341\, i$ & 0.00371099 \% \\
5 & $1.0666 -0.0974207 \, i$ & 0.00344586 \% \\
6 & $1.25916 -0.0972056 \, i$ & 0.0035555 \% \\
.. & ... & ... \\
10 & $2.02942 - 0.0967597\, i$ & 0.00339004 \% \\
\hline
\end{tabular}
\end{table}
We have shown the QNMs for different multipole numbers in Table \ref{tab:qnm_values} along with the corresponding deviations of QNMs from the Kerr BH. It is worth emphasizing that the deviations reported in Table~\ref{tab:qnm_values} arise solely from the magnetic-field deformation of the rotating BH geometry. In the limit $B \to 0$, the metric smoothly reduces to the Kerr spacetime and the corresponding QNM spectrum reproduces the well-known Kerr results. However, when the magnetic parameter $B$ is nonzero, the effective radial potential governing scalar perturbations is modified through the factor $(1+B^2 r^2)$ and the magnetically corrected horizon function $\Delta(r)$. These corrections shift both the location and the curvature of the potential maximum, which in turn alters the oscillation frequency $\omega_R$ and the damping rate $\omega_I$. The percentage deviations shown in Table~\ref{tab:qnm_values} therefore provide a direct quantitative measure of how the external magnetic field modifies the ringdown spectrum relative to the Kerr case. Although the deviations remain small in the weak-field regime ($B^2 \ll 1$), they systematically increase with the magnetic parameter and multipole number, indicating that the magnetic field introduces measurable corrections to the quasinormal spectrum of the rotating BH.

Thus the QNM spectrum of the KBR BH changes with the magnetic parameter $B$. Both the oscillation frequencies and the decay rates are shifted compared to Kerr. When $B \to 0$, the Kerr QNMs are recovered, which shows that  the method is consistent. These shifts could appear in the ringdown signal of gravitational waves. Along with BH shadow measurements, they may help probe strong electromagnetic fields around astrophysical BHs.

\begin{figure*}[htbp]
	\centerline{
		\includegraphics[width=85mm,height=65mm]{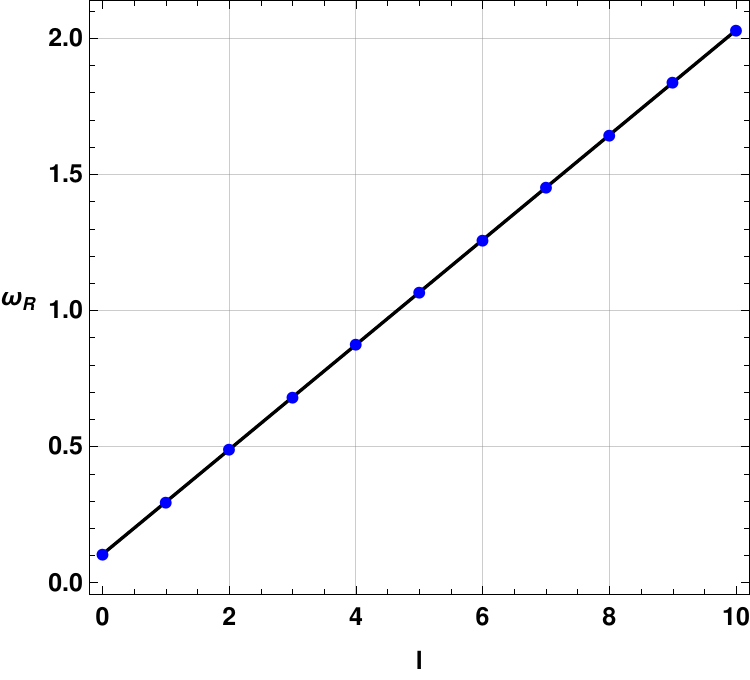}\hspace{0.5cm}
		\includegraphics[width=85mm,height=65mm]{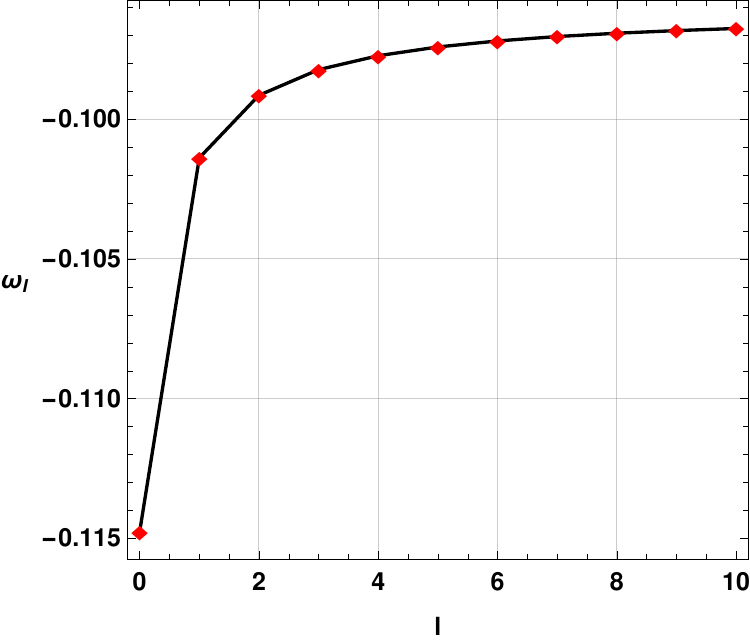}} \vspace{-0.2cm}
	\caption{Dependence of the fundamental QNM frequencies on the multipole number $l$ for a massless scalar perturbation in the KBR spacetime, assuming a weak magnetic field $B = 0.0005$. The left panel shows the real part $\omega_R$, the right panel shows the imaginary part $\omega_I$.}
	\label{QNMs00}
\end{figure*}

\begin{figure*}[htbp]
	\centerline{
		\includegraphics[width=85mm,height=65mm]{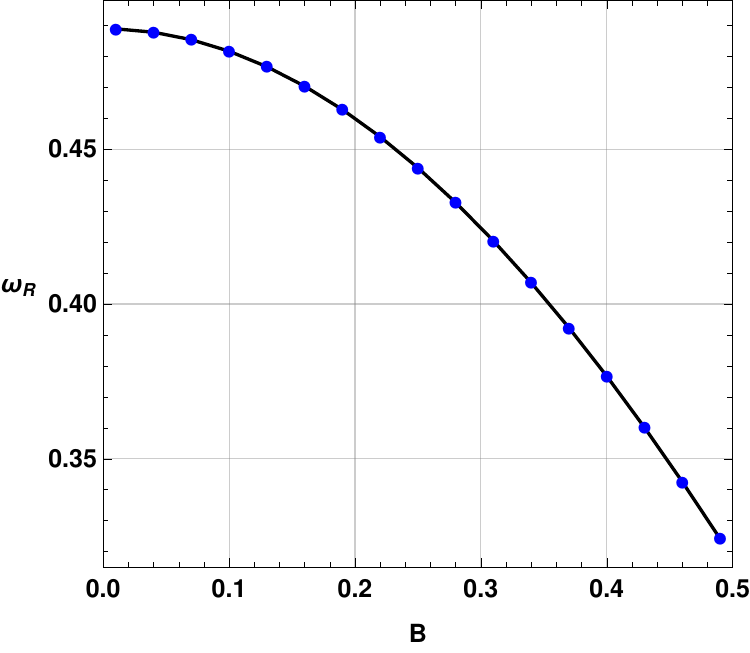}\hspace{0.5cm}
		\includegraphics[width=85mm,height=65mm]{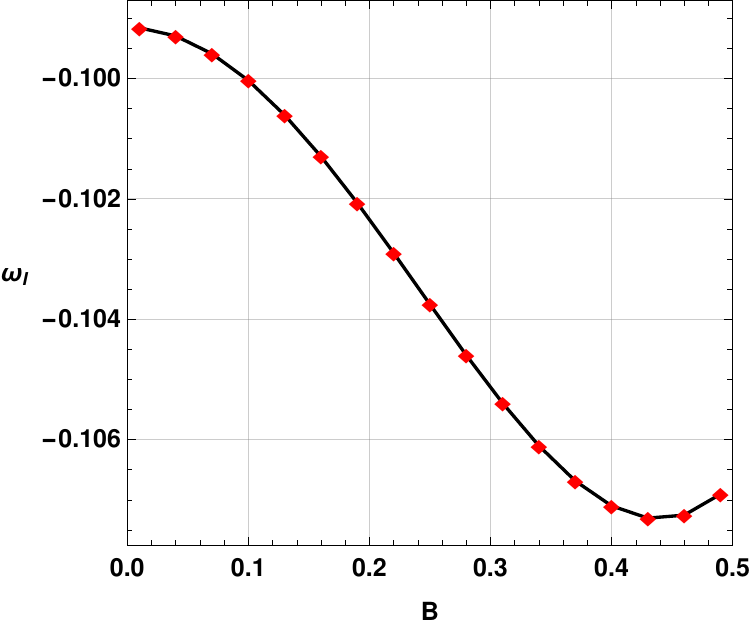}} \vspace{-0.2cm}
	\caption{Variation of the fundamental QNM frequencies with the magnetic field parameter $B$ for a scalar perturbation with $a = 0.1$, $l = 2$, $m = 1$, $n = 0$, and $M = 1$. Left shows the real part $\omega_R$ and  right shows the imaginary part $\omega_I$.}
	\label{QNMs01}
\end{figure*}

\begin{figure*}[htbp]
	\centerline{
		\includegraphics[width=85mm,height=65mm]{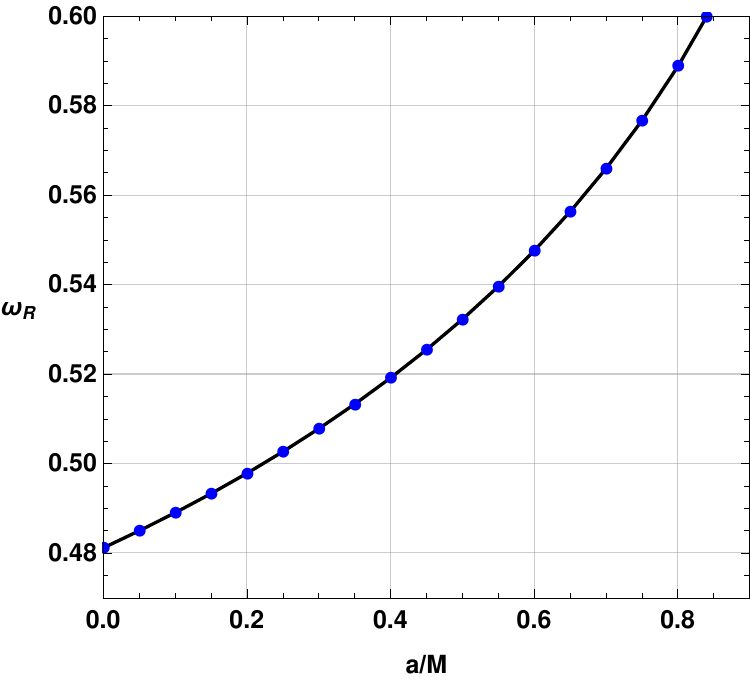}\hspace{0.5cm}
		\includegraphics[width=85mm,height=65mm]{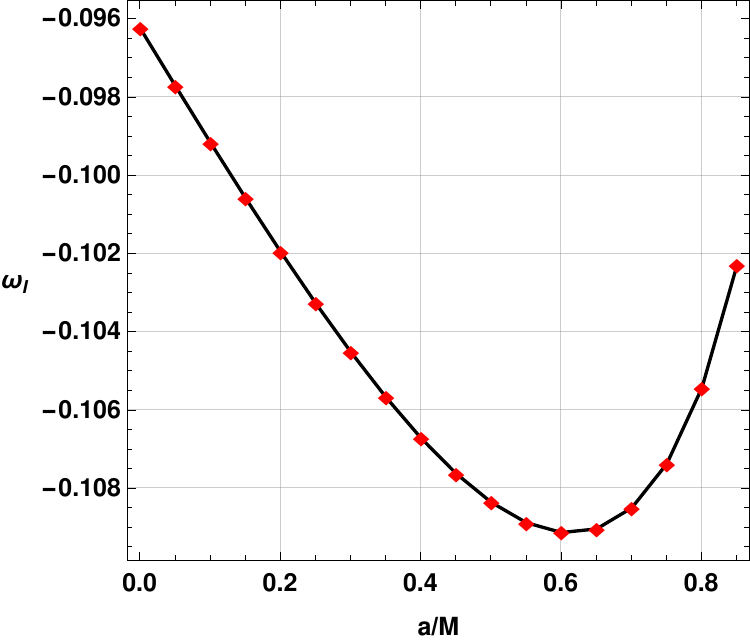}} \vspace{-0.2cm}
	\caption{Variation of the fundamental QNM frequencies with the spin parameter $a$ for a scalar perturbation with $B = 0.005$, $l = 2$, $m = 1$, $n = 0$, and $M = 1$. The left panel depicts the real part $\omega_R$a and the right panel depicts the imaginary part $\omega_I$.}
	\label{QNMs02}
\end{figure*}

In Fig.~\ref{QNMs00}, we have shown the variation of the fundamental quasinormal
mode frequencies with respect to the multipole number $l$, assuming a weak
magnetic field ($B=0.0005$). One observes that the real part of the frequency
$\omega_R$ increases almost linearly with $l$. This behavior is a generic
feature of BH perturbations in the eikonal regime and reflects the
dominance of angular momentum in shaping the effective potential barrier.
Physically, higher multipole modes correspond to perturbations that are more
strongly localized near the unstable photon orbit, resulting in faster
oscillations of the ringdown signal. The near-linear scaling of $\omega_R$ with
$l$ is consistent with the WKB prediction $\omega_R \sim l\,\Omega_c$, where
$\Omega_c$ is the angular frequency of the unstable null circular orbit.

The imaginary part $\omega_I$, which determines the damping rate of the modes,
exhibits a much weaker dependence on $l$ and approaches a nearly constant value
for large $l$. This indicates that the decay time of the fundamental mode is
primarily governed by the curvature of the effective potential at its maximum,
which becomes increasingly insensitive to $l$ in the eikonal limit. As a
result, higher-$l$ modes oscillate more rapidly but decay on comparable
timescales, a characteristic feature of quasinormal ringing.

Figure~\ref{QNMs01} illustrates the dependence of the quasinormal frequencies on
the magnetic field parameter $B$. The real part $\omega_R$ decreases monotonically
as $B$ increases, indicating that the presence of an external magnetic field
effectively lowers the oscillation frequency of scalar perturbations. This can
be understood from the form of the effective potential, where the factor
$(1+B^2 r^2)$ modifies the height and width of the potential barrier. An
increasing magnetic field weakens the restoring force experienced by the scalar
wave, leading to slower oscillations. 

The imaginary part $\omega_I$ becomes more negative with increasing $B$,
signaling enhanced damping of the QNMs. Physically, the magnetic
field strengthens the coupling between the scalar field and the spacetime
geometry, increasing the leakage of energy into the BH horizon and thus
shortening the ringdown lifetime. The smooth and monotonic variation of both
$\omega_R$ and $\omega_I$ confirms the stability of the scalar perturbations in
the considered parameter range.

In Fig.~\ref{QNMs02}, we present the variation of the quasinormal frequencies with
respect to the rotation parameter $a$. The real part $\omega_R$ increases with
$a$, reflecting the influence of frame dragging in a rotating spacetime. As the
rotation parameter grows, the effective potential peak shifts closer to the
BH and the angular velocity of the unstable photon orbit increases,
leading to higher oscillation frequencies. This behavior is consistent with
well-known results for Kerr BHs and demonstrates that the rotating
KBR geometry preserves the essential rotational features of
the Kerr spacetime.

The imaginary part $\omega_I$ shows a nonmonotonic dependence on $a$, indicating a
subtle competition between rotational effects and the magnetic-field-induced
modifications of the potential barrier. For moderate rotation, the enhanced
frame dragging increases the damping rate, while for larger values of $a$ the
potential barrier becomes broader, partially suppressing energy dissipation and
leading to a slower decay of the modes. This interplay highlights the rich
structure of QNM spectra in rotating magnetized BH
backgrounds.

Overall, the results demonstrate that while the multipole number $l$ primarily
controls the oscillation frequency, the magnetic field parameter $B$ and the
rotation parameter $a$ significantly influence both the oscillatory and damping
properties of QNMs. These effects may leave characteristic imprints
on the gravitational-wave ringdown signal and could, in principle, serve as
probes of magnetic fields and rotation in strong-gravity environments.

\begin{figure*}[htbp]
	\centerline{
		\includegraphics[width=85mm,height=65mm]{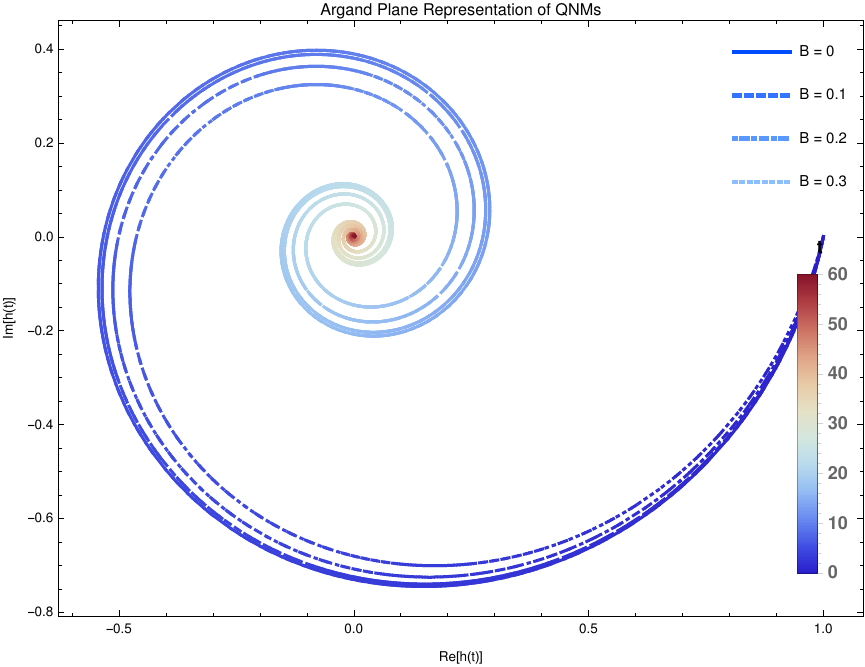}\hspace{0.5cm}
		\includegraphics[width=85mm,height=65mm]{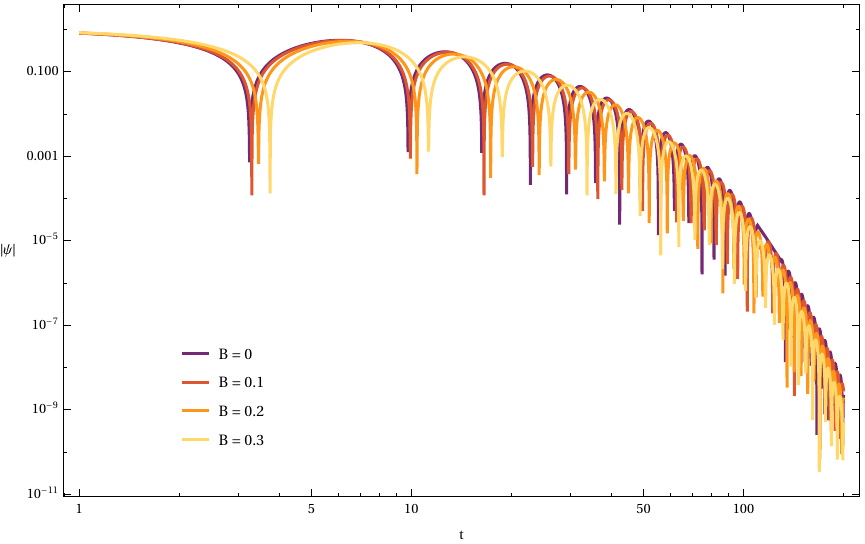}} \vspace{-0.2cm}
	\caption{Argand-plane (left) and time-domain (right) representation of the fundamental QNMs for a scalar perturbation with $a = 0.01$, $l = 2$, $m = 1$, $n = 0$, and $M = 1$, showing variation with the magnetic field parameter $B$.}
	\label{QNMs03}
\end{figure*}

\begin{figure*}[htbp]
	\centerline{
		\includegraphics[width=85mm,height=65mm]{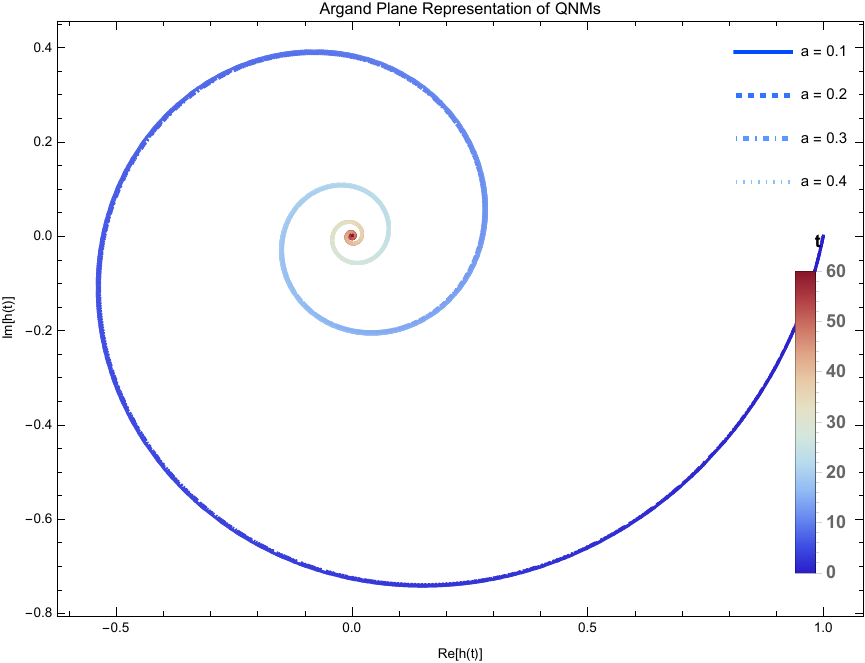}\hspace{0.5cm}
		\includegraphics[width=85mm,height=65mm]{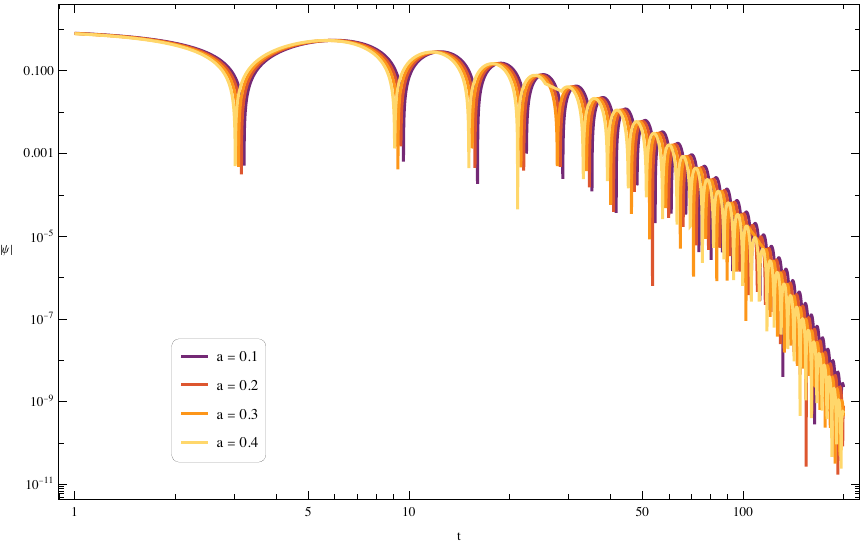}} \vspace{-0.2cm}
	\caption{Argand-plane (left) and time-domain (right) representation of the fundamental QNMs for a scalar perturbation in the KBR spacetime with $B = 0.005$, $l = 2$, $m = 1$, $n = 0$, and $M = 1$. Different curves correspond to different values of the spin parameter $a$.}
	\label{QNMs04}
\end{figure*}

\begin{figure*}[htbp]
	\centerline{
		\includegraphics[width=85mm,height=65mm]{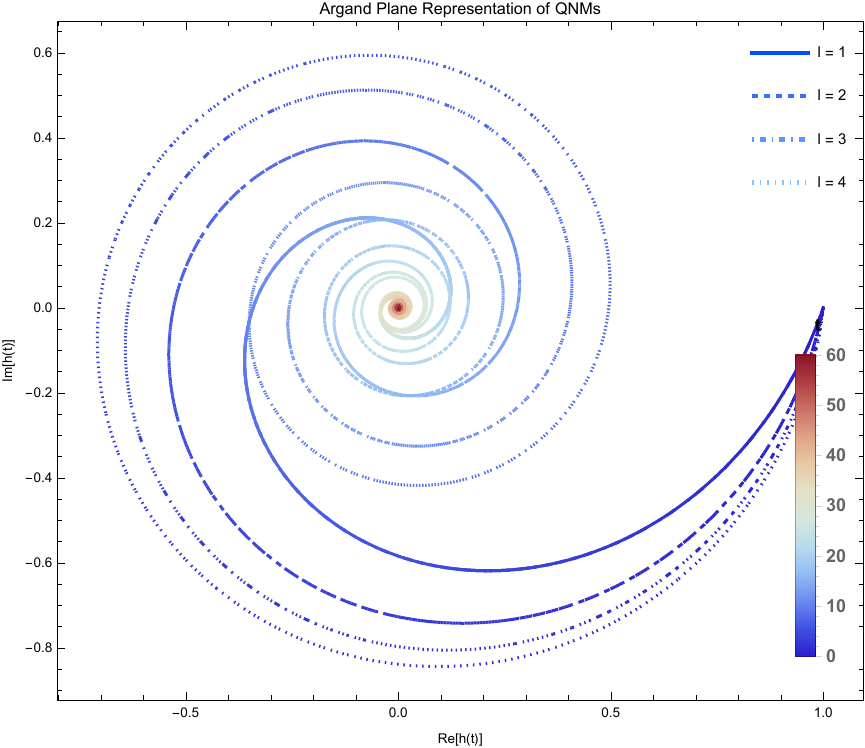}\hspace{0.5cm}
		\includegraphics[width=85mm,height=65mm]{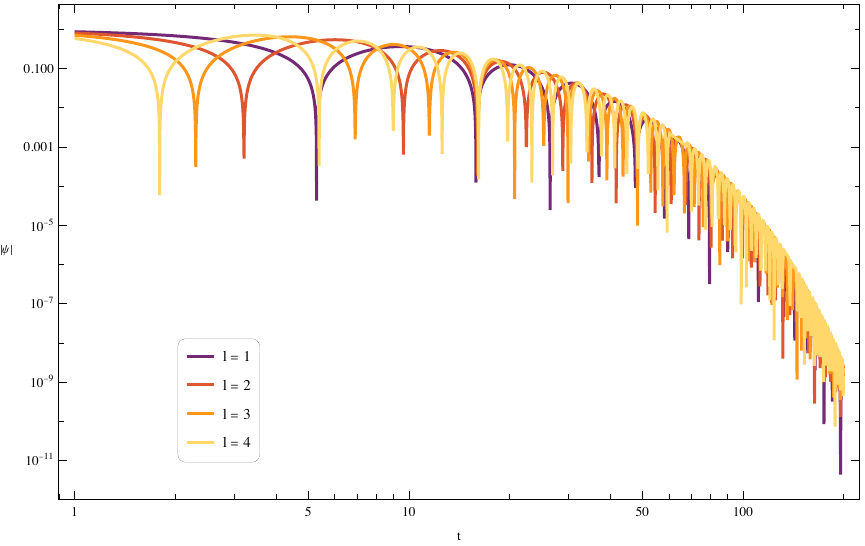}} \vspace{-0.2cm}
	\caption{Argand-plane (left) and time-domain (right) representation of the fundamental QNMs for a scalar perturbation with $B = 0.005$, $a = 0.1$, $m = 1$, $n = 0$, and $M = 1$, showing variation with the multipoles number $l$}
	\label{QNMs05}
\end{figure*}

Figs.~\ref{QNMs03}, \ref{QNMs04}, and \ref{QNMs05} provide a complementary and more intuitive
representation of the QNM spectrum by displaying the evolution of
the modes both in the complex frequency (Argand) plane and in the time domain.
These figures extend the frequency-domain analysis presented earlier in
Figs.~\ref{QNMs00}–\ref{QNMs02} and allow for a direct visualization of how the
magnetic field parameter $B$, the spin parameter $a$, and the multipole number
$l$ influence the oscillatory and damping properties of scalar perturbations.

Figs.~\ref{QNMs03}–\ref{QNMs05} provide a unified geometric and
dynamical interpretation of the QNM spectrum. While the earlier
figures quantified the parametric dependence of the real and imaginary parts of
the QNM frequencies, the Argand-plane trajectories and time-domain waveforms
presented here offer a more direct visualization of how magnetic fields,
rotation, and angular momentum shape the ringdown signal. These results reinforce
the conclusion that the magnetic field primarily enhances damping, the spin
parameter increases the oscillation frequency through frame-dragging effects,
and the multipole number controls the eikonal scaling of the modes, all while
preserving the overall stability of the rotating KBR BH.

\section{Numerical Evolution of BHL Accretion in the KBR Spacetime}
\label{Sec5}

In this section, we focus on the main physical trends revealed by the numerical simulations. In particular, we examine how the BH spin and the magnetic curvature parameter modify the morphology of the accretion flow, the formation of shock-cone and toroidal structures, and the resulting variability patterns. More detailed case-by-case descriptions are kept concise here in order to emphasize the dominant numerical results.

 The BHL accretion mechanism provides a natural framework for understanding how the surrounding spacetime of a BH shapes large-scale flow morphology and governs the time-dependent behavior of astrophysical accretion, ultimately allowing numerical results to be connected with observations. In particular, BHL accretion is crucial for interpreting wind-fed X-ray binaries, in which matter is captured from the stellar companion by gravitational focusing. The above studies have already shown that the test-particle dynamics and epicyclic frequencies around KBR BHs deviate significantly from their Kerr counterparts, indicating that the magnetic curvature parameter introduces substantial modifications in the strong-gravity interaction between matter and spacetime. We aim to both support the theoretical predictions and assess how KBR gravity may leave observable imprints on accreting systems. As in our earlier analyses for Kerr, Schwarzschild, and other modified gravity models \cite{Donmez2014MNRAS,Koyuncu:2014MPLA,Donmez2024Universe,Donmez2024JCAP,Donmez2024MPLA,Donmez2025JHEAp,Donmez2025EPJC,Mustafa2025JCAP}, we now extend our investigation to the full hydrodynamic response of matter in the KBR geometry. To this end, we solve the general relativistic hydrodynamics equations using high resolution numerical methods \cite{Donmez2004ASS,Donmez2006AMC,Donmez2017MPLA,DONMEZ2024PDU}. In this section, we therefore investigate how the KBR magnetic curvature modifies the classical Kerr shock cone, triggers flip-flop instabilities\cite{Blondin:2009bp,Donmez2012MNRAS}, and induces transitions into toroidal configurations, revealing a rich dynamical landscape that would be absent in standard GR and highlighting distinctive dynamical features of magnetically modified BH spacetimes.

\subsection{Shock-Cone Formation, Flip-Flop Instability, and Toroidal Structures} 
 
\begin{figure*}[!ht]
  \vspace{1cm}
  \center
  \psfig{file=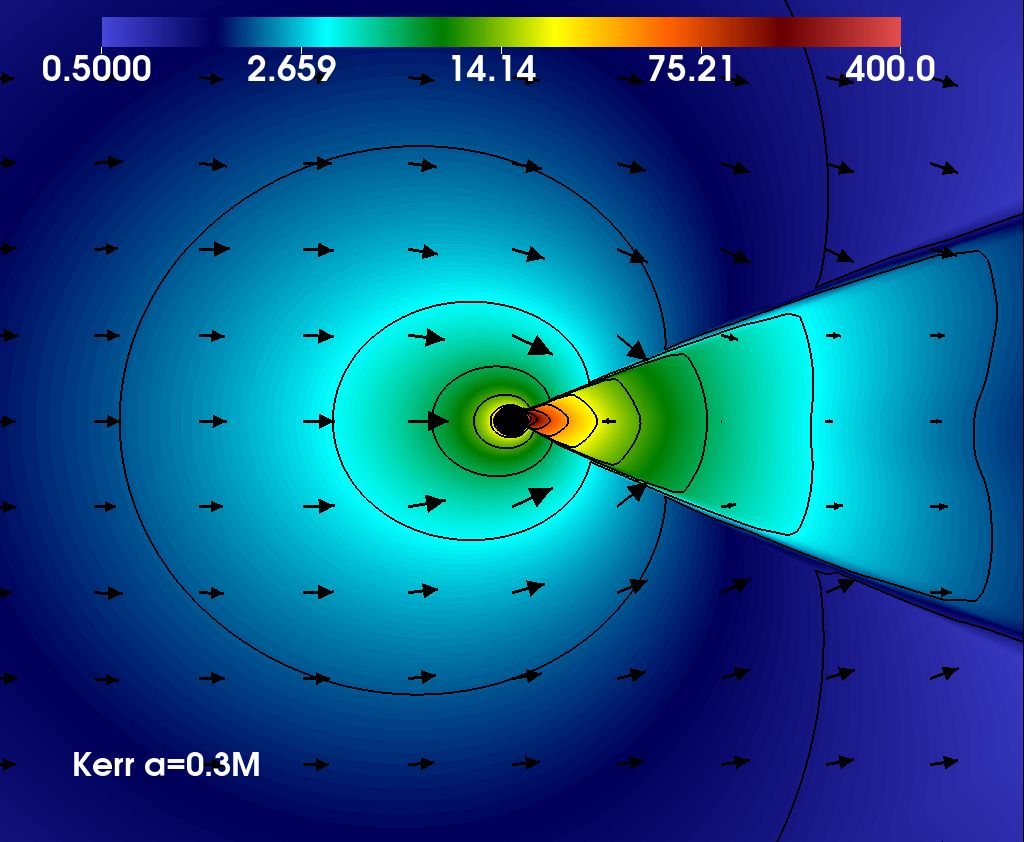,width=5.0cm,height=5.0cm}
   \psfig{file=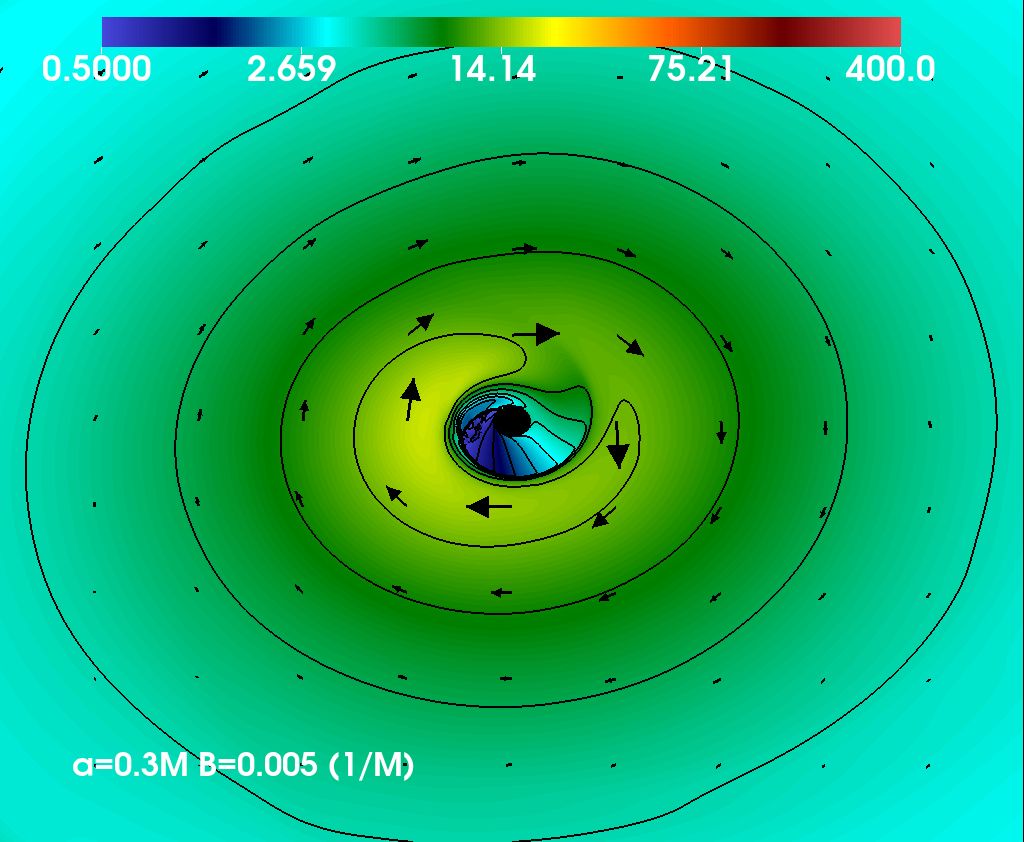,width=5.0cm,height=5.0cm}
     \psfig{file=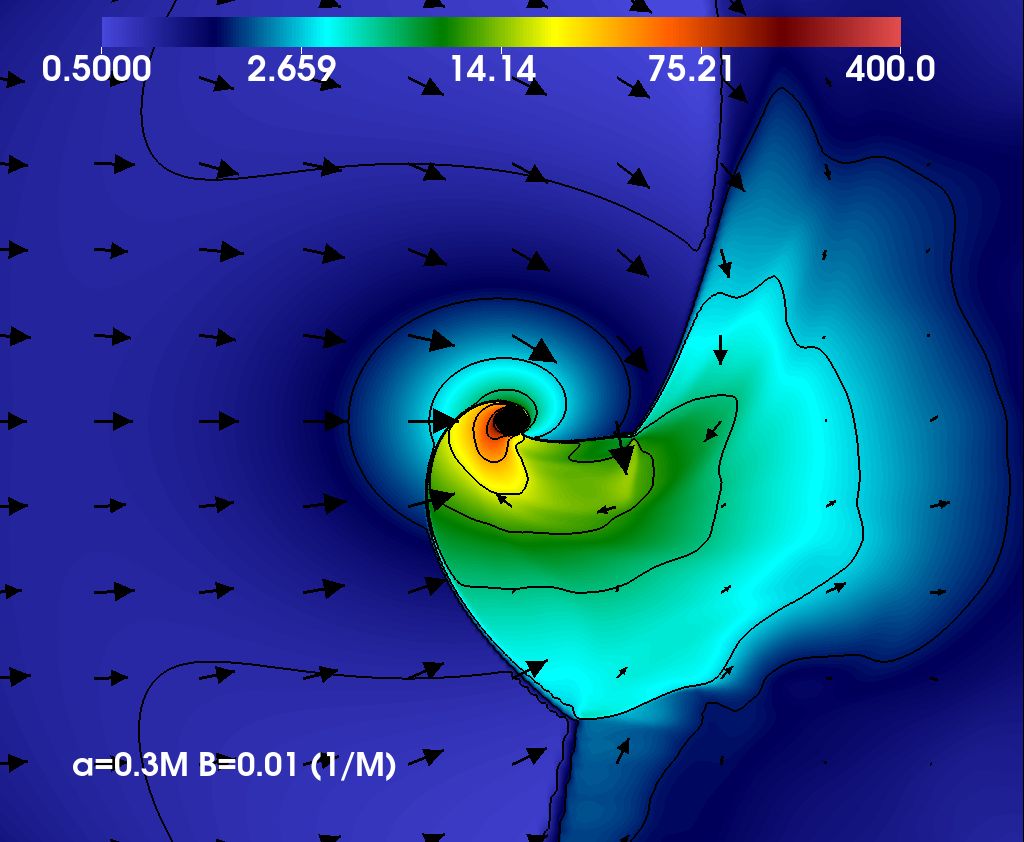,width=5.0cm,height=5.0cm}\\
  \psfig{file=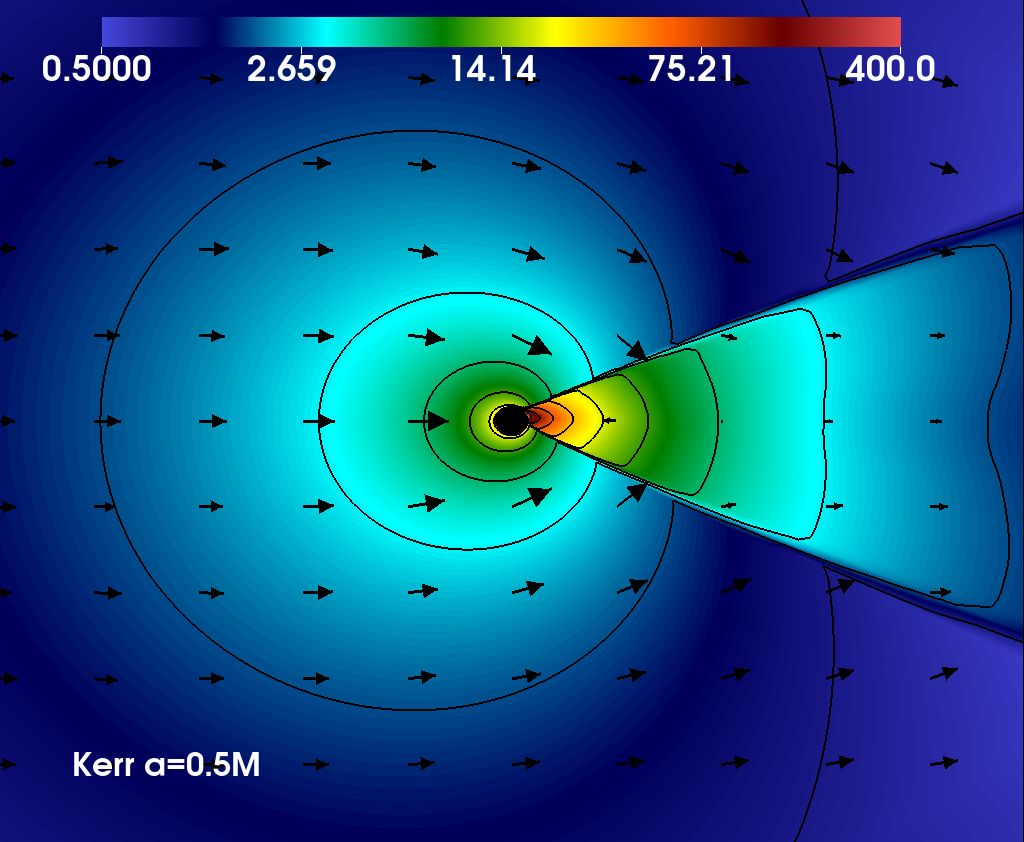,width=5.0cm,height=5.0cm}
   \psfig{file=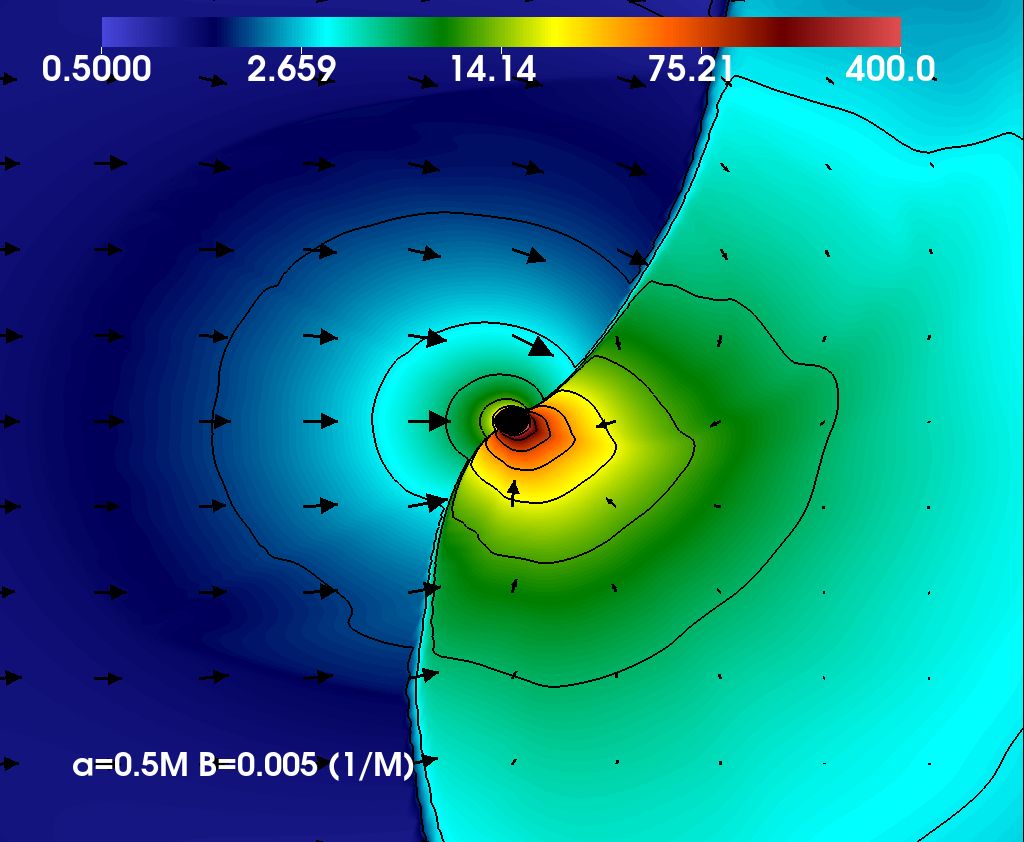,width=5.0cm,height=5.0cm}
     \psfig{file=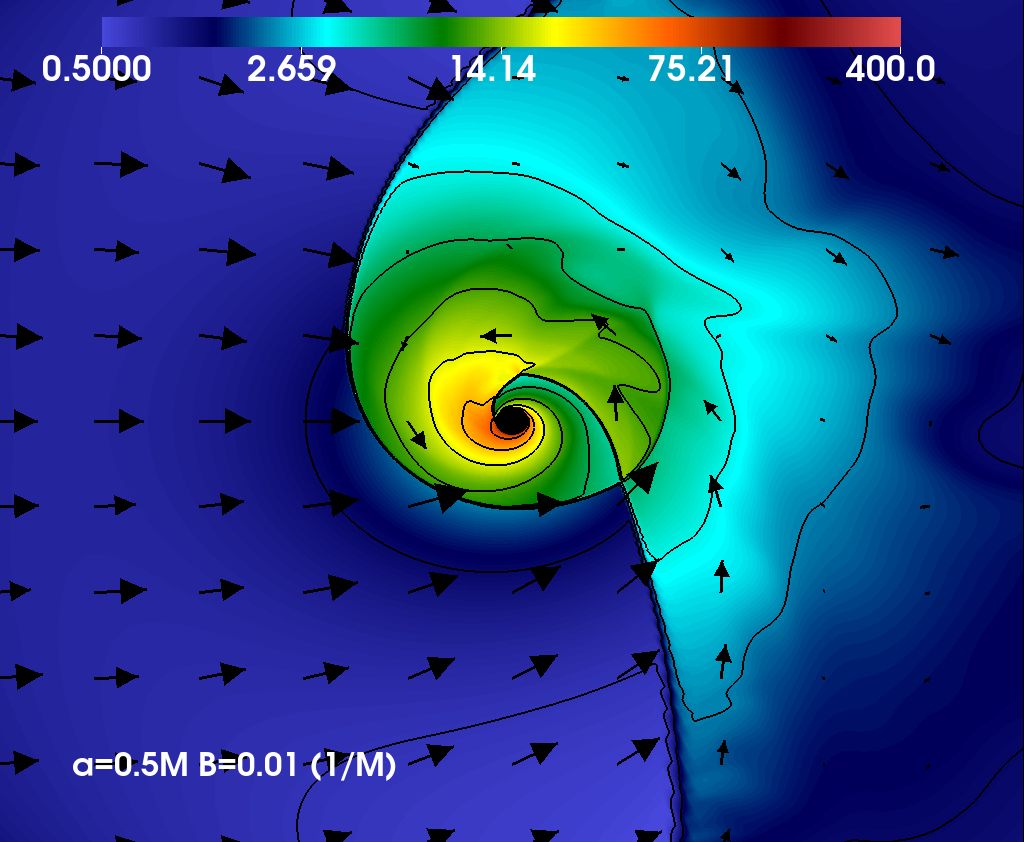,width=5.0cm,height=5.0cm}\\
  \psfig{file=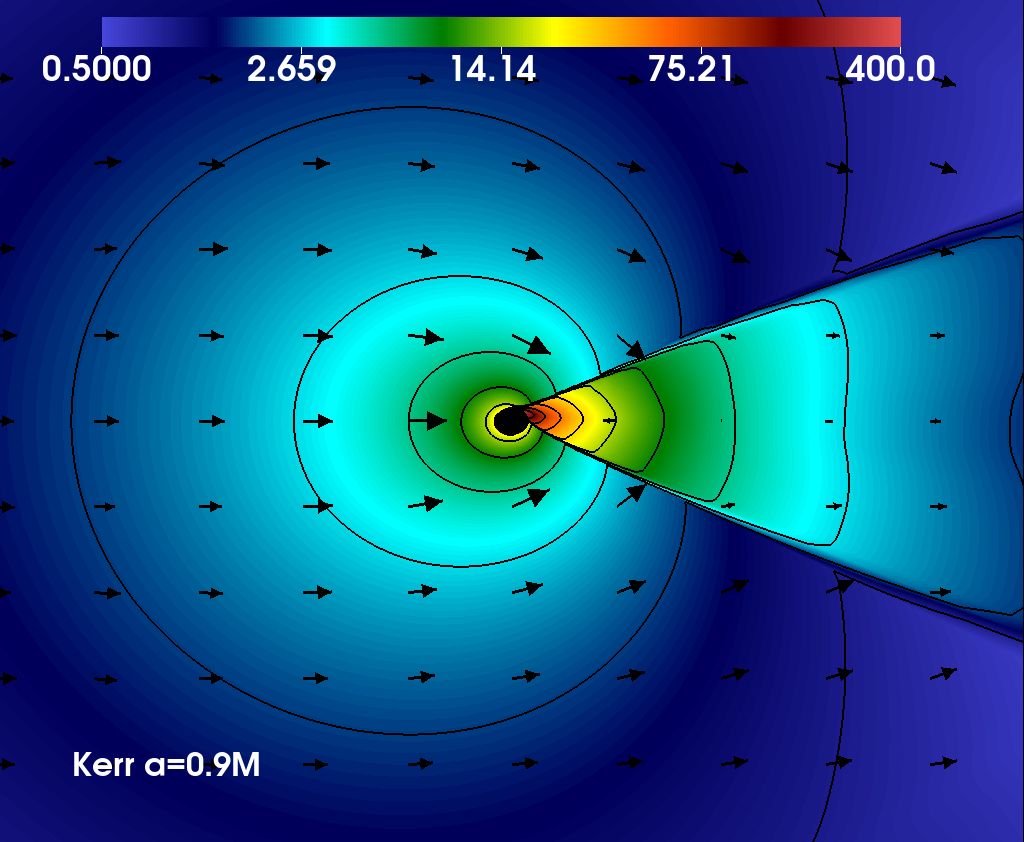,width=5.0cm,height=5.0cm}
   \psfig{file=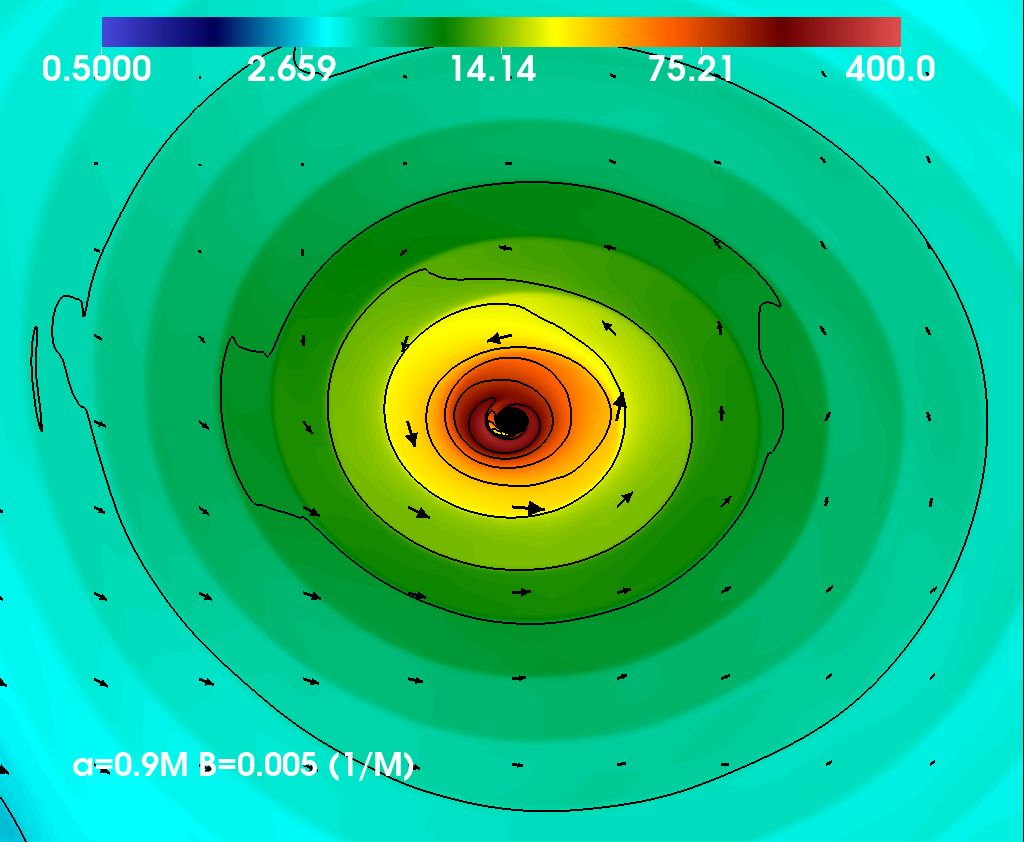,width=5.0cm,height=5.0cm}
     \psfig{file=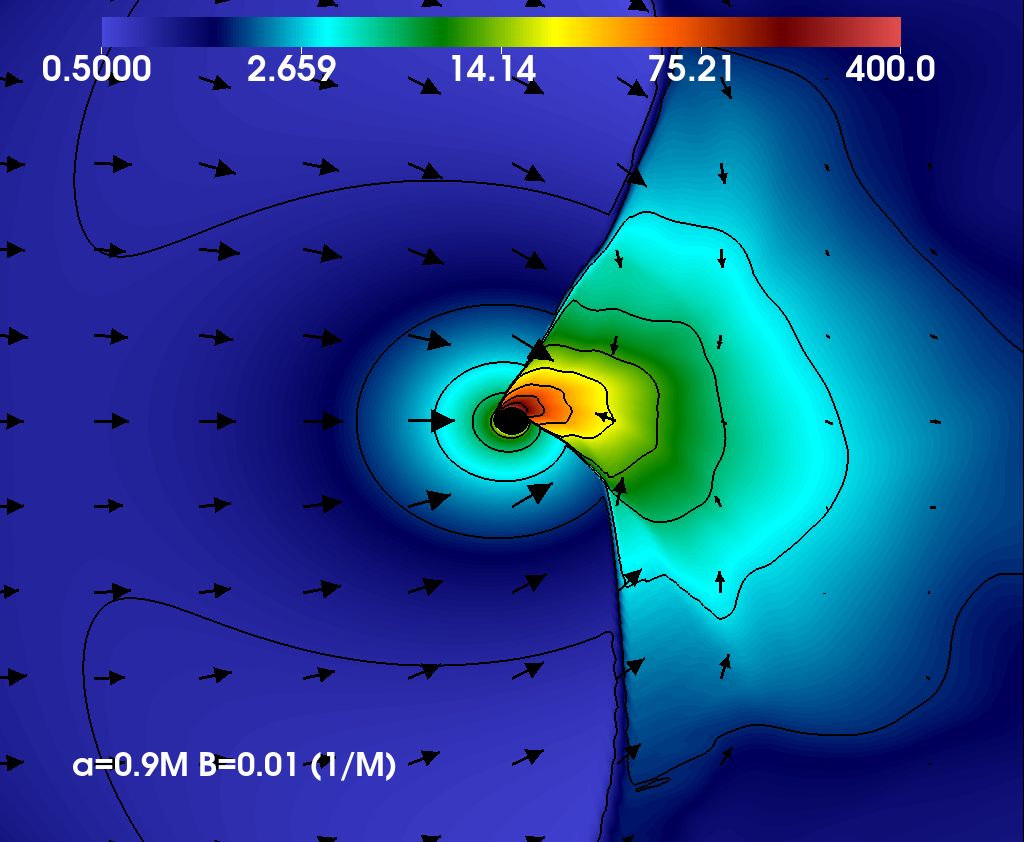,width=5.0cm,height=5.0cm}  
     \caption{ The structure of the plasma and shock cone formed around Kerr and KBR BHs with different $B$ values for various BH spin parameters is shown at approximately $t=55000M$. In the upper row, the slowly rotating BH ($a=0.3M$) case is modeled,  in the middle row, the mildly rotating BH ($a=0.5M$) case is presented, and in the bottom row, the physical mechanism formed around the rapidly rotating BH ($a=0.9M$) is illustrated. The overall morphology of the flow is found to vary systematically with the magnetic parameter $B$. It is zoomed to the range $[x_{\min}, y_{\min}] = [-70M, -70M]$ and $[x_{\max}, y_{\max}] = [70M, 70M]$.
}
\vspace{1cm}
\label{color_den_mix}
\end{figure*}

To understand how KBR gravity influences the shock cone formed in the BHL accretion mechanism, Fig.\ref{color_den_mix} presents a comparative view of the accretion structures formed around the Kerr and KBR BHs for different magnetic parameters $B$ and spin parameters $a$. Each panel shows the rest-mass density in both color and contour form, together with the velocity-field arrows that indicate the flow direction and highlight the dynamical response of the matter to the underlying gravitational field. All snapshots correspond to late-time behavior, taken long after the initial shock-cone formation phase in $t \approx 3000M$. Specifically, the morphologies shown here represent the configurations obtained in $t \sim 45000M$-$55000M$. By systematically varying $a$ and $B$ in our numerical simulations, we reveal how the accretion geometry responds to gravitational focusing, spacetime curvature, and the magnetic curvature term inherent in the KBR spacetime.

In the upper row of Fig.\ref{color_den_mix}, corresponding to the slowly rotating BH with $a = 0.3M$, the Kerr geometry ($B = 0$) produces the classical BHL shock cone in the downstream region which is  a smooth, axis aligned structure whose stability arises from the balance between gravitational focusing and the momentum of the inflowing wind. For KBR BH, even a small magnetic parameter ($B = 0.005\,M^{-1}$) noticeably alters the flow structure. The shock cone becomes asymmetric, and vortical sub-structures appear within it. The downstream region loses its axisymmetry due to changes in the effective gravitational potential caused by the magnetic curvature term.
 In the next stage of evolution for $B = 0.005(1/M)$, a toroidal structure forms around the BH. Our numerical results show that the transition from a flip-flop oscillating shock cone to a toroidal configuration is a natural scenario for accretion onto a KBR BH, and this behavior depends sensitively on both time and magnetic parameter $B$. In the right panel ($B = 0.01(1/M)$), the deformation becomes stronger and the shock cone is replaced by a highly unstable swirling configuration. The matter no longer remains confined to a downstream cone. Instead, it forms a rapidly oscillating spiral shock structure. This behavior is consistent with theoretical expectations that the combined effects of spin and the magnetic curvature term destabilize the classical Kerr shock cone.

The middle row of Figure~12 shows the mildly rotating BH with $a = 0.5M$. For Kerr, the shock cone becomes slightly broader than in the $a = 0.3M$ case, but it remains coherent and well defined. In contrast, both $B = 0.005 (1/M)$ and $B = 0.01 (1/M)$ in the KBR geometry lead to strong flip--flop oscillations. Nearly all of the structural changes observed in the $a = 0.3M$ KBR cases reappear here, demonstrating that the combined action of spin and the magnetic parameter produces similar morphological responses across these two spin values. Although physical structures may appear visually distinct between $a = 0.3M$ and $a = 0.5M$, the underlying dynamical behavior is governed by the same interplay between the angular momentum and the magnetic curvature term.

The bottom row of Fig.~\ref{color_den_mix} shows the strongest effects for the rapidly rotating BH with $a = 0.9M$. It differs from lower-spin cases due to stronger curvature and gravitational focusing, but the overall evolution is similar. The classical Kerr shock cone is destroyed by finite $B$. Depending on time, spin, and $B$, the system forms either a flip-flop shock cone or a dense toroidal structure near the horizon. In KBR cases, the matter density near the horizon increases, and the toroidal structure lies closer for model $a = 0.9M$ than for model $a = 0.3M$, providing ideal conditions to study accretion in strong gravitational fields.

\begin{figure*}[!ht]
  \vspace{1cm}
  \center
  \psfig{file=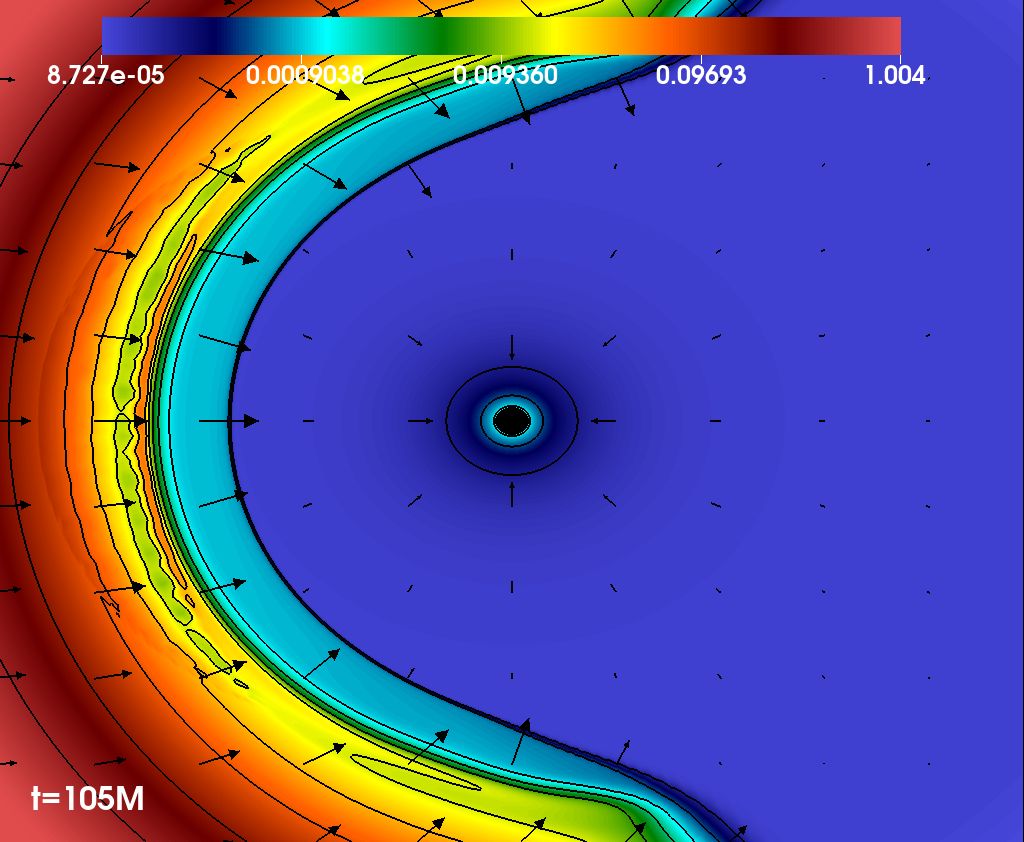,width=4.0cm,height=4.5cm}
  \psfig{file=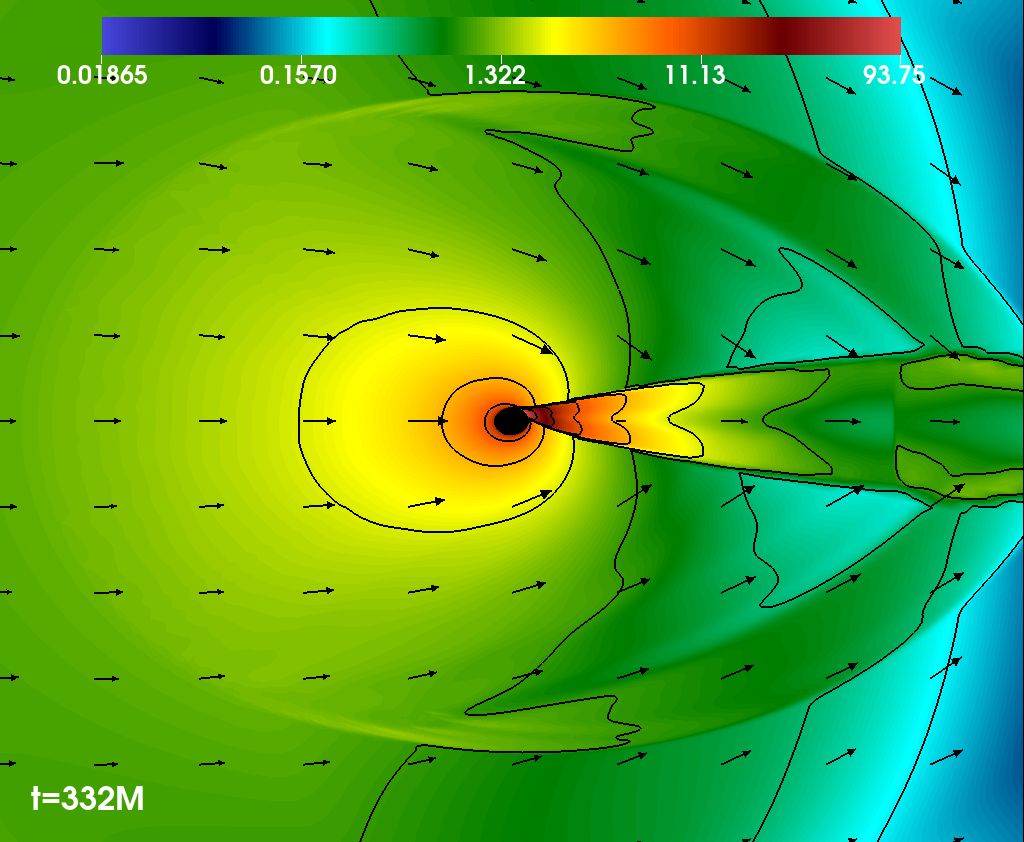,width=4.0cm,height=4.5cm}
  \psfig{file=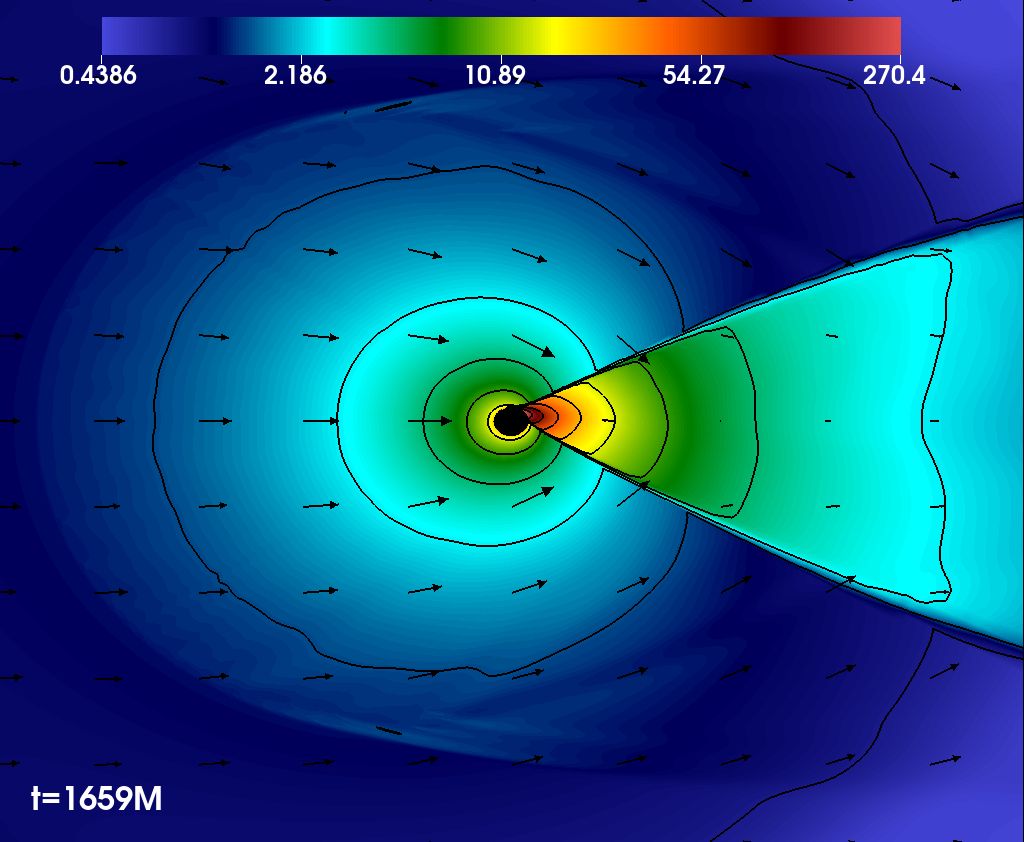,width=4.0cm,height=4.5cm}
  \psfig{file=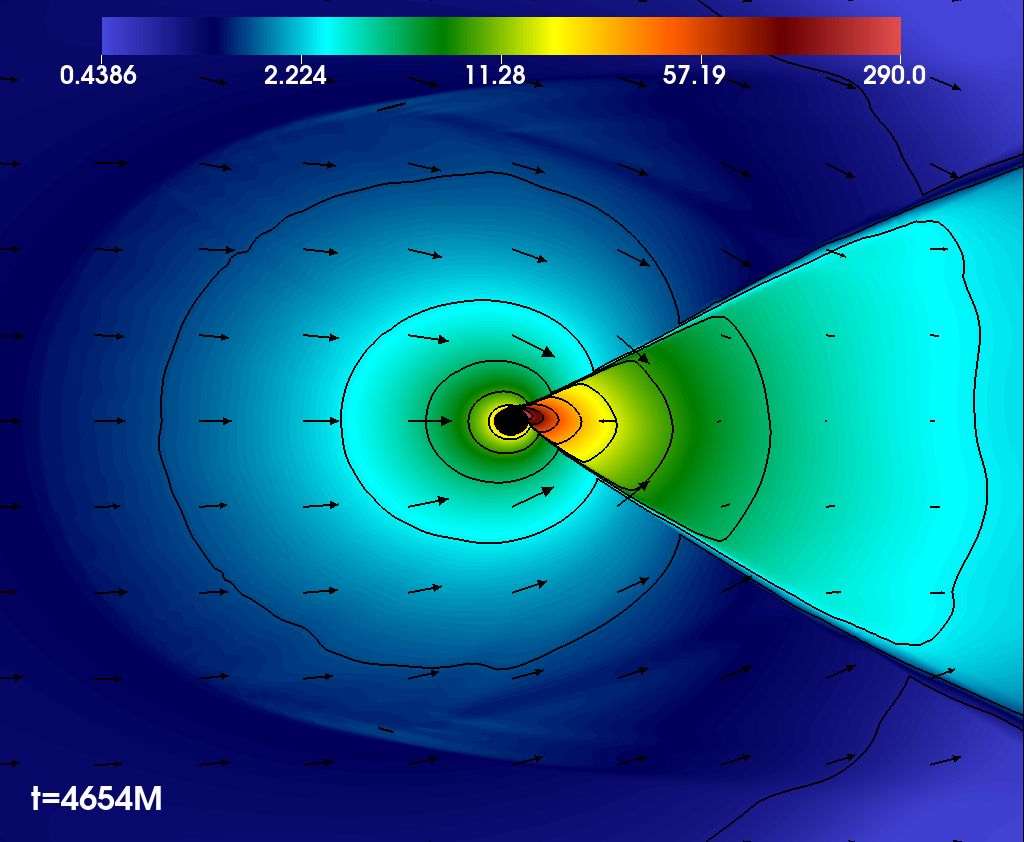,width=4.0cm,height=4.5cm}\\
  \psfig{file=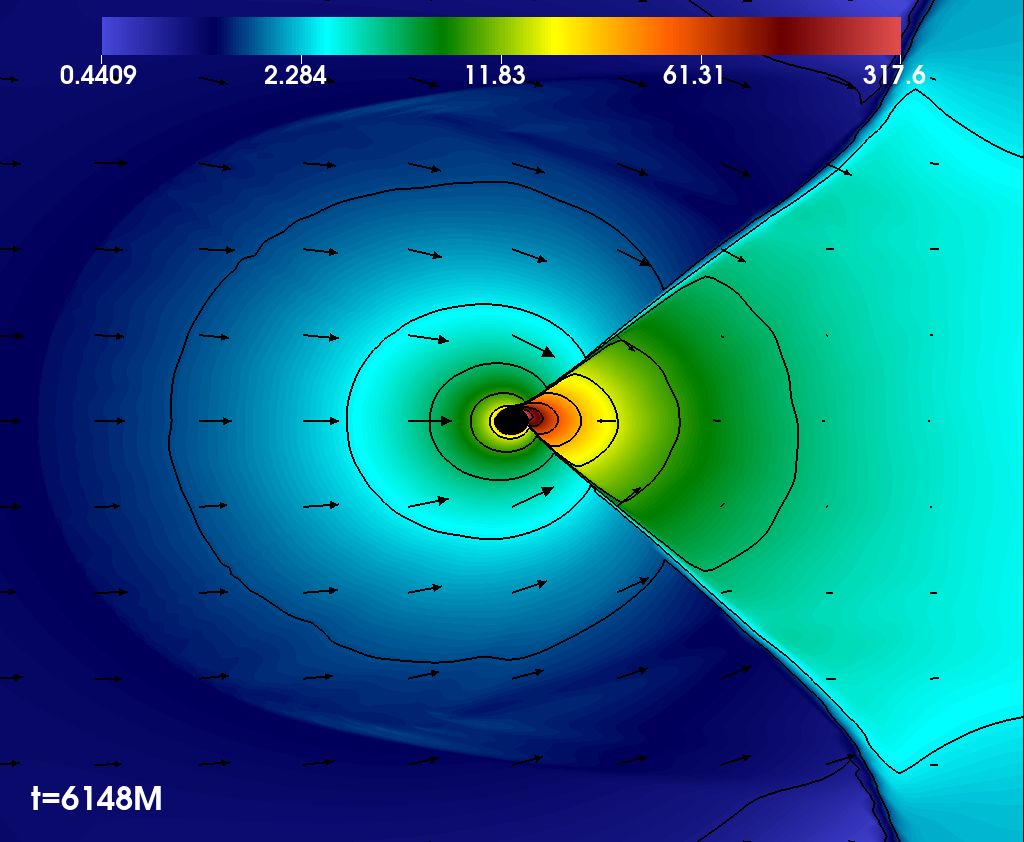,width=4.0cm,height=4.5cm}
  \psfig{file=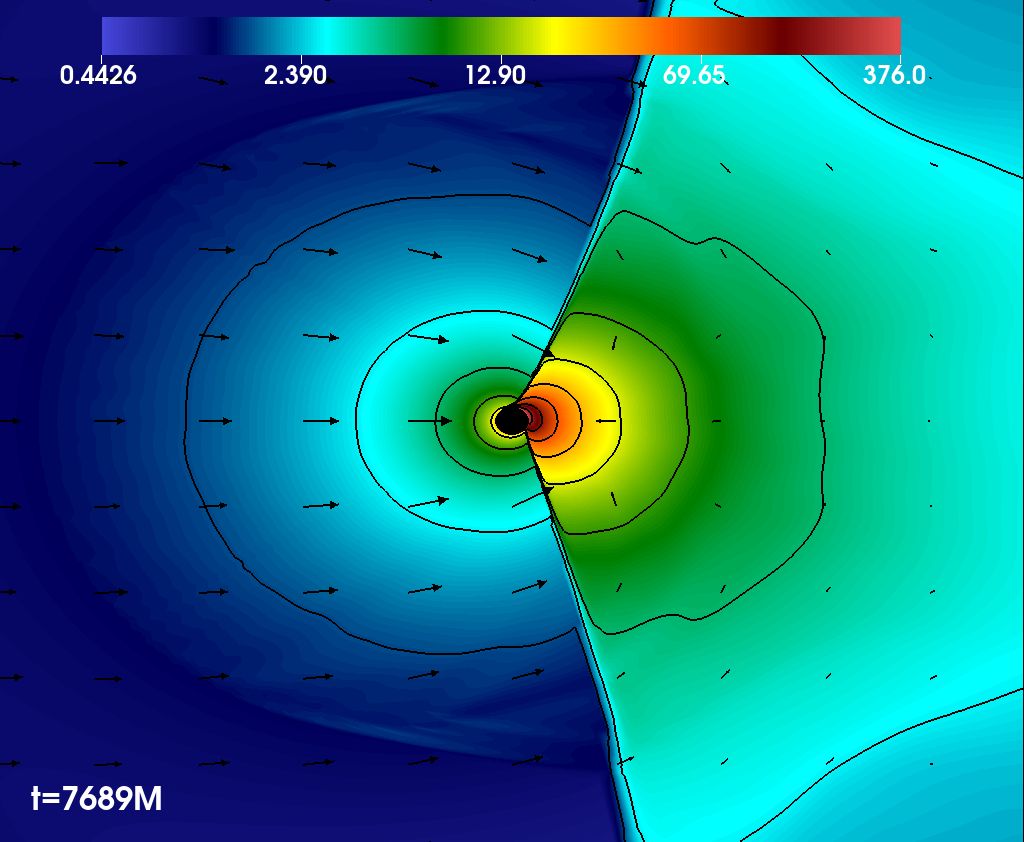,width=4.0cm,height=4.5cm}
  \psfig{file=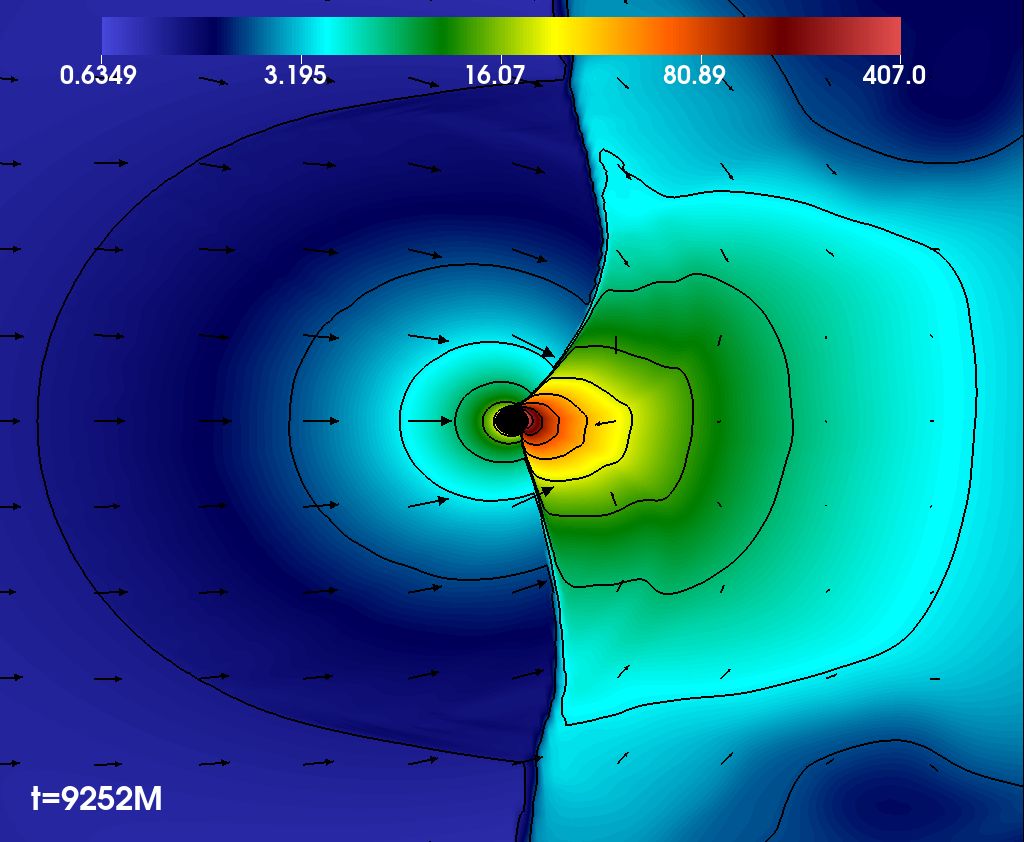,width=4.0cm,height=4.5cm}
  \psfig{file=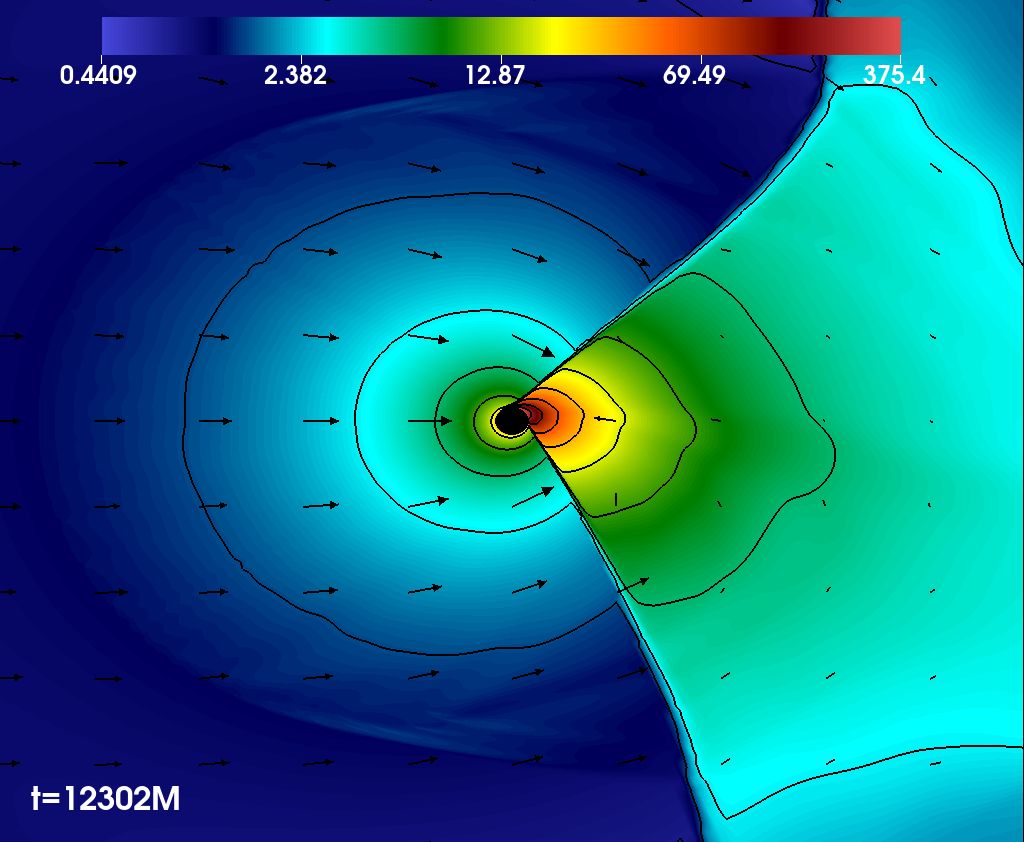,width=4.0cm,height=4.5cm}\\
  \psfig{file=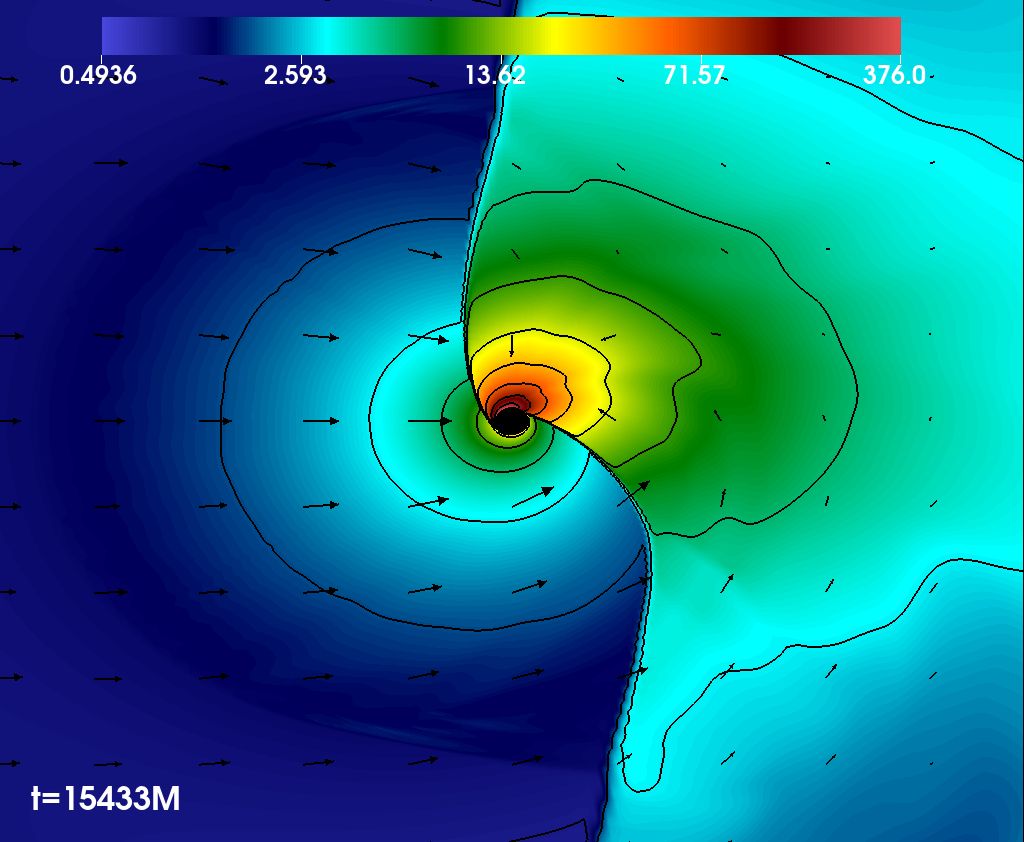,width=4.0cm,height=4.5cm}
  \psfig{file=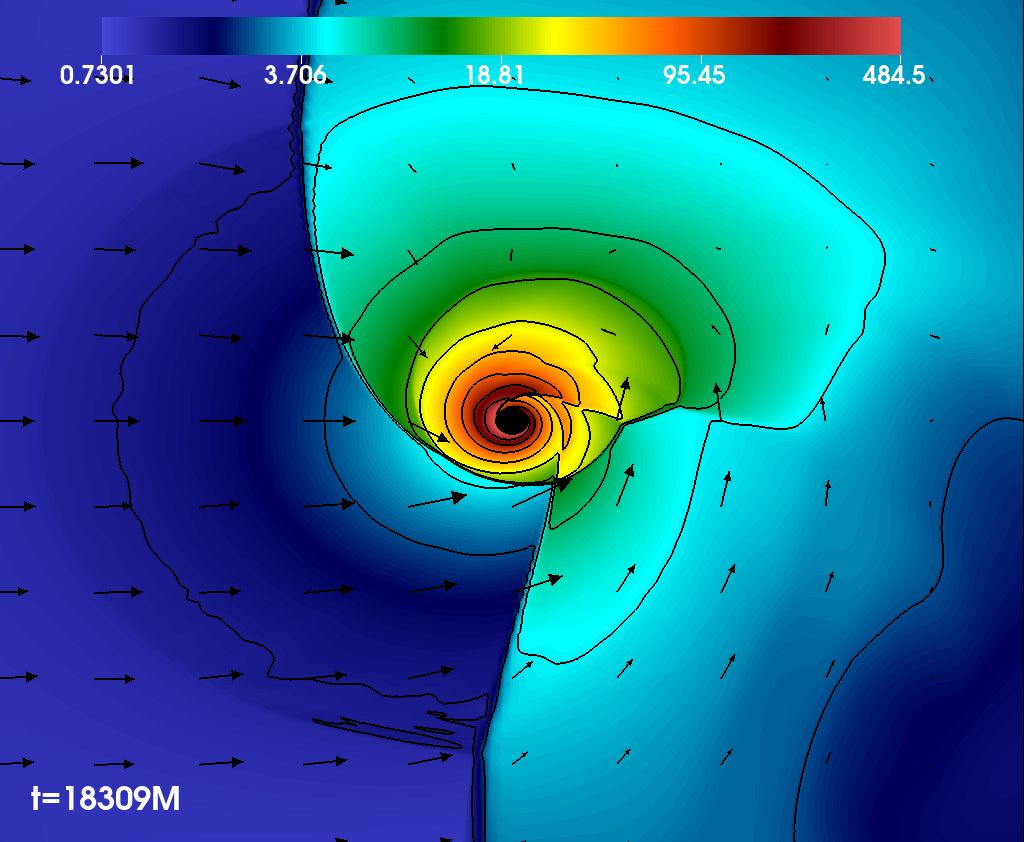,width=4.0cm,height=4.5cm}
  \psfig{file=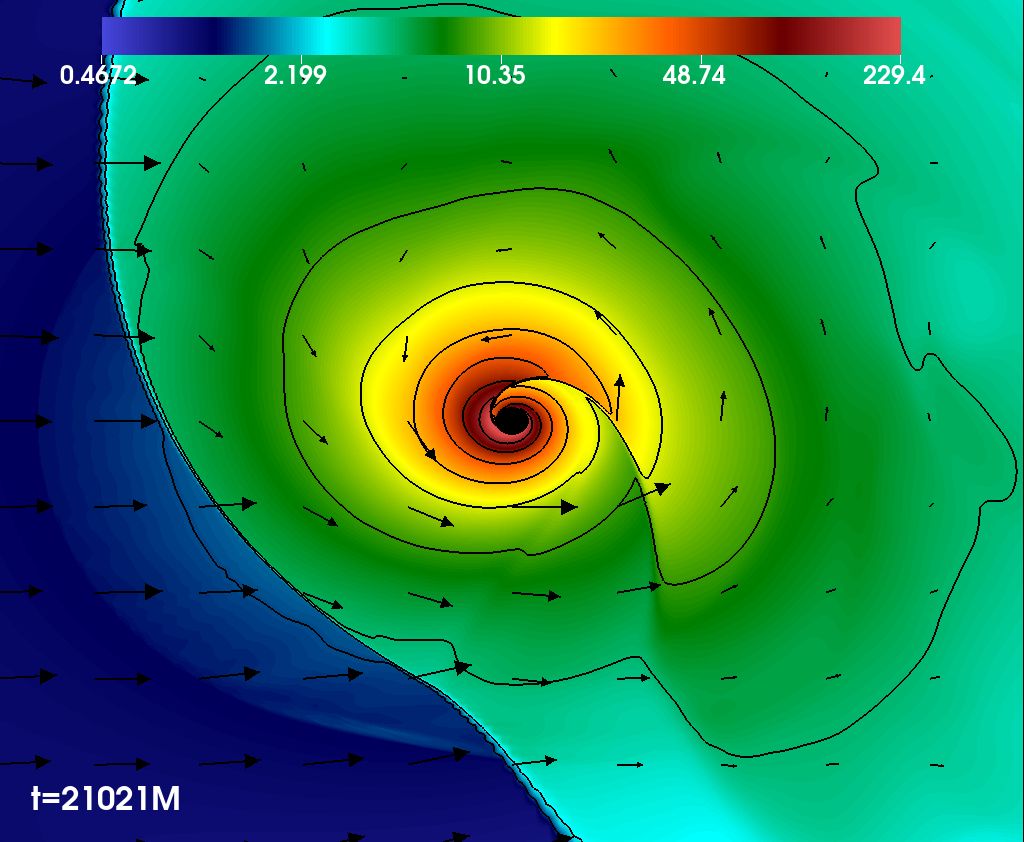,width=4.0cm,height=4.5cm}
  \psfig{file=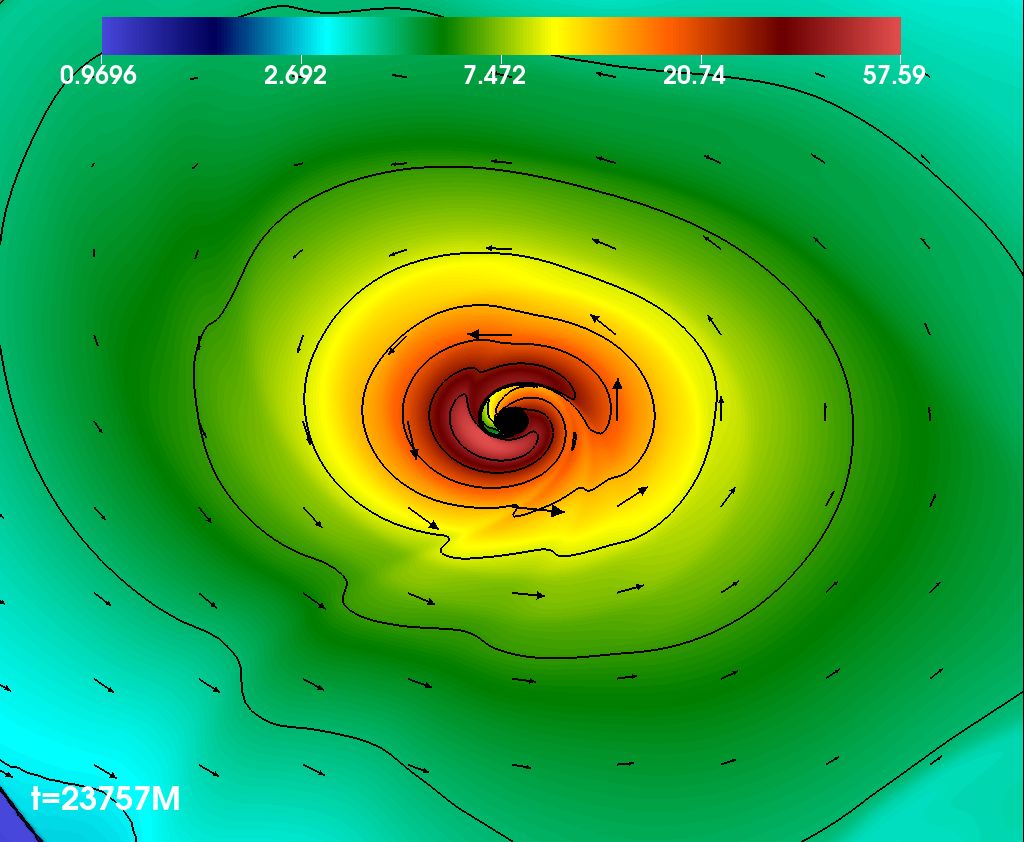,width=4.0cm,height=4.5cm}\\
  \psfig{file=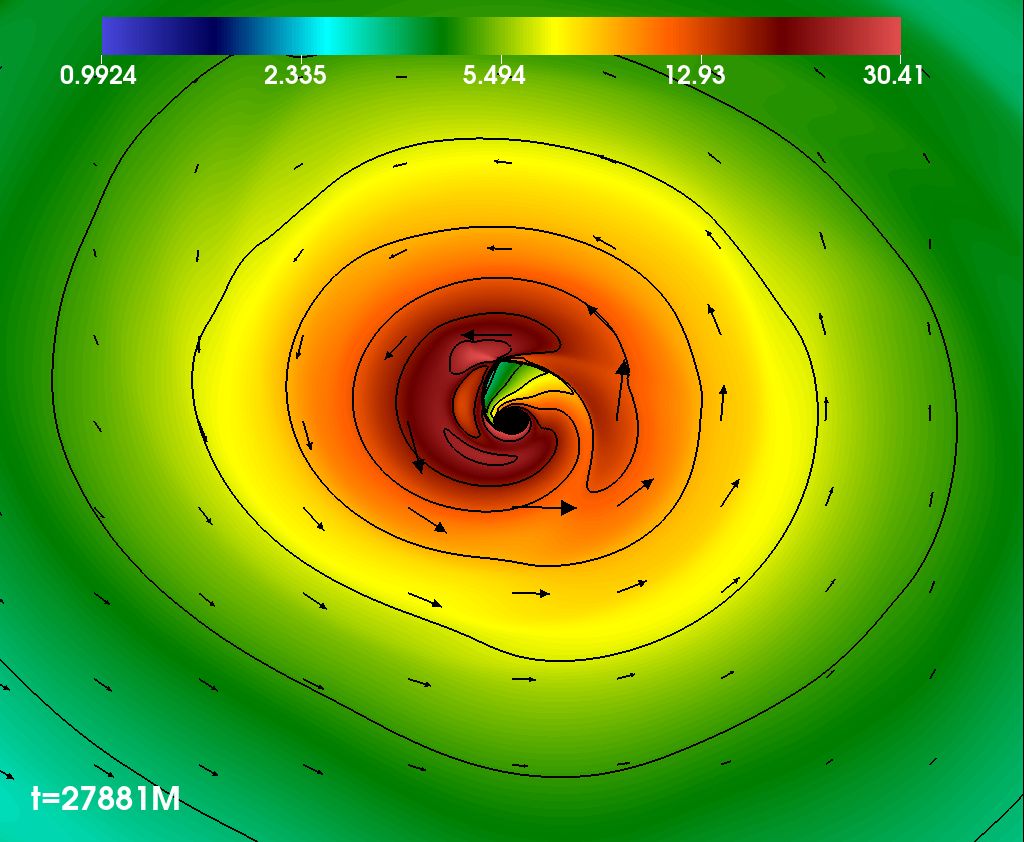,width=4.0cm,height=4.5cm}
  \psfig{file=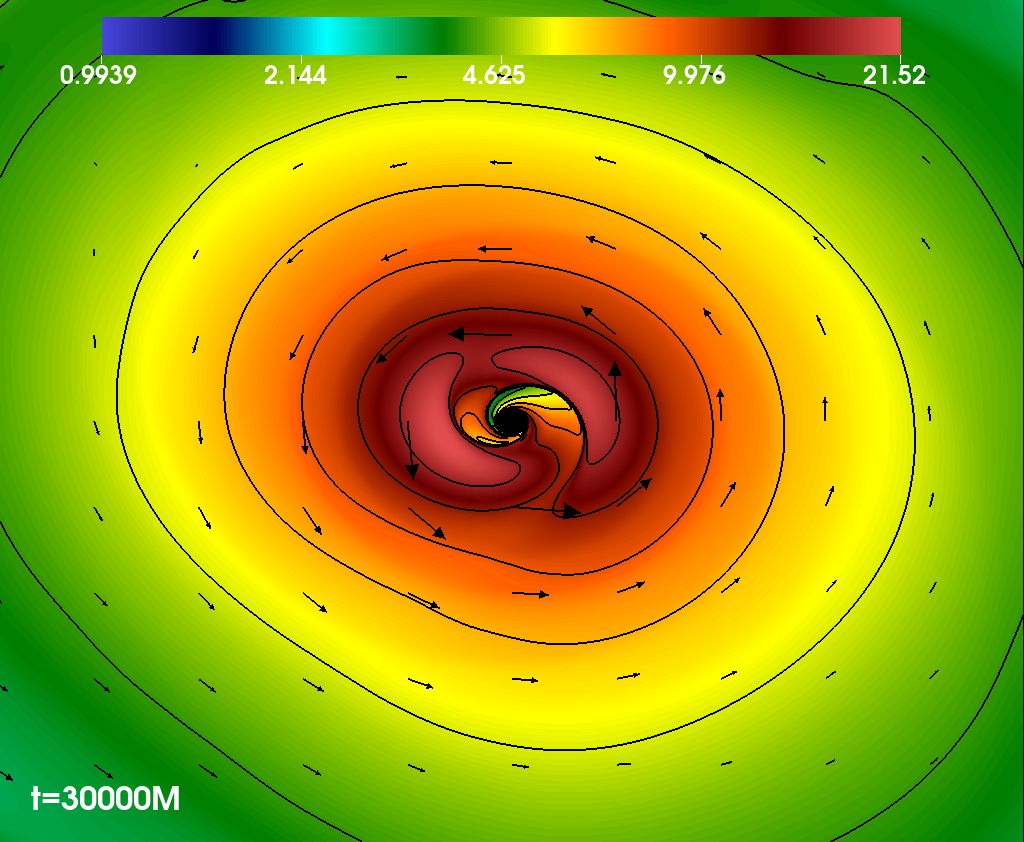,width=4.0cm,height=4.5cm}
  \psfig{file=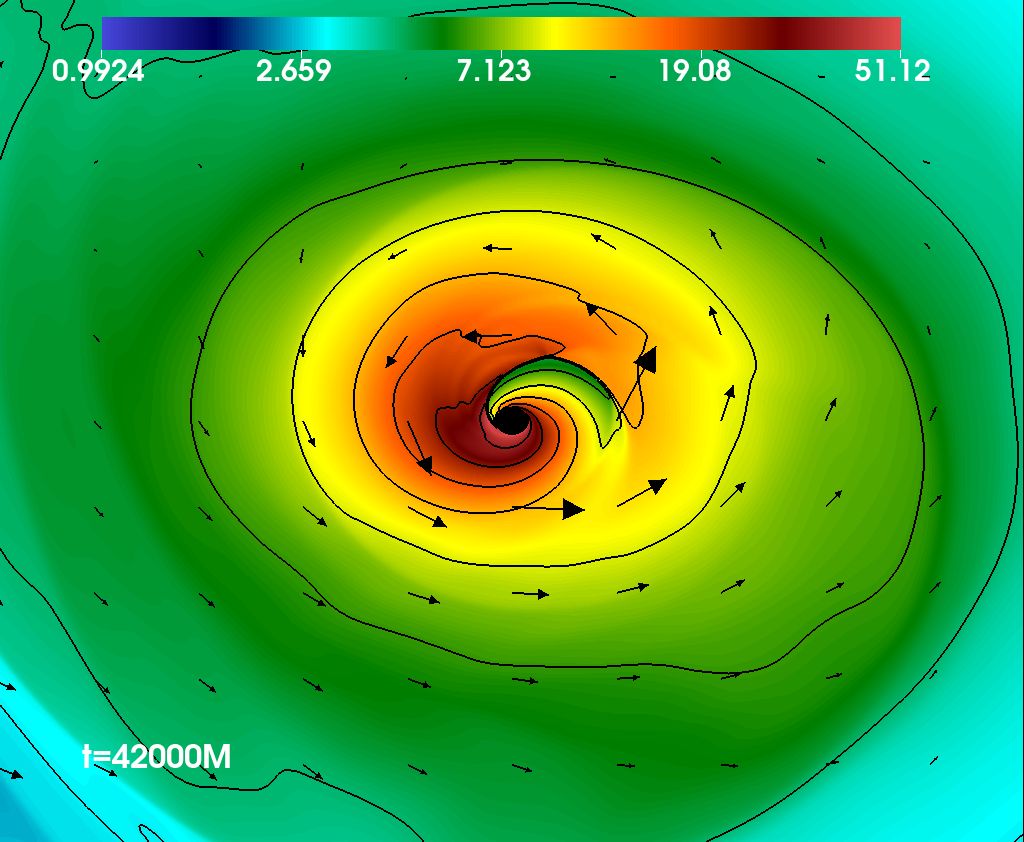,width=4.0cm,height=4.5cm}
  \psfig{file=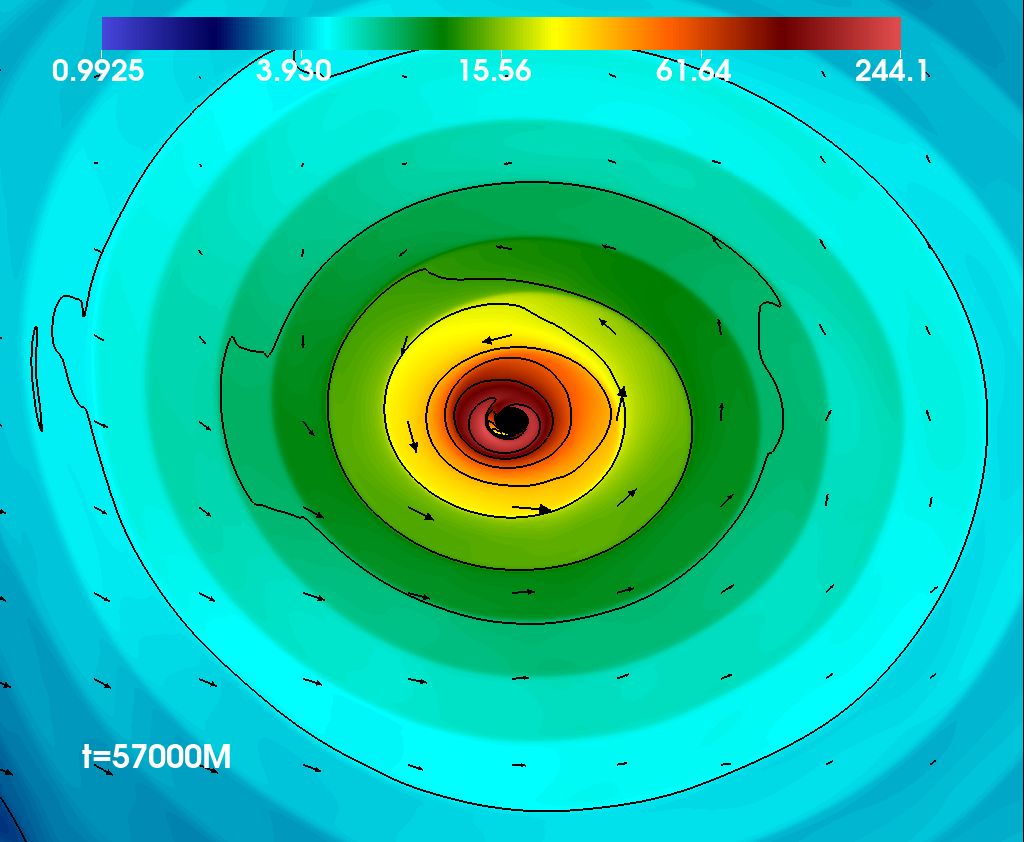,width=4.0cm,height=4.5cm}
     \caption{It illustrates the temporal evolution of the plasma structure generated by BHL accretion around a KBR BH with spin parameter $a=0.9M$ for $B=0.005(1/M)$. Initially, the inflowing matter forms a well-defined shock cone. However, as the accretion proceeds, the redistribution of angular momentum and the dynamic response of the surrounding spacetime cause the cone to become unstable and eventually dissipate, giving rise to spiral shock waves. In the later stages, the matter settles into a quasi-periodic configuration around the BH, and for $t\geq 21000\;M$, the plasma structure approaches a nearly steady-state equilibrium and it is observed that this structure is preserved up to the maximum simulation time of $t = 57000\;M$. It is zoomed to the range $[x_{\min}, y_{\min}] = [-70M, -70M]$ and $[x_{\max}, y_{\max}] = [70M, 70M]$.}
\vspace{1cm}
\label{color_dena09B0005}
\end{figure*}

\begin{figure*}[!ht]
  \vspace{1cm}
  \center
  \psfig{file=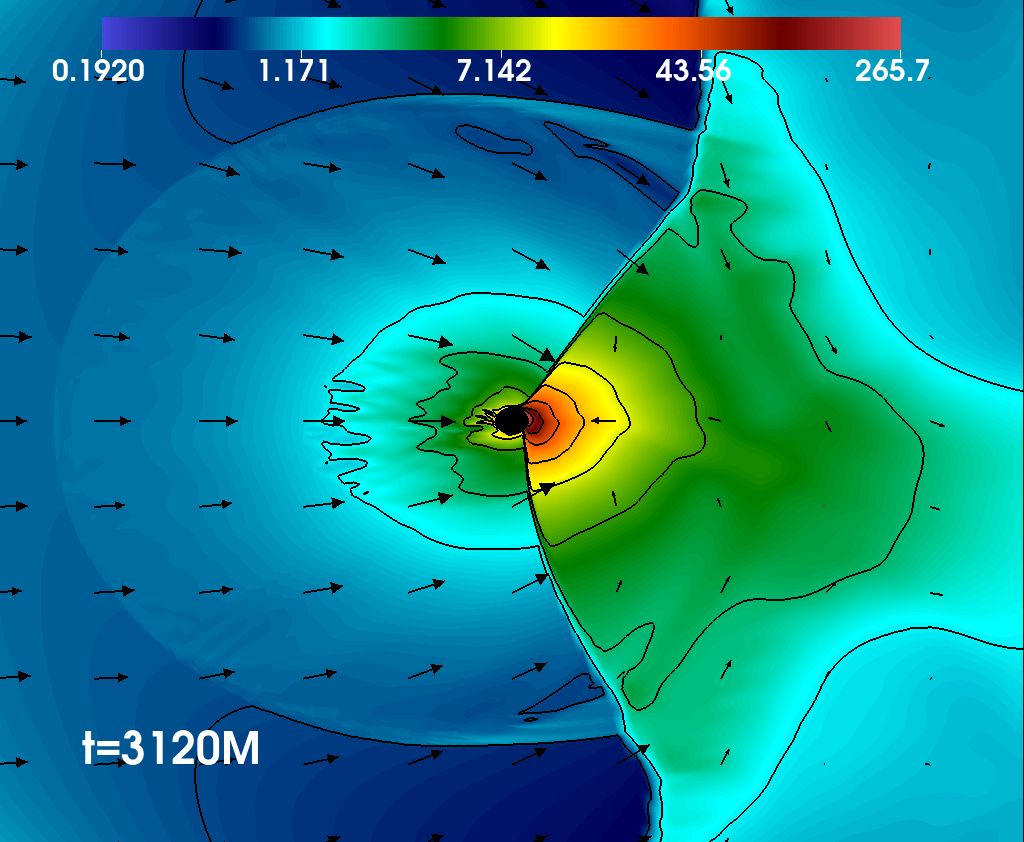,width=4.0cm,height=4.5cm}
  \psfig{file=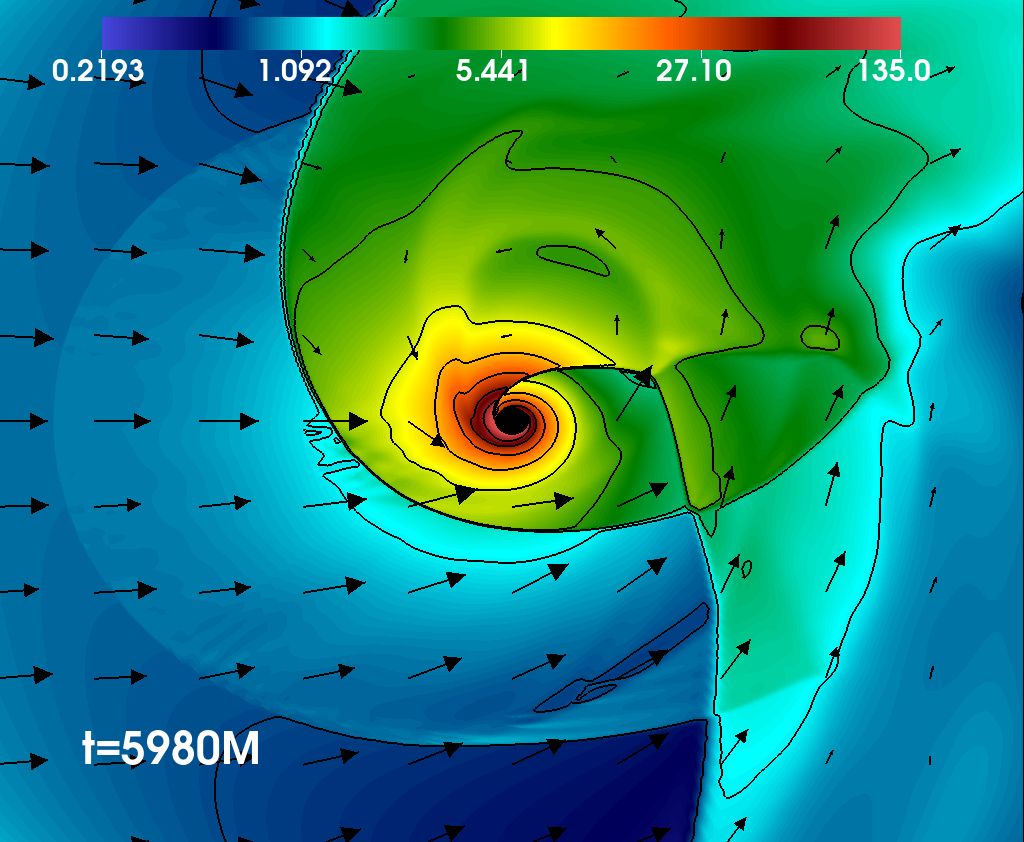,width=4.0cm,height=4.5cm}
  \psfig{file=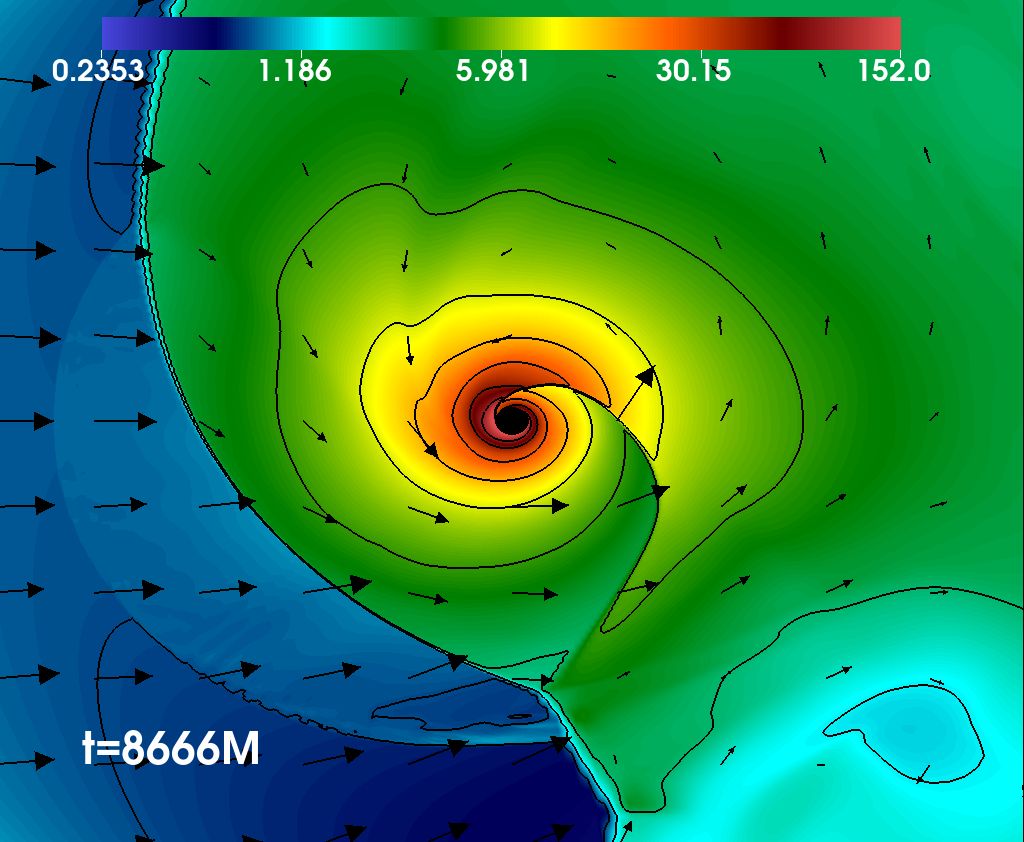,width=4.0cm,height=4.5cm}  
  \psfig{file=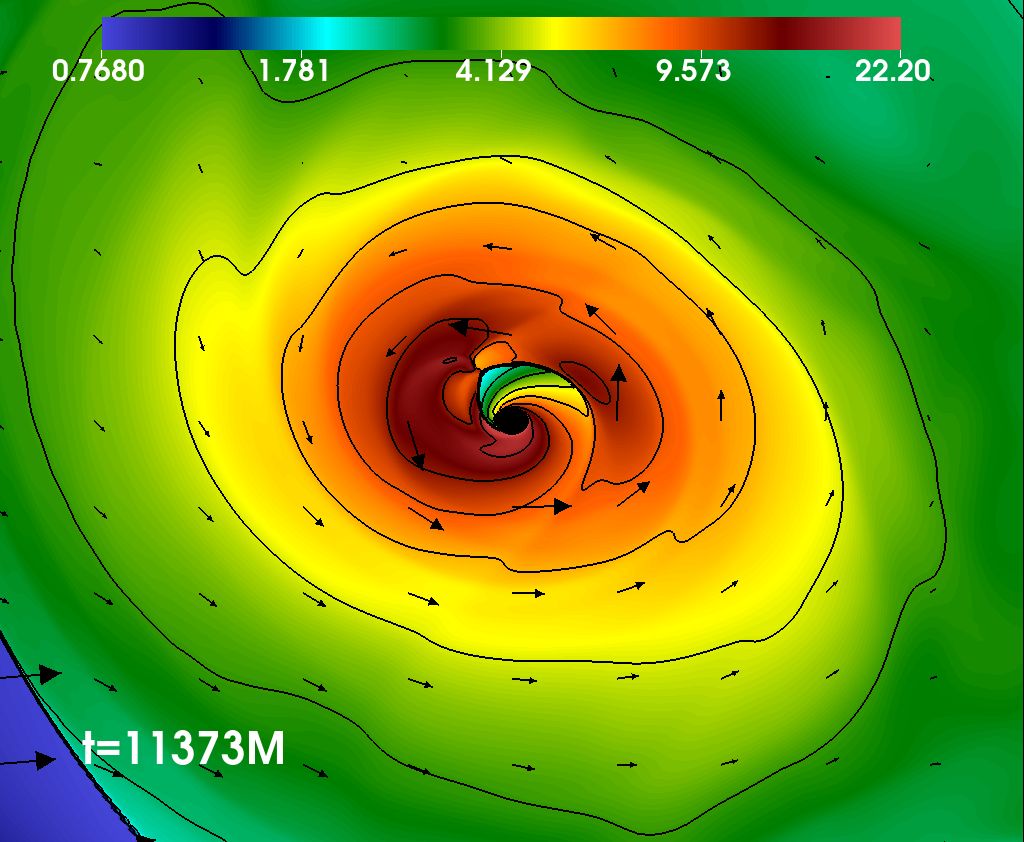,width=4.0cm,height=4.5cm}\\
  \psfig{file=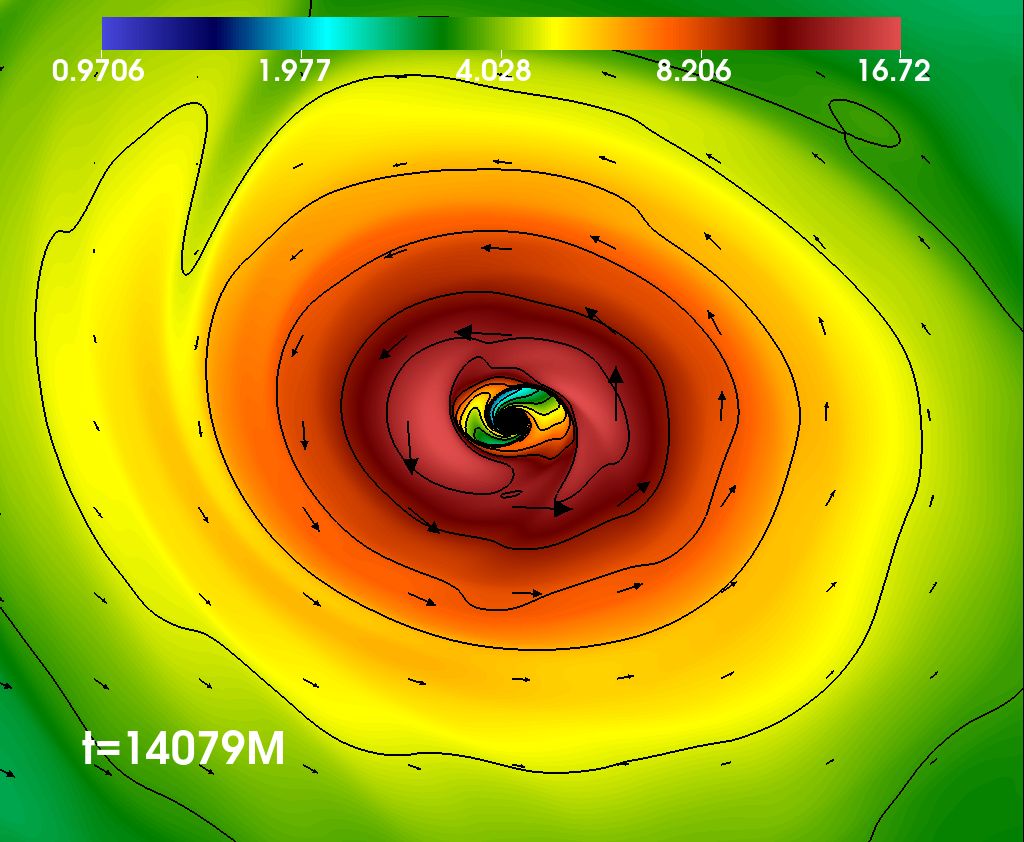,width=4.0cm,height=4.5cm}
  \psfig{file=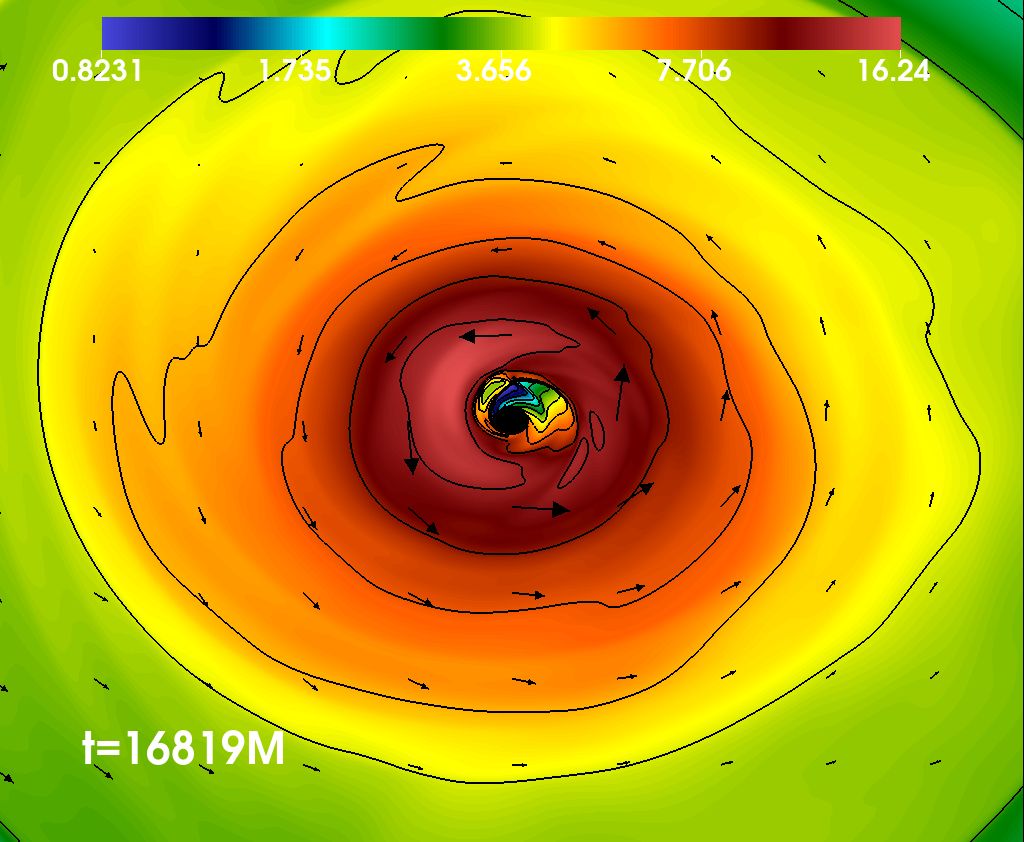,width=4.0cm,height=4.5cm}
  \psfig{file=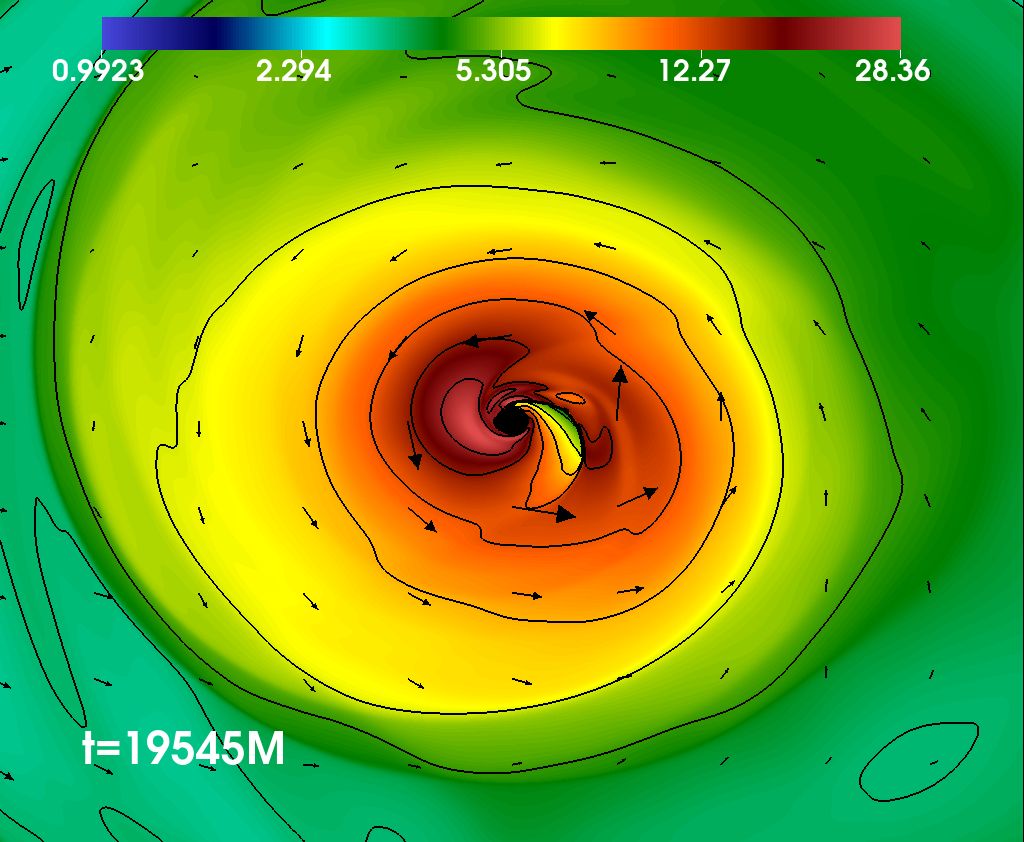,width=4.0cm,height=4.5cm}  
  \psfig{file=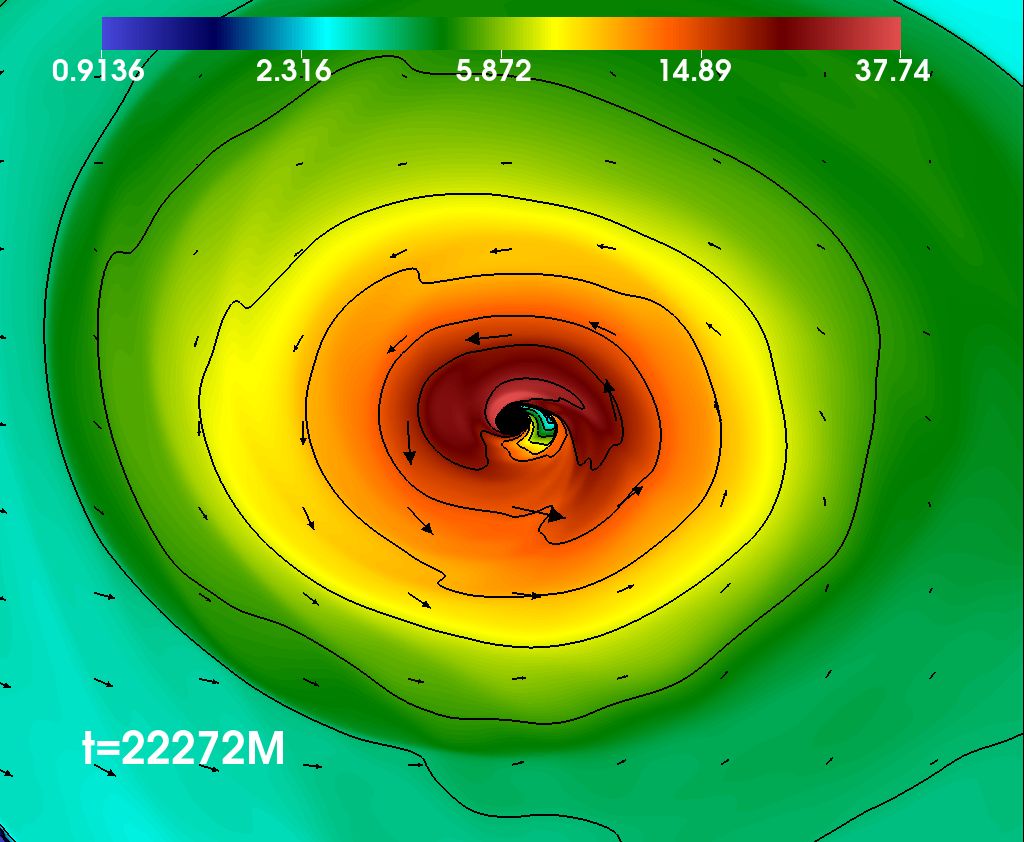,width=4.0cm,height=4.5cm}\\  
  \psfig{file=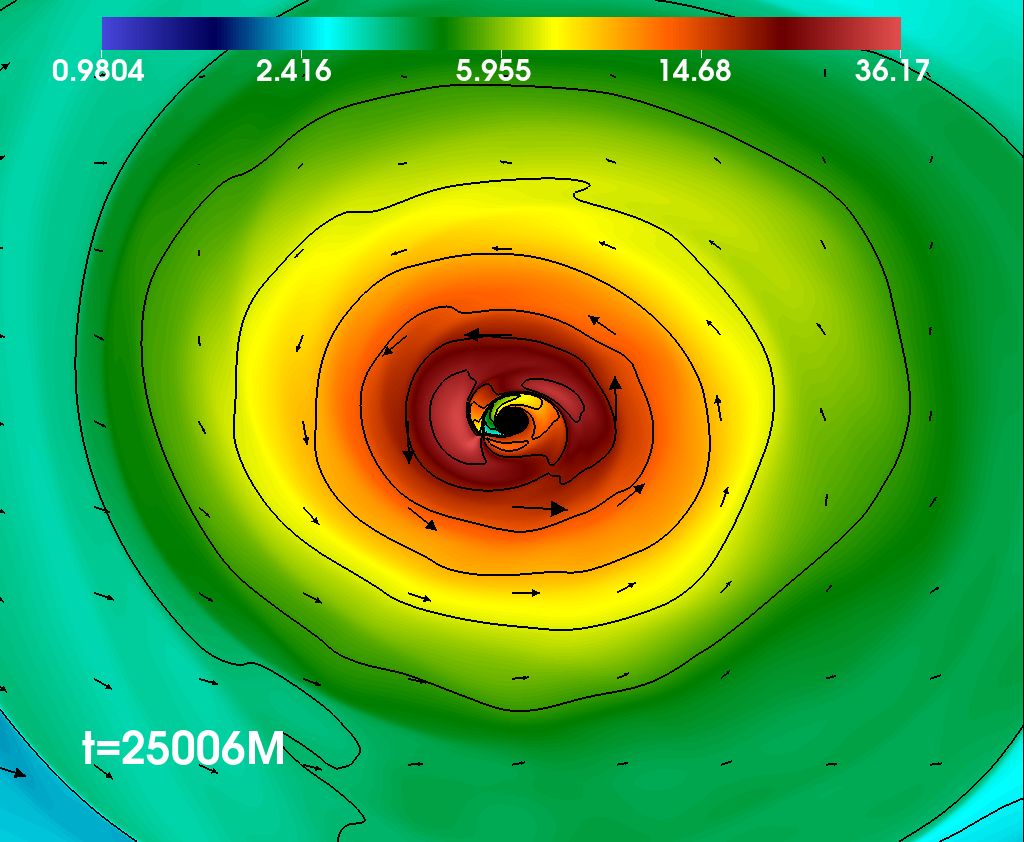,width=4.0cm,height=4.5cm}
  \psfig{file=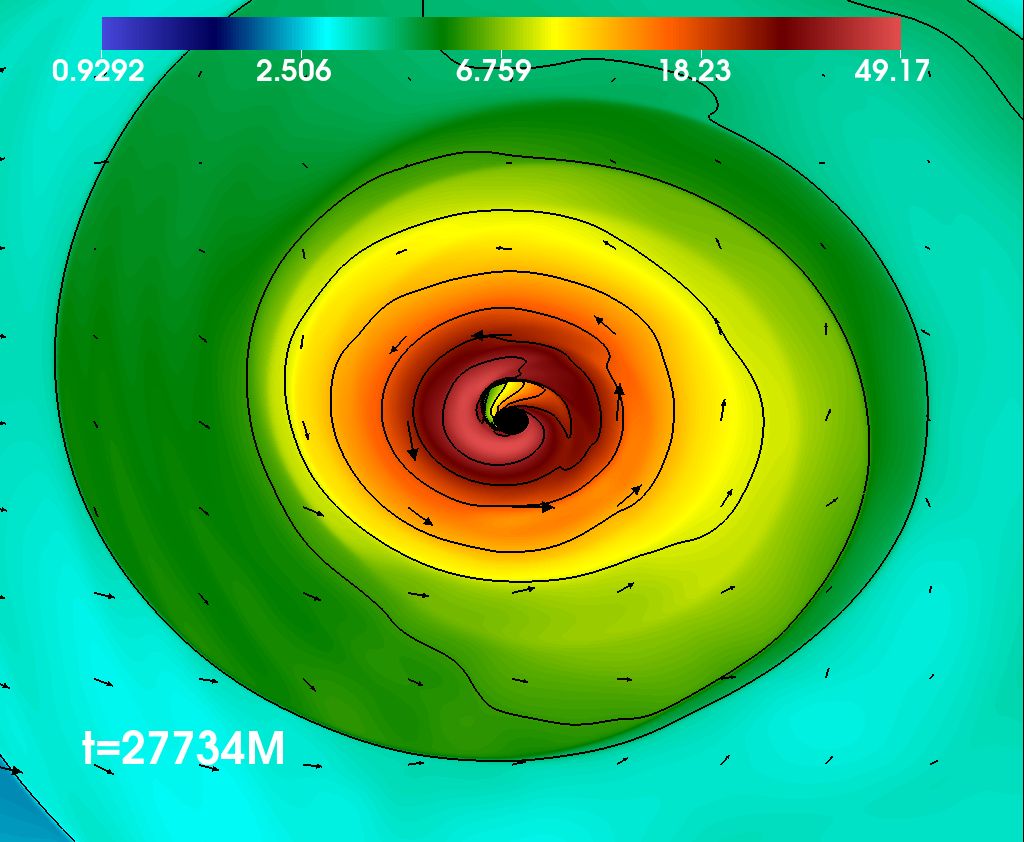,width=4.0cm,height=4.5cm}
  \psfig{file=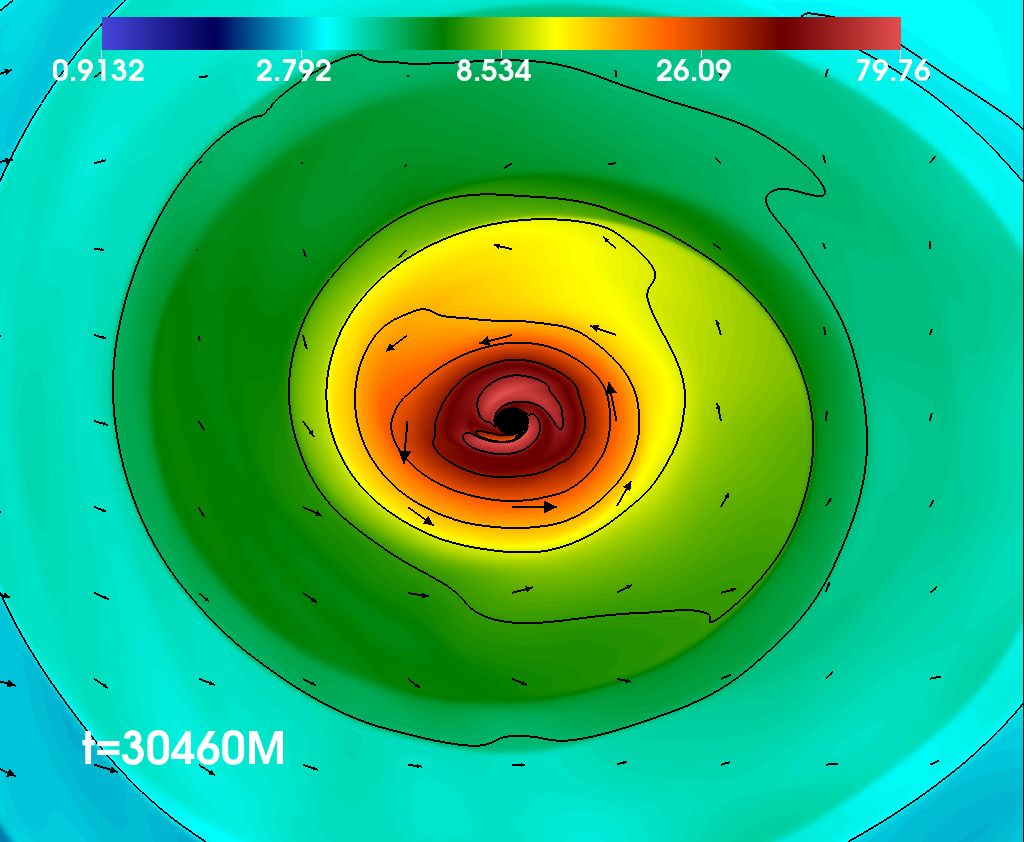,width=4.0cm,height=4.5cm}  
  \psfig{file=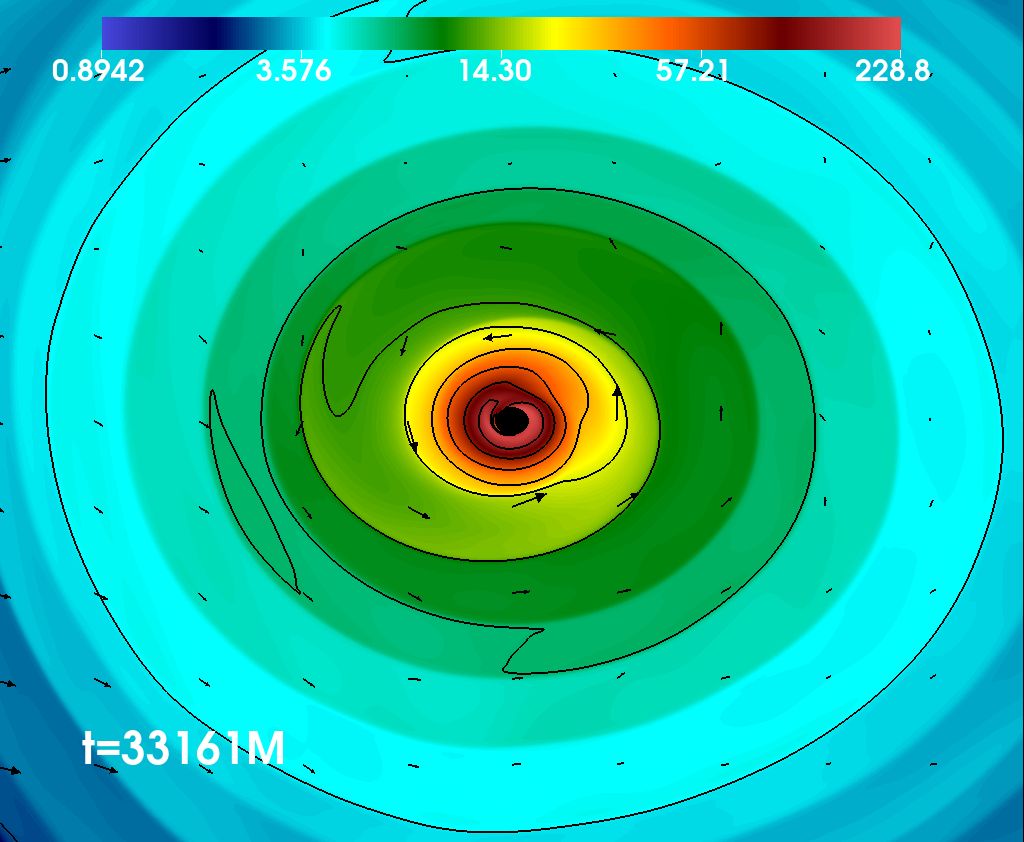,width=4.0cm,height=4.5cm}\\  
  \psfig{file=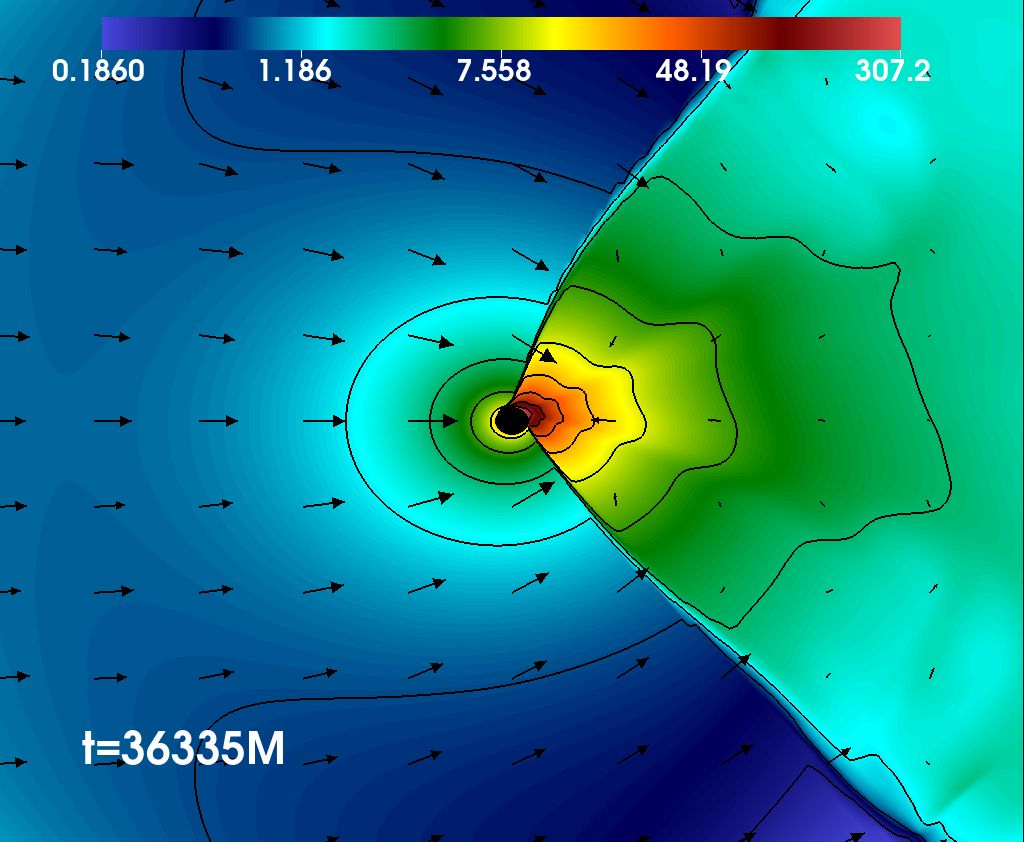,width=4.0cm,height=4.5cm}
  \psfig{file=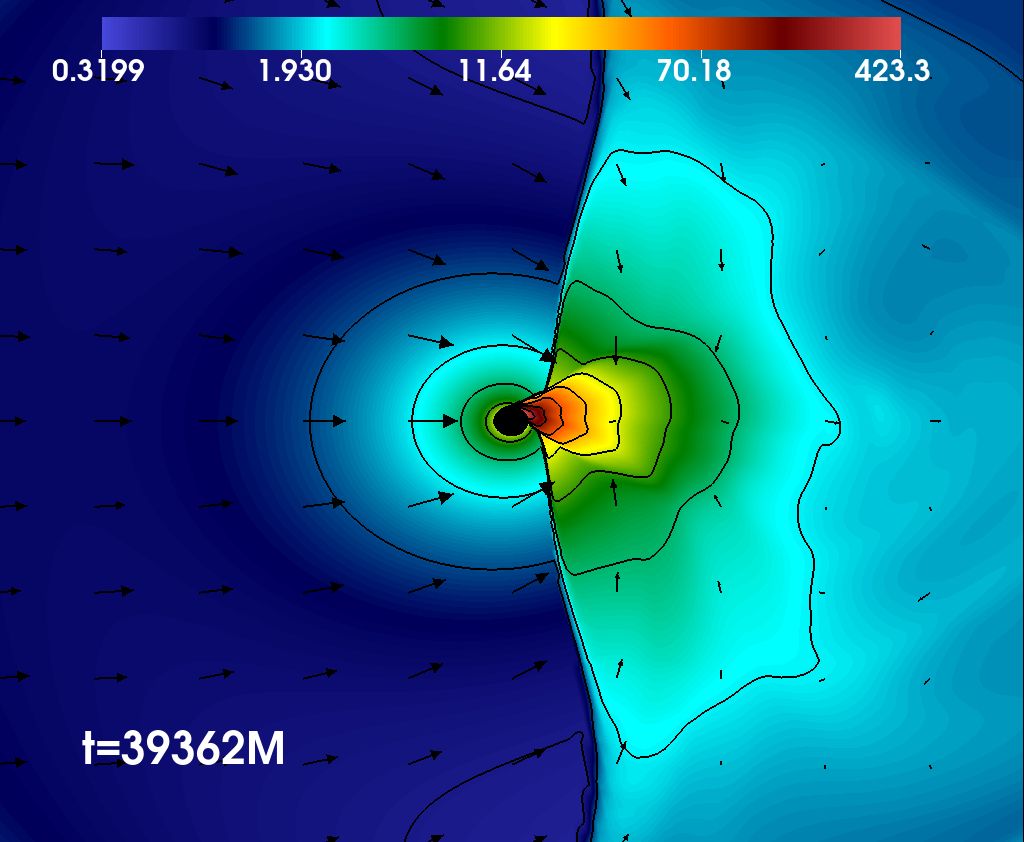,width=4.0cm,height=4.5cm}
  \psfig{file=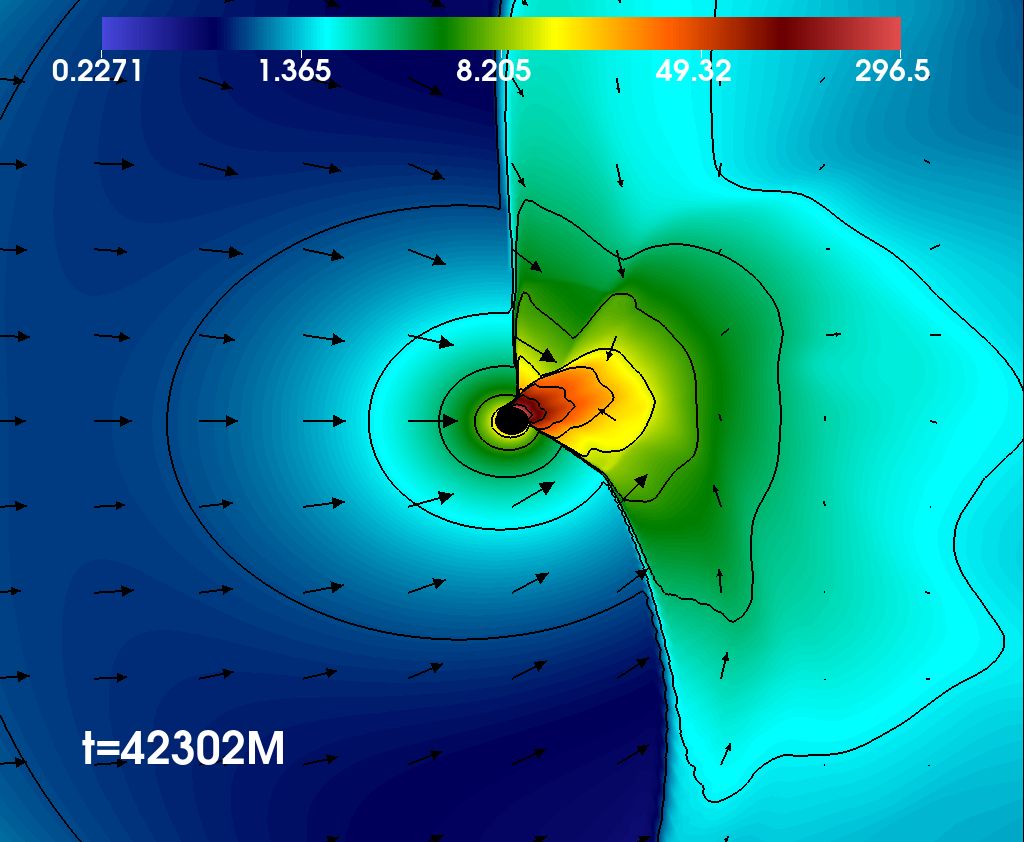,width=4.0cm,height=4.5cm}  
  \psfig{file=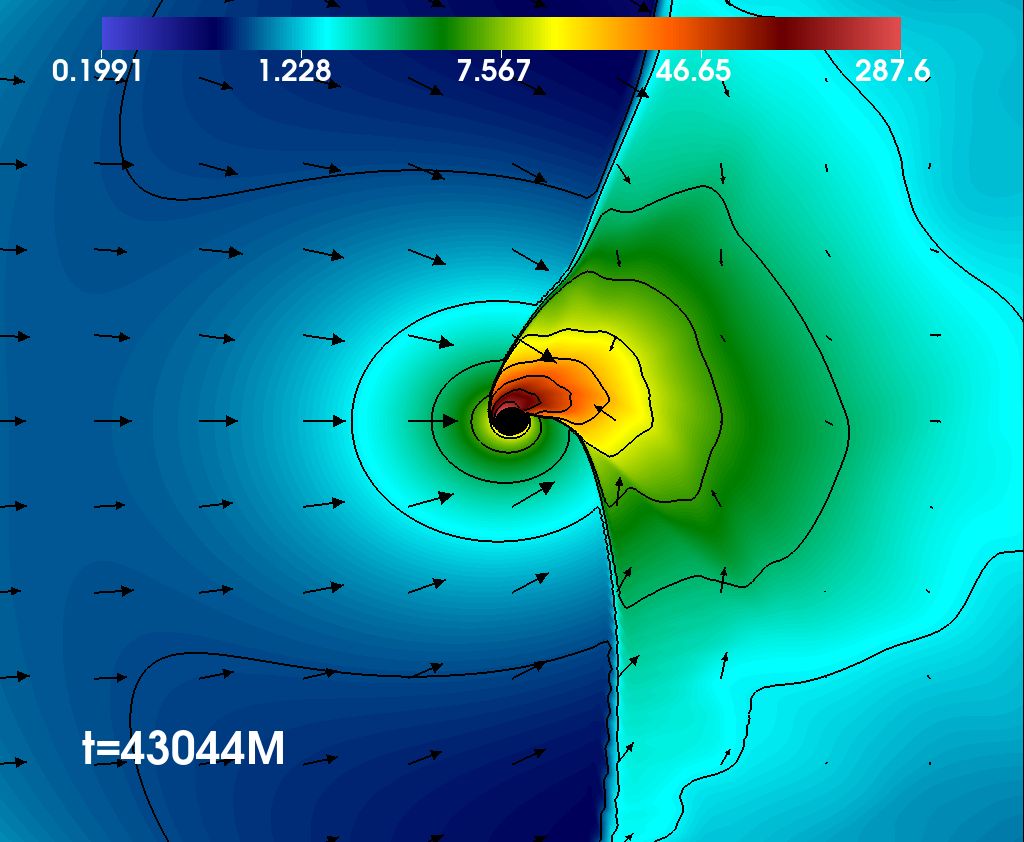,width=4.0cm,height=4.5cm}\\  
     \caption{This figure is the same as Fig.\ref{color_dena09B0005}, but in this case the time evolution of the physical structure of the accreting matter around the BH is shown for $B=0.01(1/M)$. In this model, the material accreted around the KBR BH forms a toroidal structure at approximately $t \approx 10000$. This physical mechanism then undergoes QPOs until about $t \approx 34000M$. Subsequently, as the angular momentum redistributes, the flow returns to a state similar to the early phase of the simulation, once again producing a wide-angle shock cone exhibiting flip-flop oscillations.
}
\vspace{1cm}
\label{color_dena09B001}
\end{figure*}

Overall, as shown in Figs.~\ref{color_dena09B0005} and \ref{color_dena09B001}, our results indicate that the KBR spacetime produces different structures depending on the black-hole spin $a$, the magnetic curvature $B$, and the time after the classical shock cone forms. The system either develops a strongly oscillating shock cone with a time-varying opening angle, or a toroidal structure whose boundaries and density depend on $a$ and $B$. These results highlight the departure from Kerr behavior caused by the KBR magnetic term and provide insight into accretion dynamics in magnetically modified gravitational fields.

Fig.\ref{color_dena09B0005} illustrates the temporal evolution of the accretion flow around a rapidly rotating KBR BH with the spin parameter $a = 0.9M$ for the magnetic curvature parameter $B = 0.005(1/M)$. The snapshots span the interval from the moment when the matter first begins to fall from the upstream region onto the BH due to BHL accretion, through the formation of the classical shock cone at $t \approx 3000M$, and continues until the final simulation time of $t = 57000M$, during which the accretion structure undergoes substantial dynamical changes. These rich time dependent behaviors arise from the interplay between gravitational focusing, spacetime curvature, and the magnetic curvature term inherent in the KBR spacetime.

Each panel shows the rest-mass density with color and contour lines, along with velocity arrows indicating the flow direction and the effects of gravity, angular momentum, and the magnetic curvature term. Early on, a narrow, well-defined BHL shock cone forms downstream, similar to the Kerr case, trapping matter that can undergo oscillations. As the system evolves, the magnetic curvature modifies the KBR potential, causing the cone to broaden, deform, and develop flip-flop oscillations around $t \approx 6000$--$10000M$. The velocity vectors reveal strong transverse motion at the cone edges, while matter inside oscillates, forming alternating high- and low-density regions. Spin and magnetic curvature enhance these motions, which eventually wrap around the BH to produce a spiral-like shock structure that persists until $t \approx 19000M$.

Around $t \approx 21000M$, a major structural transition occurs. The shock cone collapses, and the accretion flow reorganizes into a toroidal structure surrounding the BH. This torus is visible in the color maps as a circular ring of enhanced density with nearly closed contours, indicating a quasi-equilibrium configuration. The velocity vectors show that the matter has acquired sufficient azimuthal rotation to become trapped in the effective potential of the KBR spacetime. In this regime, the large scale flip-flop oscillation ceases, and the system transitions into a pressure supported, quasi-periodically oscillating torus with an inner radius very close to the BH horizon.

In the later stages of the simulation, from $t \approx 22000M$ to $t = 57000M$, the toroidal structure remains stable and maintains rotation around the BH. The torus persists throughout the computational domain until the final simulation time. As matter gradually falls inward due to gravitational attraction, the high density inner region of the torus becomes more compact, more circular, and exhibits quasi-periodic motion, as clearly seen in the final row of Fig.\ref{color_dena09B0005}. The velocity field continues to show a coherent rotational pattern, and the absence of large deformations or oscillations indicates that the magnetic parameter $B = 0.005 (1/M)$ places the system in a regime where long-term torus stability is favored.

However, as discussed in detail in Fig.\ref{color_dena09B001}, the final snapshots of Fig.\ref{color_dena09B0005} reveal that the matter begins to accrete closer to the horizon over time. Between $t = 42000M$ and $t = 57000M$, the density in the inner region increases by nearly a factor of five. Since the wind-pressure force coming from the upstream region becomes more dominant than the forces generated by the angular momentum around the BH, it is expected that the torus eventually loses stability and transitions back into a flip–flop oscillation regime. In fact, this behavior was observed at higher magnetic strengths. Thus, the evolution shown in Fig.\ref{color_dena09B0005} highlights the delicate balance between rotation, magnetic curvature, and gravitational focusing that governs accretion in the KBR spacetime.

In cases $a = 0.9M$ and $B = 0.01(1/M)$, the temporal evolution of the accretion morphology shown in Fig.\ref{color_dena09B001} corresponds to a situation in which the magnetic curvature parameter is twice as large as in Fig.\ref{color_dena09B0005}. As clearly illustrated in Fig.\ref{color_dena09B001}, the combined interaction between gravitational focusing, BH rotation, magnetic curvature effects, and the wind pressure of the upstream inflowing material gives rise to multiple distinct dynamical phases.

Before $t \approx 3000M$, the traditional BHL mechanism creates a shock cone during the first part of the simulation. Similar to the Kerr scenario, a high--density zone forms inside the downstream cone. However, in contrast to the Kerr model, the strong combined influence of the previously indicated physical phenomena causes the shock cone in this KBR arrangement to quickly evolve into an unstable state. The cone shows a distinct lateral asymmetry from the very beginning of the simulation, and the oscillatory displacement eventually takes on the distinctive flip-flop behavior. Strong transverse components are also visible in the velocity field close to the shock boundaries, indicating that the downstream morphology is extremely sensitive to the spacetime curvature caused by the parameter $B$.

The shock cone opening angle keeps widening as the structure experiences more intense oscillations between around $t = 3000M$ and $t = 6000M$. The density distribution inside the cone splits into alternating high and low density shear layers as a result of these oscillations, which also cause the shock front to repeatedly swing across the downstream region. The cone gradually loses coherence as the instability intensifies, making it impossible for the traditional BHL shock structure to remain contained in the downstream area. During this epoch, the severe frame dragging associated with the spin parameter $a = 0.9M$ steers the matter azimuthally around the BH.  The redistributed angular momentum enhances the rotational motion of the inflowing gas, an effect further amplified by the $B$-dependent modification of the effective potential, which introduces significant anisotropy in the fluid trajectories near the horizon.

A major morphological transition occurs between $t = 6000M$ and $t = 10000M$. In this interval, the highly unstable shock cone collapses, giving rise to a toroidal structure around the BH. As seen clearly in Fig.\ref{color_dena09B001}, this physical configuration forms an almost circular high density ring whose inner radius is very close to the BH horizon. The formation of this torus is driven by the buildup of centrifugal support. Once the inflowing gas accretes sufficient specific angular momentum, a permanent pressure supported torus forms, replacing the fragmented shock cone. At this stage, the velocity field vectors exhibit primarily circular motion, suggesting that the torus supported by angular momentum now governs the global morphology and inhibits the large-scale flip-flop instability.

Between $t \approx 10000M$ and $t \approx 33000M$, the toroidal structure shows QPOs while remaining globally stable. Inside the torus, radial pulsations, mild azimuthal distortions, and transient spiral shocks develop. These oscillations are supported by the interplay of centrifugal forces, pressure gradients, and the continuous wind pressure supplied by the upstream region. Over time, matter slowly drifts inward towards the BH, causing the density near the horizon to increase, the torus cross–section to become increasingly compact, and the overall structure to evolve toward a more coherent quasi–steady–state configuration. Despite ongoing accretion, the torus preserves a well–defined and nearly axisymmetric morphology throughout this quasi–periodic phase, and no classical downstream shock cone reappears.

A second major transition occurs at approximately $t  \approx 34000M$. As the density near the horizon increases, the angular–momentum distribution adjusts, weakening the centrifugal barrier. As the BHL wind pressure increases near the horizon, it eventually dominates, causing the torus to collapse. The closed density contours open, and the high-density ring breaks apart. This allows the broad shock cone seen earlier to reappear. The new shock cone oscillates more strongly than before, with a pronounced flip-flop motion. The shock front swings widely across the downstream region, producing large azimuthal displacements. Spiral shocks reemerge as the flow reorganizes, and this dynamic oscillatory state continues until the end of the simulation.

The quasi-periodic toroidal state and the powerful flip-flop shock mechanism seem to be natural physical results of accretion in KBR spacetime, as also shown in Fig.\ref{color_dena09B0005} and further supported by our study of Fig.\ref{color_dena09B001}. These two physical mechanisms may alternatively occur around a KBR BH, depending on the initial setup, thermodynamic circumstances, and especially the value of the magnetic curvature parameter $B$. The coexistence of various QPO frequencies (such as LFQPOs and HFQPOs) seen in X-ray binaries during several epochs from the same astronomical source may be explained theoretically by this behavior. This interpretation is further supported by a thorough comparison with observational data provided in the literature's later part.

In many modified gravity models studied so far, BHL accretion generically leads to the formation of a downstream shock cone \cite{Donmez2012MNRAS,Donmez2024Universe,Donmez2024JCAP,Donmez2025EPJC}, whose morphology and stability are quantitatively modified by the underlying gravitational theory. Typically, such modifications manifest as changes in the opening angle, oscillation amplitude, or growth rate of instabilities, while the overall shock-cone structure remains intact. In contrast, accretion in the KBR spacetime exhibits a qualitatively different phenomenology controlled by the magnetic curvature parameter B. Our simulations show that an initially formed shock cone can be completely destabilized and destroyed, giving rise to a sequence of distinct accretion states, including flip–flop oscillating shock cones and long-lived toroidal structures. Remarkably, all three configurations, classical shock cones, flip–flop oscillations, and toroidal accretion, can occur within the same spacetime for a fixed value of B, emerging at different stages of the evolution depending on the strength of the magnetic curvature. This behavior has not been reported in previous studies of BHL accretion in modified gravity and highlights the role of magnetic curvature as a genuinely new control parameter that drives qualitative transitions in accretion morphology and dynamics. Such diverse behaviors may help explain why QPO frequencies inferred from the same astrophysical source can vary over time.

The main result of this section is that the background magnetic curvature significantly affects the accretion morphology and flow dynamics. Depending on the model parameters, the system develops distinct shock-dominated and torus-dominated configurations, which provide the physical basis for the variability analysis discussed below.
\subsection{Mass Accretion Rate as a Diagnostic of Accretion Dynamics}
\label{Sec6}

The mass accretion rate is one of the most fundamental diagnostics of the accretion dynamics around BHs \cite{Done:2007nc,Orh1_1}. Through the temporal evolution of the mass accretion rate, one can directly quantify the efficiency with which matter is captured by the gravitational field of the BH, the transfer of angular momentum throughout the flow, and the degree of hydrodynamical instabilities that arise within the accretion structure. $dM/dt$ is a useful tool in numerical simulations for detecting and describing transitions between various accretion stages, including the growth or collapse of torus-like structures, the oscillatory motion of the shock structure, and the production of a shock cone. There are direct observable analogies for these dynamical transitions.   The luminosity variations and QPOs observed in astrophysical systems are intimately linked to fluctuations in the mass accretion rate. For these reasons, examining the time dependence of  $dM/dt$ and its variation with the spacetime parameters that govern the underlying gravitational geometry and the spin of the BH is crucial for testing the gravitational model employed in this work and for understanding the dynamical response of accreting matter.

\begin{figure*}[!ht]
  \vspace{1cm}
  \center
  \psfig{file=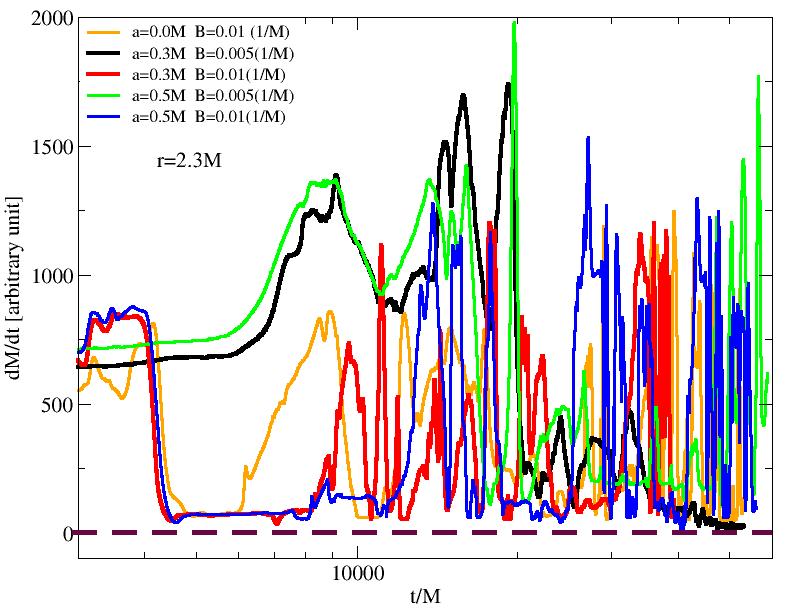,width=14.0cm,height=7.5cm}\\
   \psfig{file=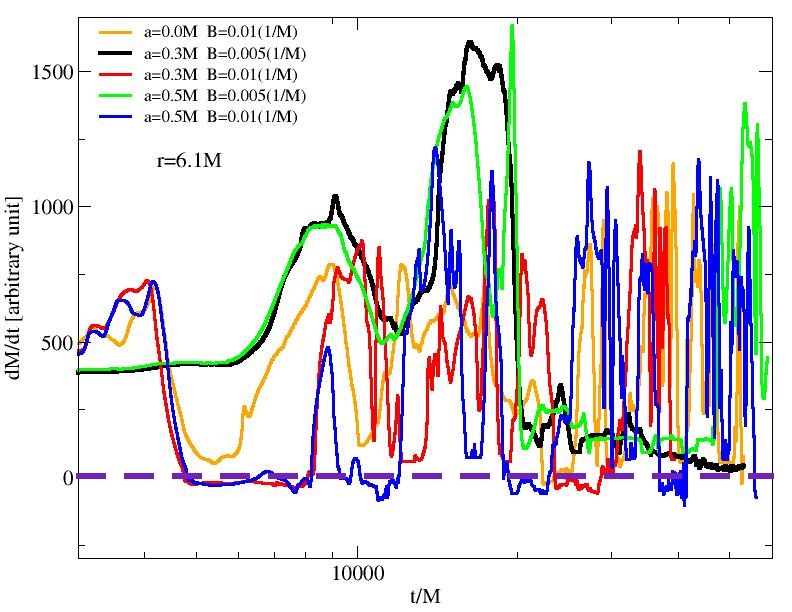,width=14.0cm,height=7.5cm}\\
    \psfig{file=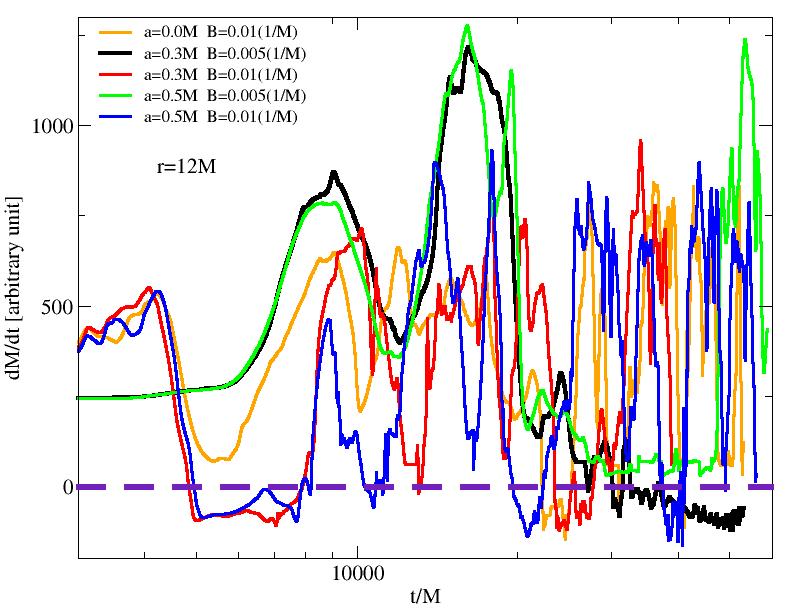,width=14.0cm,height=7.5cm}  
     \caption{The temporal evolution of the mass accretion rates for different model cases is shown at three distinct radial locations around the non-rotating and rotating KBR BHs. The point at $r = 2.3M$ represents the accretion rate in the innermost region of the flow, closest to the KBR BH horizon, whereas $r = 6.1M$ corresponds to the variation of the accretion rate around the ISCO. The point at $r = 12M$ illustrates the behavior of the mass accretion rate in a region where the gravitational field is relatively weak. In each model case, a highly dynamic behavior of the surrounding matter is observed due to the transfer of the spacetime angular momentum to the accreting material around the KBR BH.
}
\vspace{1cm}
\label{mass_acc}
\end{figure*}

To study the effects of black-hole spin $a$ and magnetic curvature $B$ on accretion, Fig.~\ref{mass_acc} shows the mass accretion rate at three radii: $r = 2.3M$ (near the horizon), $r = 6.1M$ (close to the ISCO), and $r = 12M$ (weaker gravity). For non-rotating and slowly rotating KBR BHs, the accretion rate initially declines smoothly following the onset of BHL inflow. Later, the system becomes dynamically active and the accretion rate fluctuates significantly. Clear differences appear between $B = 0.005\,M^{-1}$ and $B = 0.01\,M^{-1}$. After the classical shock cone forms, the response of the flow varies: for $B = 0.01\,M^{-1}$, the accretion rate drops rapidly at all spins and radii, while for $B = 0.005\,M^{-1}$, it rises during the same interval.

The mass accretion rate has a strong oscillatory characteristic after the initial drop, signifying the start of hydrodynamical instabilities. The smaller $B$ example likewise exhibits similar oscillations, but they happen later. This clearly shows that, regardless of the black-hole spin parameter $a$, the magnetic curvature parameter $B$ significantly affects the time of occurrence of various states, amplitudes, and topologies of the resulting variability. The greater anisotropy of the effective gravitational potential in the KBR spacetime is directly linked to the improved dynamical response of the flow at a bigger $B$.

Furthermore, for all values of $a$, the temporal variations in the mass accretion rate remain correlated at all three radial locations, indicating that oscillations in the shock structure propagate coherently through the inner and intermediate regions of the flow. The influence of spin becomes more evident at the innermost radius as shown for $r = 2.3M$, the mass accretion rate remains positive at all times. The strong gravitational field and spacetime curvature ensure continuous inward motion of matter toward the BH, while at larger radii negative values of the accretion rate appear. This indicates that, locally, the matter undergoes quasi-periodic behavior largely independent of the global accretion state, consistent with the presence of shock driven or torus driven variability. Overall, for the cases shown in Fig.\ref{mass_acc} in which the BH is non-rotating or slowly rotating, it is clearly observed that the magnetic curvature parameter plays a significant role in determining both the onset time of the physical mechanisms and the strength of the oscillations that develop in the accretion flow.

In contrast, as clearly seen in the upper-left panels of Figs.\ref{QPOs_a09B0005} and \ref{QPOs_a09B001}, the behavior of the mass accretion rates around rapidly rotating BHs exhibits a far more dramatic and distinct character for different values of the magnetic curvature parameter $B$. In both cases, the early phase of the simulation shows that once the classical shock cone forms, it quickly transitions into a strong flip-flop oscillation. However, whether the system remains in this flip-flop state or instead transitions to a torus-like structure is predominantly determined by the value of $B$. In the smaller case $B$ shown in Fig.\ref{QPOs_a09B0005}, the shock that undergoes flip-flop oscillations disappears and the toroidal structure emerges around $t \approx 18000M$, whereas in the larger case $B$ presented in Fig.\ref{QPOs_a09B001}, this transition occurs significantly earlier, at approximately $t \approx 7000M$. Furthermore, the upper-left panels of Figs.\ref{QPOs_a09B0005} and \ref{QPOs_a09B001} clearly demonstrate that the influence of the black-hole spin parameter is felt most strongly in the innermost region, particularly at $r = 2.3M$ where the gravitational field is strongest. The spin not only modifies the amplitude of the oscillations but also increases the overall amount of matter falling toward the BH.

To study the effects of black-hole spin $a$ and magnetic curvature $B$ on accretion, Fig.~\ref{mass_acc} shows the mass accretion rate at three radii: $r = 2.3M$ (near the horizon), $r = 6.1M$ (close to ISCO), and $r = 12M$ (weaker gravity). For non-rotating and slowly rotating KBR BHs, the accretion rate initially declines smoothly following the onset of BHL inflow. Later, the system becomes dynamically active and the accretion rate fluctuates significantly. Clear differences appear between $B = 0.005\,M^{-1}$ and $B = 0.01\,M^{-1}$. After the classical shock cone forms, the response of the flow varies: for $B = 0.01\,M^{-1}$, the accretion rate drops rapidly at all spins and radii, while for $B = 0.005\,M^{-1}$, it rises during the same interval.

\subsection{Azimuthal Evolution of Accretion Structures}

The azimuthal evolution of the accreting plasma provides a clear diagnostic of the dynamical response of the flow to the underlying spacetime and magnetic curvature. Tracking the temporal variation of the azimuthal modulation of density and shock features at fixed radial locations allows one to follow the formation, destabilization, and reorganization of accretion structures. This representation highlights how initially coherent shock patterns evolve into more complex configurations as the accretion flow adjusts to the KBR geometry.

\begin{figure*}[!ht]
  \vspace{1cm}
  \center
  \psfig{file=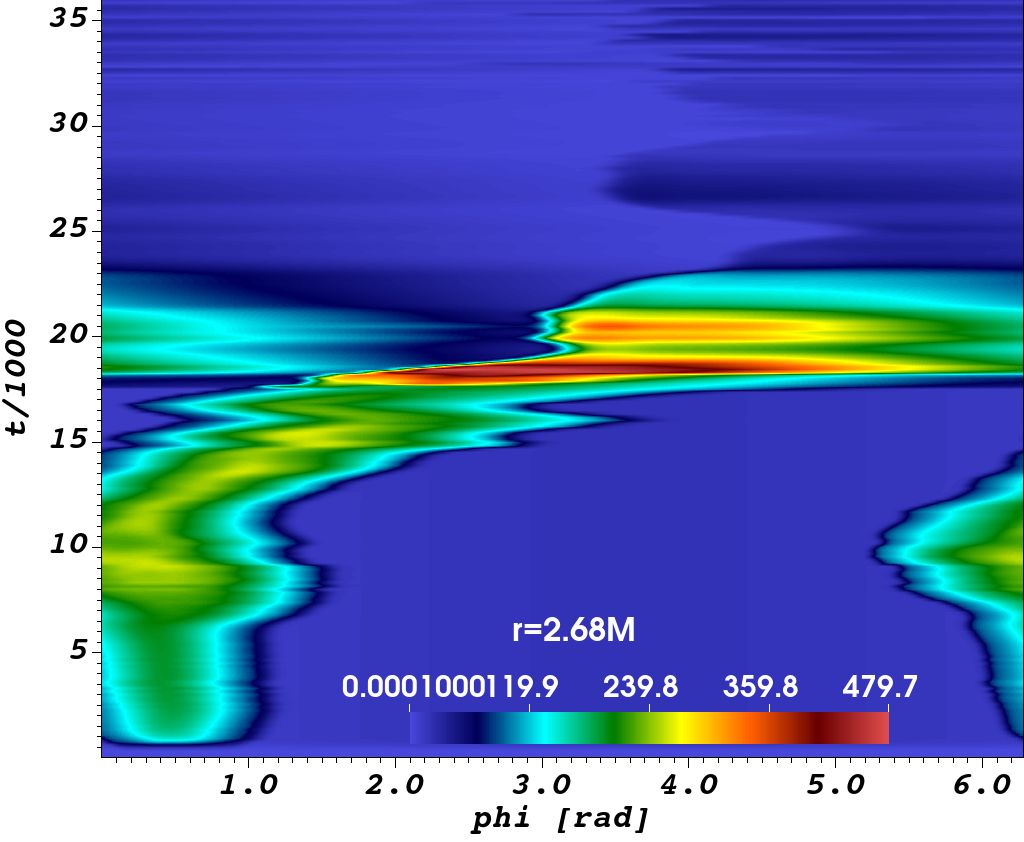,width=7.5cm,height=7.0cm}
  \psfig{file=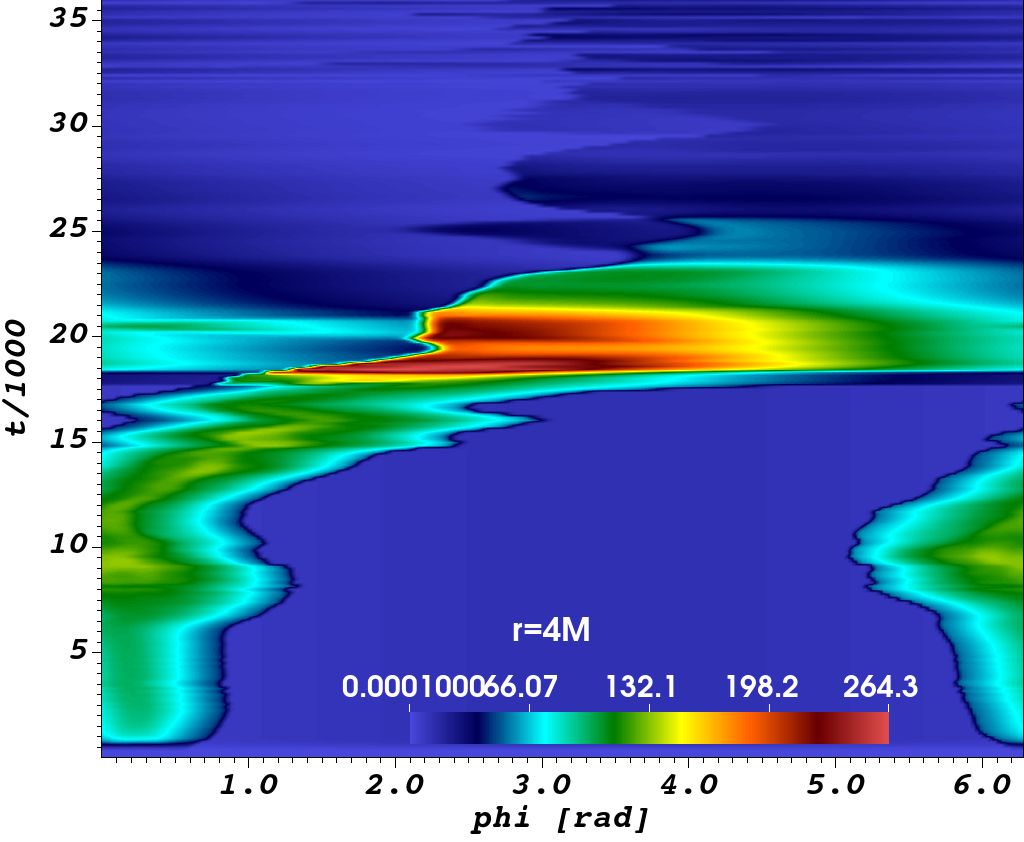,width=7.5cm,height=7.0cm}\\
  \psfig{file=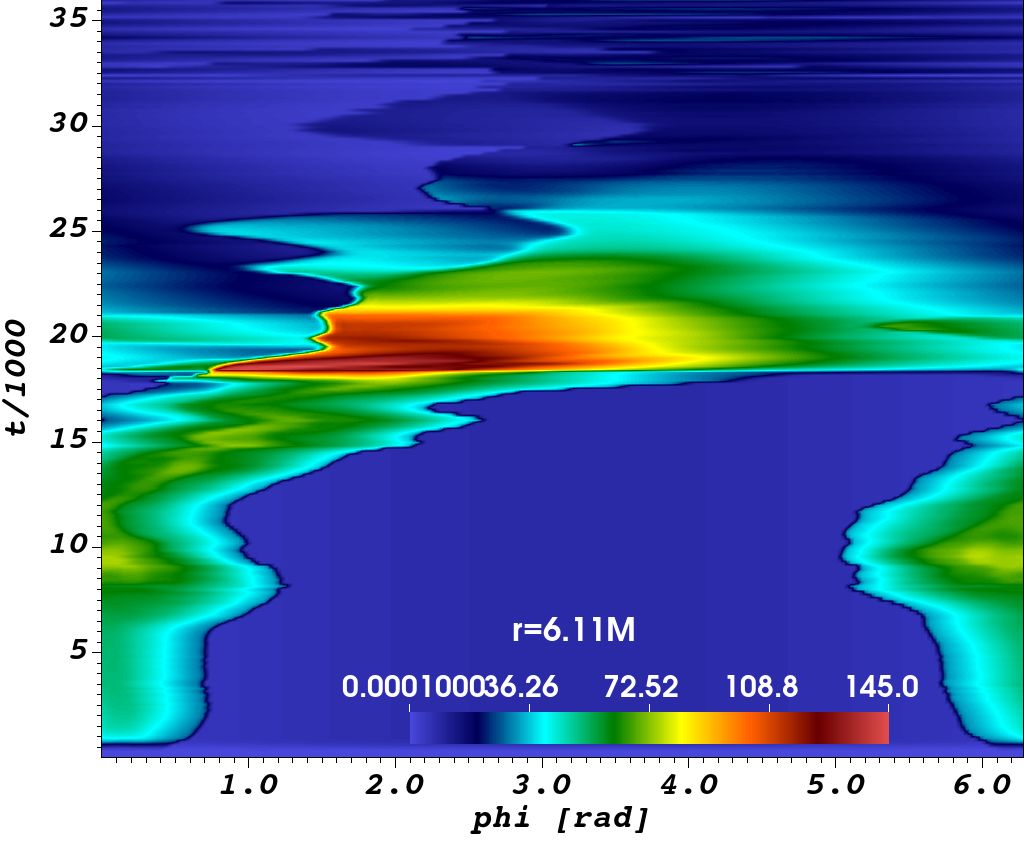,width=7.5cm,height=7.0cm}
  \psfig{file=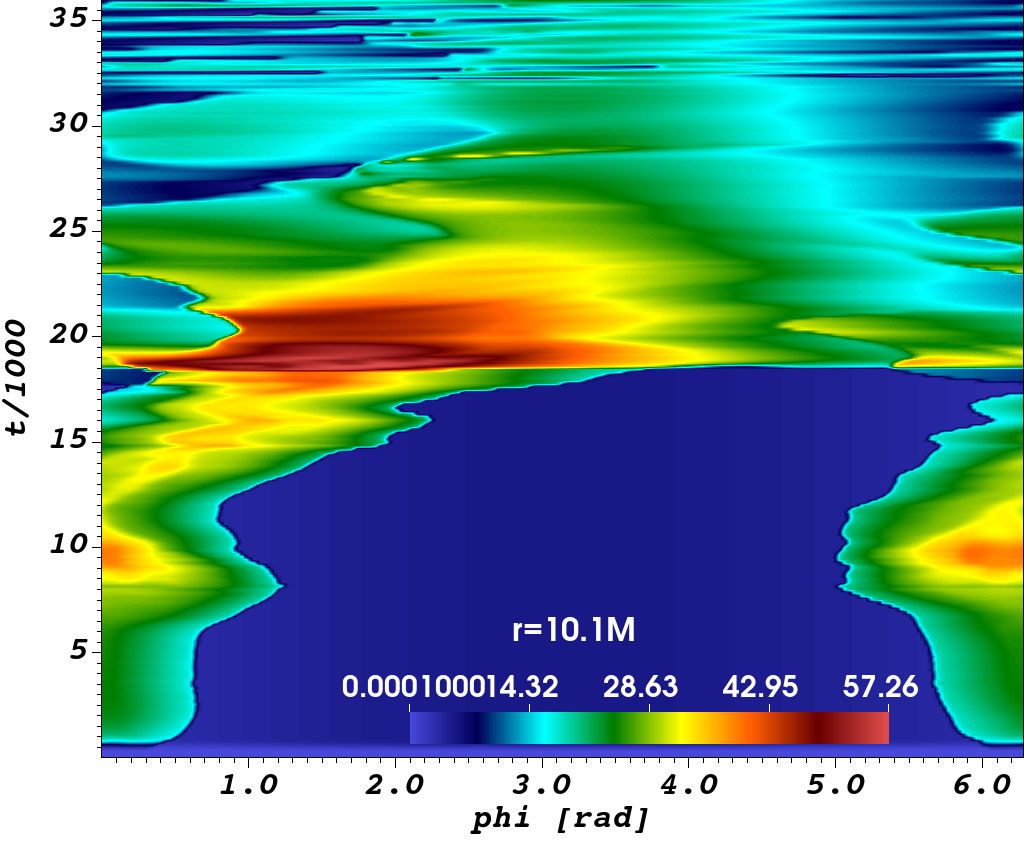,width=7.5cm,height=7.0cm}\\
  \psfig{file=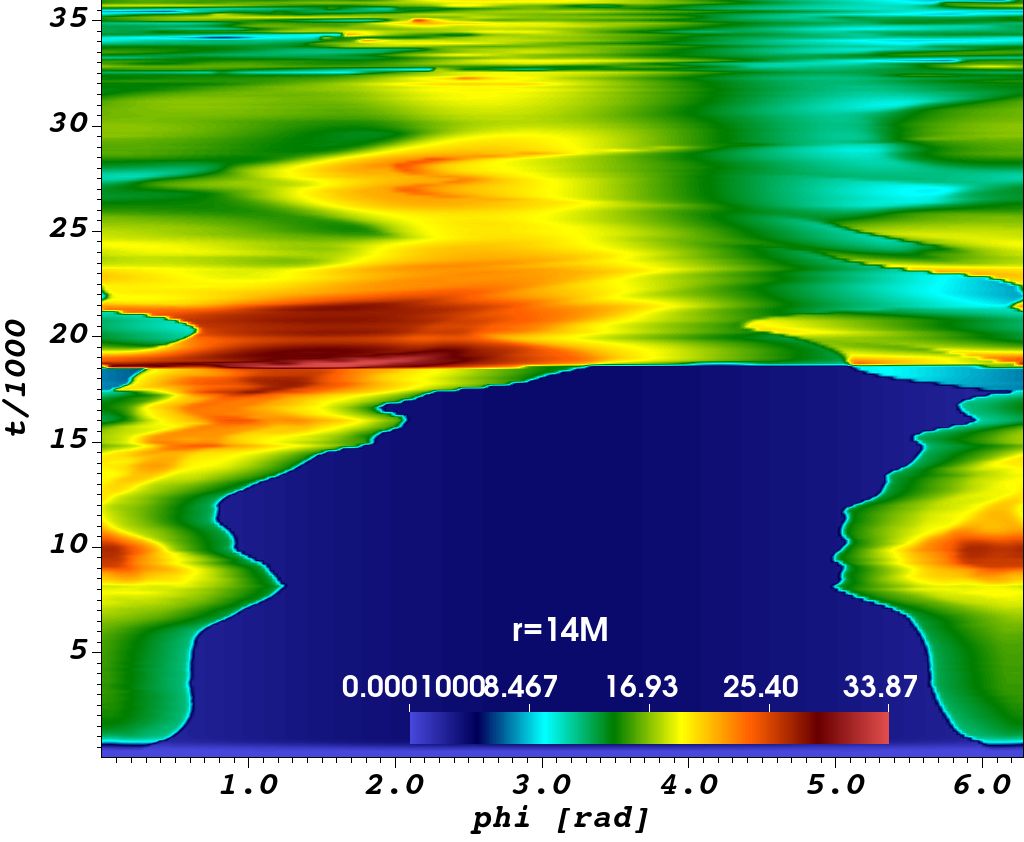,width=7.5cm,height=7.0cm}
  \psfig{file=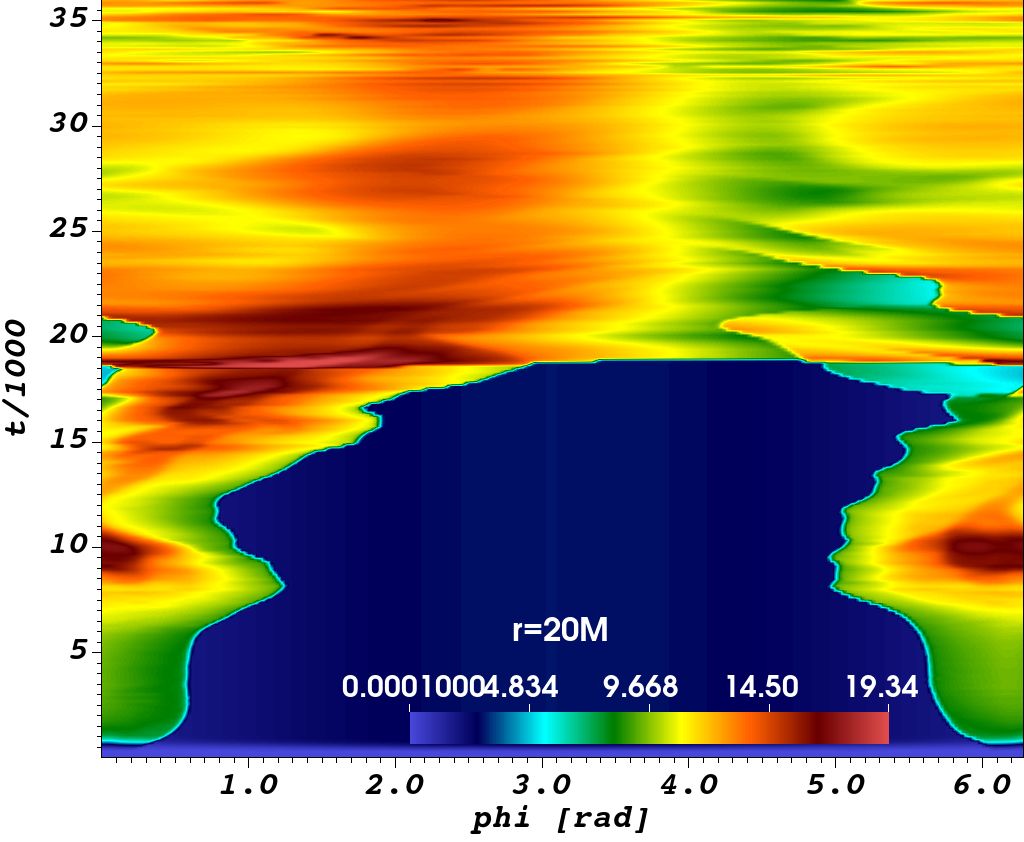,width=7.5cm,height=7.0cm}  
     \caption{  Visualization of the azimuthal variation of the plasma and shock structure formed by the infalling matter toward the BH over time around the KBR BH with spin parameter $a=0.9M$ with $B=0.005(1/M)$, measured at different fixed radial positions. The horizontal axis represents the azimuthal coordinate, while the vertical axis shows the normalized time. 
}
\vspace{1cm}
\label{fixed_t_phi_B0005}
\end{figure*}

At the radii closest to the horizon in Fig.\ref{fixed_t_phi_B0005}, $r=2.68M$ and $r=4M$, the first Kerr type shock cone forms rapidly and is followed by a high density region trapped inside the cone. Between $t=3000M$ and $t=20000M$, this high density stripe undergoes large lateral excursions in $\phi$, appearing as a slanted and intermittently bifurcating band. This reflects the strong flip-flop oscillation of the shock cone in the near horizon region. After $t=20000M$, the pattern changes qualitatively. The dense region broadens and becomes far less localized in azimuth, gradually forming an almost ring like structure with weaker embedded spiral modulations. This marks the transition of the flow within these radii into a torus dominated accretion state. However, due to the strong gravitational field near the horizon at $r=2.68M$ (and to a lesser extent at $r=4M$), the material continues to fall inward, causing the density to drop sharply and effectively erasing the torus signature at these inner locations. However, the behavior at $r=4M$ shows that the density begins to increase again for $t\gtrsim21000M$, indicating that the outer layers of the torus start to contribute to the density at this radius.

This increase becomes more pronounced beginning at $r=6.11M$ given in Fig.\ref{fixed_t_phi_B0005}, clearly showing that the torus--like structure forms between approximately $4M$ and $6M$, i.e., near the ISCO or slightly interior to it. At $r=6.11M$, where ISCO scale physics dominates, transitions between different dynamical mechanisms appear most distinctly. During the shock cone epoch, the dense azimuthal band is present but becomes increasingly fragmented.  The multiple high density patches appear and the band intermittently splits, demonstrating that by the ISCO region the flip-flop instability generates strong shear and secondary shocks. Peak density excursions demonstrate the oscillatory shock front's outward propagation and occur later in time than at the inner radii. Nevertheless, coherent shock-driven structures continue to exist at the ISCO radius, as evidenced by the density distribution's non-uniformity in azimuth.

As seen in Fig.\ref{fixed_t_phi_B0005}, the flip-flop oscillation phase and the early classical shock cone are observable from $r \geq10.1M$ until about $t=20000M$. Beyond this time, however, the torus that has fully developed in the inner region begins to engulf the BH, forming broad azimuthally extended high density structures. These radii are located inside the outer torus or its outer boundary, according to the behavior in $r=10.1M$, $r=14M$, and $r=20M$. The quasi-circular density ring is no longer clearly defined, but the inflow is still modulated by the spiral shocks and torsional motions. The pattern becomes more diffuse at even larger radii, indicating the dominance of outward propagating disturbances coming from the torus and inner shock structures, as well as the diminishing gravitational influence.

Generally speaking, the density is first concentrated in a small area at $r=2.68M$ and $r=4M$, creating a coherent azimuthal band that clearly demonstrates the existence of the conventional downstream shock cone. Strong flip-flop oscillations are confirmed to exist in the near-horizon area when this band later exhibits substantial lateral excursions and occasional splitting. On the other hand, the same band splits into several high-density patches at $r=6.11M$, indicating that as the oscillations spread outward, shear-driven secondary shocks appear, and the shock-induced variability diminishes. The early flip-flop oscillations are still discernible at wider radii, specifically $r=10.1M$, $r=14M$, and $r=20M$, but the density enhancements become less coherent and broader, suggesting significant azimuthal smearing . After $t=20000M$, the outer disk surrounding the BH becomes dominant, producing nearly continuous high-density sectors associated with the fully developed torus and its spiral shock extensions.Since densities at smaller radii decrease due to continuous infall, while densities at and beyond the ISCO increase and change into a torus-dominated, quasi-antisymmetric configuration, the combined comparison confirms that the inner radius of the torus must lie between roughly $4M$ and $6M$.

These results show that the variability of the simulated flow is not random, but reflects the underlying structure of the accretion state. The radial and azimuthal evolution consistently indicate that different dynamical regimes develop in time, with clear transitions between shock-dominated and torus-dominated behavior. 

\subsection{Power Spectral Analysis of Accretion Variability Across Spin and Magnetic Configurations}
\label{Sec7}

\begin{figure*}[!ht]
  \vspace{1cm}
  \center
  \psfig{file=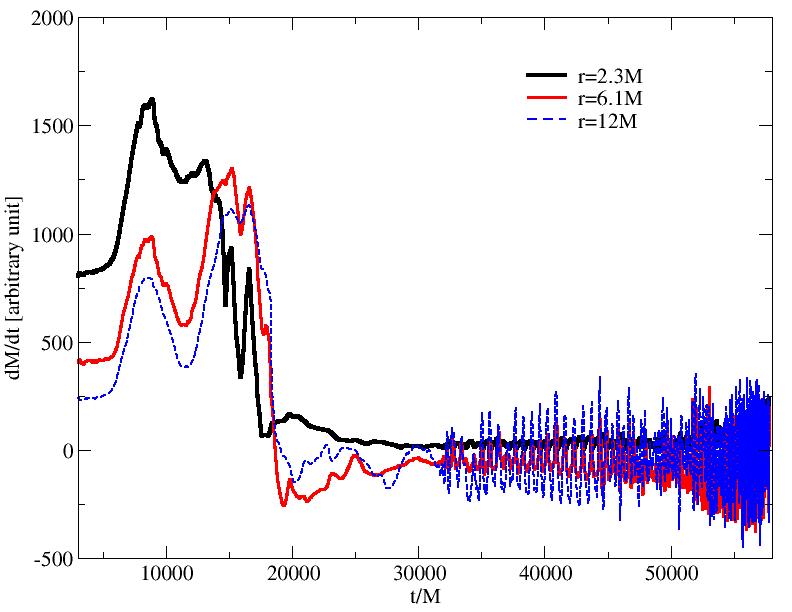,width=7.5cm,height=7.0cm}
  \psfig{file=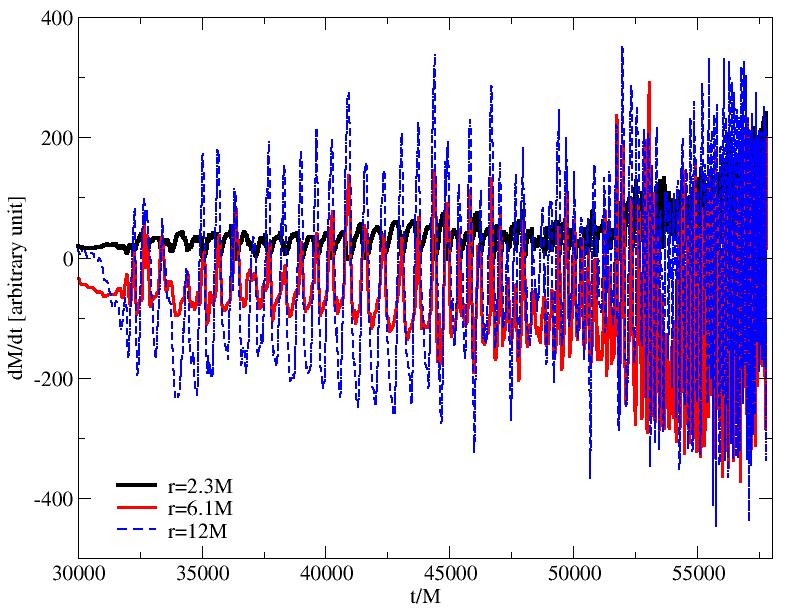,width=7.5cm,height=7.0cm}\\
  \psfig{file=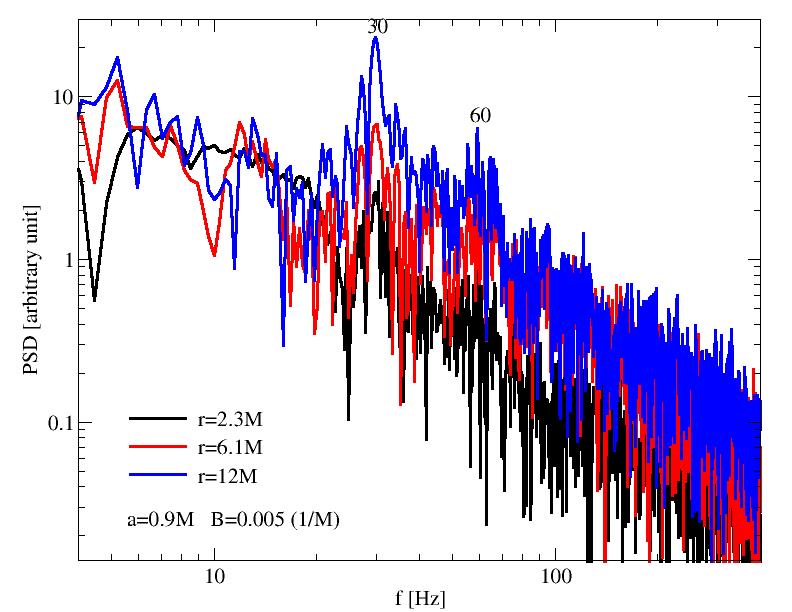,width=7.5cm,height=7.0cm}
  \psfig{file=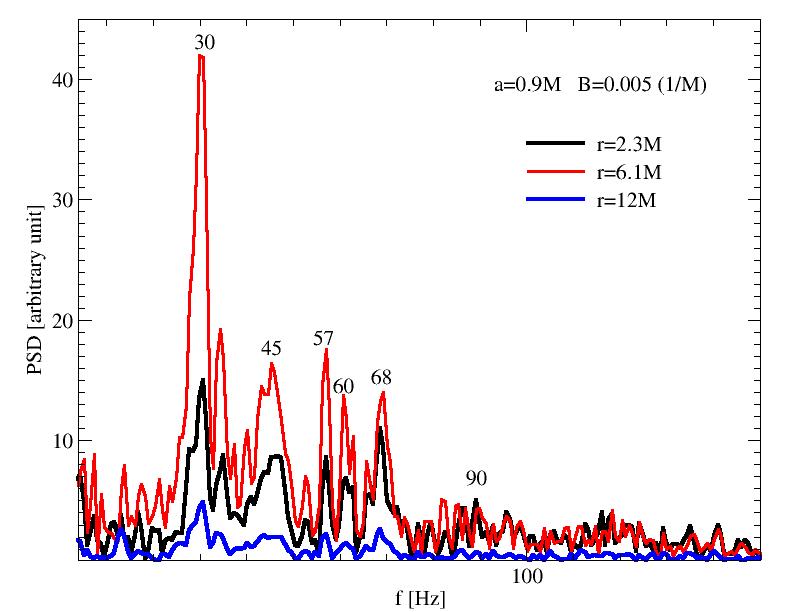,width=7.5cm,height=7.0cm}
\caption{Upper-left panel: Time evolution of the mass accretion rate for the rapidly rotating KBR BH with $a = 0.9M$ and $B = 0.005 (1/M)$, measured at three radial locations ($r=2.3M$,$r=6.1M$, and $r=12M$). The curves show the accretion dynamics from the moment the first shock cone forms (around $t \approx 3000M$) until the end of the simulation at $t \approx 58000M$. Bottom left panel: PSD of the accretion rate fluctuations shown in the upper-left panel, revealing several prominent QPO peaks associated with the oscillatory motion of the shock structure. Top-right panel: Mass accretion rate at the same radii after the toroidal density structure forms around $t \approx 30000M$, followed again until the end of the simulation. Bottom right panel: Corresponding PSD computed from the upper-right time series, highlighting the dominant frequency components generated by the QPOs of the inner torus and the reformed spiral shock wave.
}
\vspace{1cm}
\label{QPOs_a09B0005}
\end{figure*}

The mass accretion rate serves as a dynamical diagnostic of relativistic flows, providing direct information about the accretion mechanism and the physical processes operating in the system. Both the local hydrodynamic instabilities and the global oscillation modes of the disk-shock-cone arrangement imprint themselves as oscillations in the accretion rate, making such information accessible.  In Fig.\ref{QPOs_a09B0005}, for a rapidly rotating BH with $a=0.9M$ and the magnetic curvature parameter $B=0.005(1/M)$, we analyse the long-term and short-term accretion rate time series, compute the corresponding PSDs, and physically interpret the numerically obtained QPO frequencies. Throughout all QPO analyses in this work, the BH mass is set to $M=10M_{\odot}$. The upper-left panel of Fig.\ref{QPOs_a09B0005} shows the full evolution of the accretion rate from the moment the magnetic curvature destroys the classical shock cone and triggers the flip-flop instability at $t\approx3000M$, up to the end of the simulation. In this time, the flow changes structurally from a flip-flop shock cone that was first formed and was dominated by lateral oscillations and advected acoustic perturbations to a quasi-stationary toroidal configuration whose dynamics are controlled by inertial acoustic and epicyclic-type oscillation modes. Consequently, the PSD obtained from the long-term accretion rate exhibits a broad spectrum containing low- and intermediate-frequency components associated with global sloshing and vortex shedding motions of the shock cone, as well as higher frequency features produced by the radial breathing, vertical sloshing, and trapped p-mode oscillations of the inner torus. On the other hand, the peaks become much sharper and distinct QPOs occur at $f=30, 45, 57,$ and $68$~Hz when the PSD is only calculated during the time where a quasi-periodically oscillating torus exists ($t=30000M$-$55000M$). These frequencies are understood as combinations of the nonlinear couplings between the radial and azimuthal epicyclic frequencies. The presence of these frequencies at several radii ($2.3M$, $6.1M$, $12M$) shows that pressure waves and spiral density perturbations triggered by the torus or shock structure spread throughout the flow, resulting in synchronous modulation of the accretion flux even far from the oscillation region. There is also a physical significance to the variation in peak amplitudes throughout the radii. The long-term PSD has greater power at $r=12M$ during the shock-dominated phase because the outer flow is more strongly perturbed, while the short-term PSD has higher amplitudes at $r=6.1M$ in the torus regime due to the concentration of kinetic and compressional energy close to the pressure maximum. 

 The QPO frequencies obtained in both regimes lie entirely within the observed ranges of type-C LFQPOs ($0.1-30$Hz) and the lower HFQPO band ($40-70$Hz). Therefore, mode excitation in magnetically curved spacetimes such as the KBR metric can naturally account for the multi-epoch, multi-frequency timing properties observed in X-ray binaries, including the persistence of specific QPO peaks across spectral states and luminosity changes.

\begin{figure*}[!ht]
  \vspace{1cm}
  \center
  \psfig{file=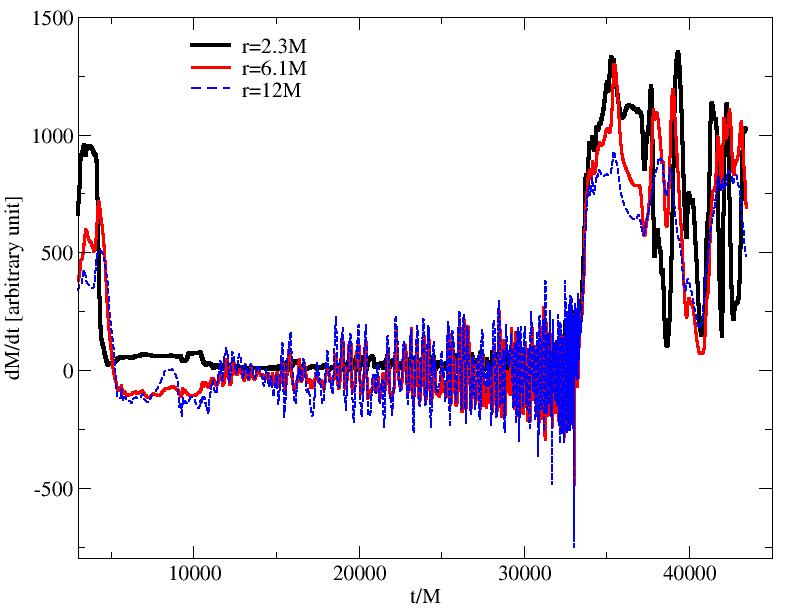,width=5.5cm,height=7.0cm}
  \psfig{file=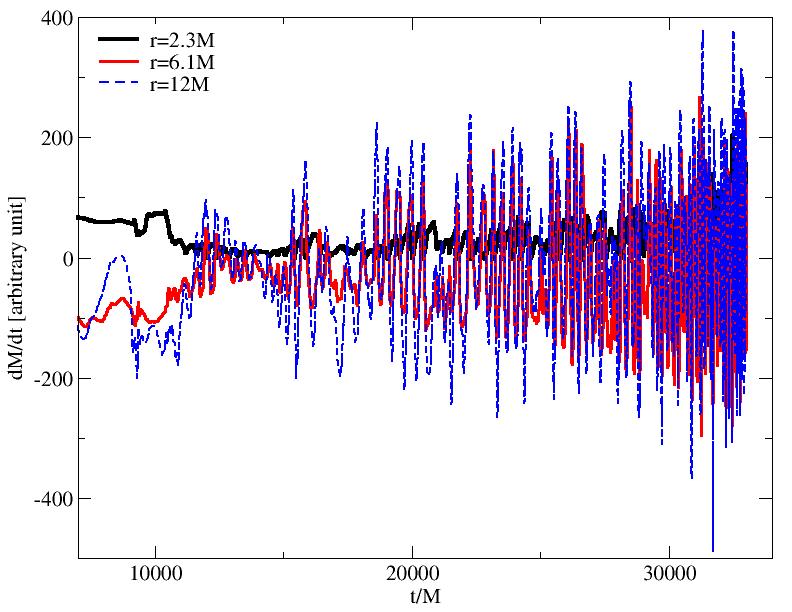,width=5.5cm,height=7.0cm}
  \psfig{file=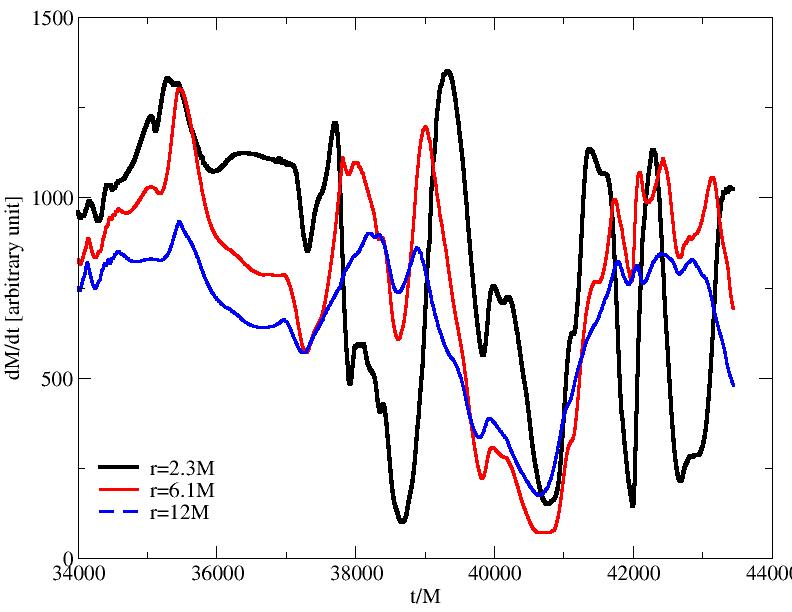,width=5.5cm,height=7.0cm} \\
  \psfig{file=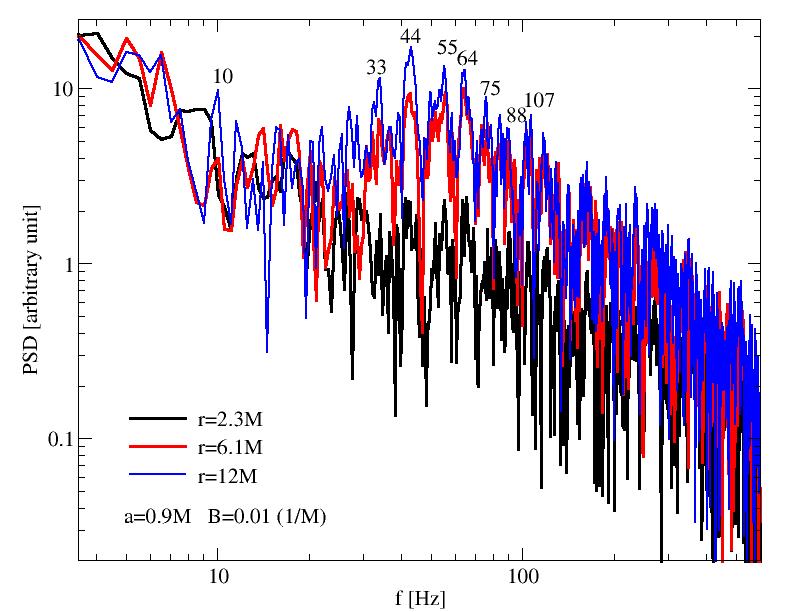,width=5.5cm,height=7.0cm}
  \psfig{file=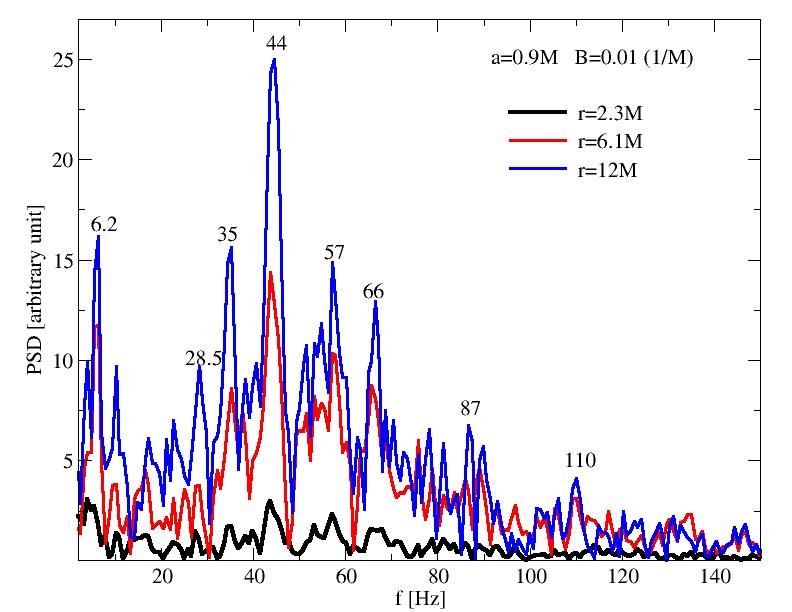,width=5.5cm,height=7.0cm}
  \psfig{file=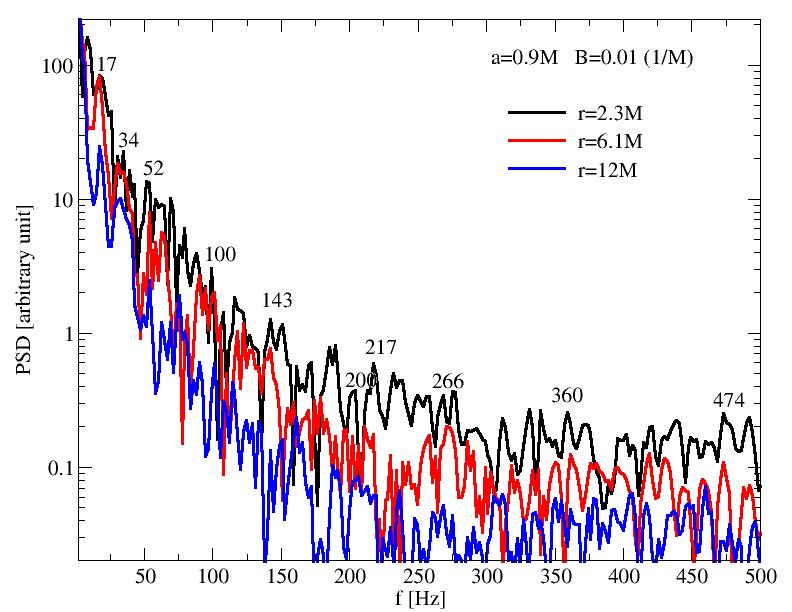,width=5.5cm,height=7.0cm}   
\caption{This figure presents the same type of analysis as Fig.\ref{QPOs_a09B0005}, but now for the case $B = 0.01(1/M)$. Since the formation of the first shock cone is again completed at approximately $t = 3000M$, the upper-left panel shows the mass accretion rate from $t = 3000M$ until the end of the simulation at $t = 43000M$. The two upper panels on the right display the accretion rate evolution in two distinct intervals: an intermediate phase between $t = 7000M$ and $t = 34000M$, and a later phase between $t = 34000M$ and $t = 43000M$. The PSD analyses shown in the lower panels are computed directly from their corresponding accretion rate time series above, illustrating the dominant quasi--periodic components active during each evolutionary stage.
}
\vspace{1cm}
\label{QPOs_a09B001}
\end{figure*}

In Fig.\ref{QPOs_a09B001}, the behavior of the mass accretion rate computed at different radii around the rapidly rotating KBR BH with $a=0.9M$ and the magnetic curvature parameter $B=0.01(1/M)$ is shown, together with the corresponding PSD analyses obtained from each time interval. The upper-left panel presents the full evolution of the accretion rate from the moment the classical shock cone forms until the end of the simulation (from $t=3000M$ to $t=44000M$). As seen in this figure, a strong flip-flop oscillation appears in the early phase, but after approximately $t=7000M$ a toroidal structure forms around the BH. This quasi-periodically oscillating torus persists between $t=7000M$ and $t=34000M$, after which the flow returns to a second flip-flop phase in which the opening angle and the strength of the oscillations again show large variations. The mass accretion rates corresponding to these different dynamical phases are shown in the upper middle and upper-right panels of Fig.\ref{QPOs_a09B001}.

The bottom row displays the PSDs computed from each of these accretion rate data sets. These PSDs demonstrate that the accretion rate signals, each containing different physical mechanisms, produce broad sets of QPO peaks resulting from the combined behavior of the two distinct oscillation regimes. In the middle PSD (corresponding to the toroidal structure), the coherent torus oscillates quasi-periodically, leading to sharper peaks that trace intrinsic torus modes such as radial breathing, trapped inertial acoustic modes, and epicyclic type oscillations imposed by the KBR spacetime. The fact that the same QPO frequencies also appear in the long-term PSD (computed from the full accretion rate evolution) shows that these modes are robust, governed by the spacetime geometry of the KBR BH, and persist even when the accretion morphology changes. Whenever the torus forms, these modes are re-excited and generate the same characteristic frequencies. This indicates that these QPOs are not produced by transient instabilities associated with the transitions between physical states, but instead arise from persistent global spacetime modes, making them suitable candidates for observational identification. This is consistent with the fact that real X-ray binaries often show the same QPO frequencies recurring across different epochs and spectral states. The dominant frequencies obtained from the final flip-flop phase highlight the characteristic oscillations of the renewed shock structure. 

These peaks represent global sloshing of the shock cone, lateral excursions, and spiral shock reformation, and they connect naturally to the other two PSDs because the long-term PSD contains contributions from both the initial and final flip-flop phases. This again demonstrates that the shock oscillations remain modulated by the same underlying relativistic characteristic frequencies of the spacetime. As seen in all three PSDs, the same QPO frequencies occur at different radii ($2.3M$, $6.1M$, $12M$). This shows that global modes excited by the torus or the shock cone generate pressure waves and spiral density perturbations that propagate outward and coherently modulate the mass accretion rate, producing synchronized QPO signals even at radii far from the oscillation region. Finally, the stronger peak amplitudes at $r=12M$ in the first two PSDs indicate that the dominant perturbation energy resides in the outer flow during the shock dominated early stage and the intermediate torus-plus-spiral extension phase. In these regimes, the outer flow lies directly in the path of large scale shock sloshing and spiral arms, causing strong density and velocity fluctuations as the perturbations sweep through the weak field region and therefore producing larger PSD amplitudes. In contrast, although a strong response near the ISCO might be expected in a purely torus dominated state, for the $B=0.01$ case, the outer spiral and shock structures remain dynamically significant. This explains why the peaks at $r=12M$ remain strong. Overall, the persistent global modes responsible for the QPOs arise directly from how the KBR spacetime modifies the classical shock cone. The large oscillations observed in the accretion rate (and therefore in luminosity) are physically meaningful, observable quantities, and the resulting QPOs are well suited for interpreting multi epoch, multi frequency QPO phenomenology in X-ray binaries, including the coexistence and recurrence of low-frequency like shock modes and high-frequency like torus/epicyclic modes over time.

\begin{figure*}[!ht]
  \vspace{1cm}
  \center
  \psfig{file=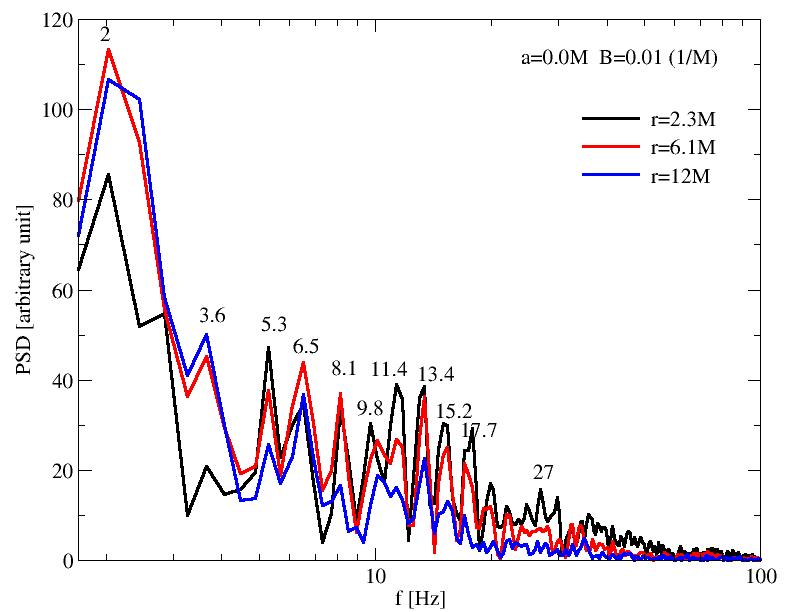,width=15.0cm,height=7.0cm}
  \psfig{file=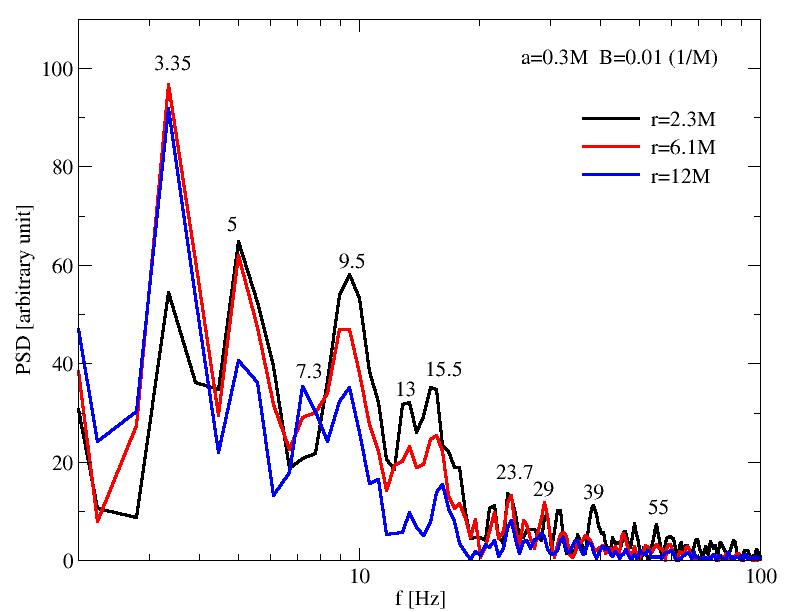,width=15.0cm,height=7.0cm}
  \psfig{file=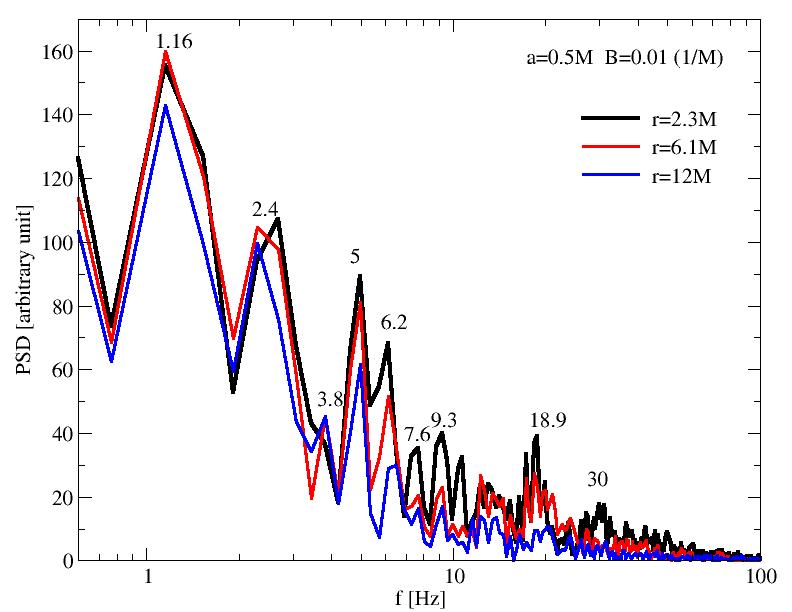,width=15.0cm,height=7.0cm}  
\caption{Using the mass accretion rates shown in Fig.\ref{mass_acc}, the PSD analyses at different radial locations are presented for the case in which the KBR BH is non-rotating or slowly rotating, with the magnetic parameter fixed at $B = 0.01 (1/M)$. The upper panel corresponds to the non-rotating case $a = 0.0M$, showing the PSD computed from the accretion rate fluctuations during the interval immediately after the formation of the shock cone and continuing until the end of the simulation. The middle and bottom panels display the PSD results for the slowly and mildly rotating BH cases with $a = 0.3M$ and $a = 0.5M$, respectively.
}
\vspace{1cm}
\label{QPOs_low_a}
\end{figure*}

In Fig.\ref{QPOs_low_a}, the PSD analyses of oscillation modes excited around the KBR BH with $B = 0.01(1/M)$ are presented for different BH spin parameters, namely $a=0M$, $a=0.3M$, and $a=0.5M$, corresponding to the non-rotating or slowly rotating cases. PSDs are computed using the mass accretion rates shown in Fig.\ref{mass_acc}. Each PSD peak is obtained from the accretion rate fluctuations calculated from the moment the classical (Kerr-like) shock cone forms until the end of the simulation. The resulting QPO frequencies form a rich structure, generally appearing at or below $\sim30$Hz. These QPOs arise entirely from the sloshing motion of the shock structures formed around the BH because in these models the BH either does not rotate or rotates slowly, meaning that epicyclic oscillations driven by strong gravitational frame dragging are not dominant, and neither shock nor torus structures generate gravity driven epicyclic modes. As shown in Fig.\ref{QPOs_low_a}, increasing the spin parameter slightly modifies the peak structure, making it more complex and altering the spacing between consecutive peaks. Even though the spin values are small, the curvature of spacetime changes the effective gravitational potential, which in turn affects the stability and oscillation behavior of the physical mechanism. Spin also causes some peaks to shift slightly toward higher frequencies because the post-shock flow acquires modest angular momentum due to frame dragging and asymmetric accretion. Overall, a slowly rotating BH produces QPO frequencies over a broader spectral range, enhances harmonic content, and slightly increases characteristic frequencies. For a fixed spin parameter, the QPOs computed at different radii occur at the same frequencies, which shows that the oscillations are global. Once excited, shock sloshing and spiral shock structures launch pressure waves that propagate outward and modulate the accretion rate at every radius. Even in non-rotating or slowly rotating BH cases, the KBR spacetime under BHL accretion excites global oscillation modes. As discussed in detail in Figs.\ref{QPOs_a09B0005} and \ref{QPOs_a09B001}, rapidly rotating BHs strongly influence the resulting physical structures and the associated QPO behavior.

In summary, the power spectral analysis shows that the numerical simulations naturally generate multiple QPO-like components associated with different flow configurations. The lower-frequency features are mainly linked to large-scale shock oscillations, whereas the higher-frequency peaks become stronger when a toroidal structure forms. This confirms that the simulated variability carries clear signatures of the changing accretion morphology.

\section{Comparison with Observed QPO Variability in X-ray Binaries and Physical Interpretation of Model Transitions}
\label{Sec8}

The variability analysis shows that the simulated accretion flow naturally produces multiple QPO-like components with distinct frequency ranges. In the following discussion, we focus on the main phenomenological trends and their possible astrophysical relevance, while more detailed remarks are kept concise for readability.

Before discussing individual source classes, we emphasize the scope of the present comparison. The frequencies extracted from our numerical simulations are used here only in a qualitative and phenomenological sense, namely to examine whether the dynamical timescales generated by the KBR accretion flow fall within ranges relevant to observed QPO behavior in X-ray binaries. We do not claim a direct fit to observational data, nor do we attempt to establish a unique identification between the simulated modes and specific observed QPO peaks. Such an inference would require a substantially more complete framework including emission modeling, source-specific parameter estimation, timing statistics, and a careful treatment of observational and theoretical degeneracies.

In the numerical simulations, two different shock-dominated physical configurations are observed which are a wide opening shock structure that exhibits strong flip-flop oscillations, and a toroidal configuration that undergoes QPOs. Owing to the strong interaction between the BH spin and the magnetic curvature term, both of these shock driven mechanisms appear in the rapidly rotating BH scenarios, and their morphological properties have been discussed in detail above. Depending on the value of $B$, these two distinct configurations arise in different time intervals, and each of them produces characteristic QPO frequencies in the PSD of the mass accretion rate. These frequencies change according to which mechanism dominates. In general, in the toroidal configuration we numerically obtain LFQPOs together with moderate amplitude HFQPOs (typically associated with the radial and azimuthal epicyclic frequencies and their nonlinear couplings), whereas in the purely strong flip-flop phase only HFQPOs are present. Those HF frequencies are mainly produced by modes trapped within the wide angle cone and oscillating together with the cone structure. Observationally, X-ray binary systems show two broad QPO families,  LFQPOs in the $0.1$-$30$Hz range, which arise mostly in hard and hard-intermediate states from the precession of the disk or inner hot flow in the strong gravity region, and rarer HFQPOs in the $40$-$450$Hz range, often appearing in frequency ratios such as 3:2 \cite{Orh1}. In particular, the sources GRS~1915+105, XTE~J1550-564, GRO~J1655-40 and H1743-322 each exhibit recurrent QPOs at multiple distinct frequencies from the same object. The observed signals can include simultaneous LFQPOs and HFQPOs. For example, GRS~1915+105 shows LFQPOs in the $0.1$-$10$~Hz band together with additional QPOs in $34$, $41$ and $67$Hz \cite{Orh2,Orh2_1}. GRO~J1655-40 displays QPOs at $300$-$450$ Hz, while XTE~J1550-564 shows QPOs at $184$-$276$ Hz \cite{Orh2}. H1743-322 exhibits HFQPOs at approximately $160$ and $240$Hz that are detected only in intervals where $0.1$-$20$Hz QPOs are present \cite{Orh3}. 

These findings show that at least two families of oscillations can be found in a single source: a low-frequency family associated with the global disk/corona geometry and a high-frequency family associated with motion close to the ISCO, whose frequencies are extremely consistent throughout spectral states and outbursts.   In this context, our numerical results for the KBR BH with spin parameter $a=0.9M$ provide a concrete hydrodynamic realization of such multi frequency behavior. The type-C LFQPO band (with a mass $M=10M_\odot$, about $0.1$-$30$ Hz) is populated by low-frequency peaks produced by flip-flop shock oscillations and large-scale spiral shock waves, whereas the torus-dominated frequencies in the range $30$-$70$ Hz correspond to moderate HFQPOs. For instance, the pair $41/67$ Hz seen in GRS~1915+105 is congruent with the pair $44/66$Hz seen in the simulations.  Because these numerically found frequencies are not produced by transient states during transitions between different mechanisms, but instead by epicyclic and internal inertial acoustic modes set by the spacetime geometry, the same frequencies reappear whenever a toroidal or flip-flop shock structure is reestablished. This naturally leads to the repeated occurrence of LFQPOs and HFQPOs at nearly fixed frequencies in different epochs, as observed in GRS~1915+105, XTE~J1550-564 and H1743-322 \cite{Orh4}. We also showed that the same frequencies are present at different radii, implying that in shock--dominated phases larger radii exhibit higher amplitudes, whereas in torus-dominated phases the strongest response can occur near the ISCO. This prolonged, global character is reflected in the common relationship of LFQPOs with broad band continuum variability, whereas HFQPOs are associated with increased energy output and shocks or oscillations created in the inner strong gravity zone.

A rich QPO frequency morphology appears when $a=0.9$ and $B=0.01(1/M)$. QPOs are calculated in the toroidal structure with quasi-periodic behavior during the early flip-flop phase and again during the late time phase when a strongly oscillating shock cone reforms. The peaks obtained in the PSD of the middle panel of this model correspond to sharp intrinsic torus frequencies. In contrast, the PSD computed from the final stage of the simulation during the renewed flip-flop phase highlights the global modes of wobbling spiral shock waves. Since the dominant frequencies produced in the numerical simulations are consistent with those appearing in the long-term PSD, the low-frequency components generated by the shock and the high-frequency modes generated by the torus arise from the same KBR spacetime effects and are present in all dynamical stages. This behavior provides a natural explanation for why HFQPOs in several BH binaries appear only in specific spectral states but always at nearly the same frequencies, and why LFQPOs can persist or reappear with frequencies similar to the source cycles between hard, intermediate, and soft states \cite{Orh3}.

The frequencies obtained in the present work should not be interpreted as direct observational predictions of astrophysical QPOs. It is well known that the interpretation of observed QPOs remains highly model-dependent and is subject to important experimental and theoretical limitations. In particular, the commonly used mapping between geodesic or epicyclic frequencies and observed QPO peaks is not unique, and precession-based interpretations, although physically informative, face known inconsistencies and incompletenesses when confronted with the full complexity of the data. Additional effects related to the internal structure and spin of the orbiting matter, anharmonic corrections in the strong-field region, radiative processes, and ambiguities inherent in simplified QPO models may all affect the observed frequencies and their interpretation \cite{Bianchini:2025zvp, Giambo:2025ukm, Boshkayev:2022haj}. Moreover, robust constraints from QPO observations generally require dedicated statistical inference frameworks, such as Bayesian or MCMC analyses, together with a careful treatment of systematics and parameter degeneracies \cite{Boshkayev:2023rhr}. For these reasons, the numerical frequencies obtained here should be regarded as providing qualitative insight into how the KBR deformation modifies the characteristic dynamical timescales of the accretion flow. A quantitative validation against observational QPO data lies beyond the scope of the present work and is left for future investigation.

\section{Conclusions}\label{Sec9} 

This study presents a comprehensive analysis of particle dynamics and epicyclic oscillations in KBR spacetime, with particular emphasis on the influence of BH parameters on particle trajectories, accretion processes, and potentially observable signatures. The parameters $a$ and $B$ exert a significant and nontrivial influence on key physical quantities, including the specific energy, angular momentum, and effective potential of the test particles. An increase in the magnetic field and spin of the BH has been found to reduce the energy of bound orbits. However, Kerr BHs systematically permit higher particle energies than their rotating KBR counterparts. The radial dependence of the oscillation eigenfrequencies \(\nu_j\), corresponding to small harmonic perturbations of neutral test particles moving on circular orbits around a rotating KBR BH, has been analyzed as a function of the  radial coordinate for different values of the dimensionless spin parameter \(a\) and the magnetic field parameter \(B\), as measured by an asymptotic observer. In the non-rotating limit (\(a=0\)), the radial and vertical epicyclic frequency profiles coincide, whereas for \(a \neq 0\) these two frequency branches split and exhibit qualitatively distinct radial behavior. Moreover, it is found that, in the KBR geometry, the radial epicyclic frequency profiles are displaced toward larger radii relative to those obtained for the Kerr spacetime with the same spin parameter.

The behavior of the magnetic parameter $B$ differs between the numerical simulations and the theoretical test-particle analysis, and this difference motivates the use of distinct ranges of $B$ in the present work. In the fully relativistic BHL accretion scenario, the inflowing matter must fall toward the BH in order for accretion to take place. This allows a physical shock mechanism to develop around the BH through accretion. Our numerical simulations show that, for accretion to occur around the BH via the BHL mechanism, the magnetic curvature parameter must satisfy $B \lesssim 0.01\,(1/M)$. When such values of $B$ are adopted in the numerical simulations, the matter is successfully captured by the gravitational field of the BH and accretes toward it in a stable manner. As a consequence of this accretion process, we observe in the simulations the successive emergence of a shock cone, followed by flip--flop oscillations and transitions to toroidal accretion structures.  
In contrast, for larger values of $B$, the magnetic curvature term becomes dominant in the effective potential, preventing the accretion of matter toward the BH. In this regime, the accretion flow is suppressed and the system evolves toward an effectively empty configuration, which is not astrophysically realistic for accreting BHs. For these reasons, values of $B \lesssim 0.01\,(1/M)$ are adopted in the numerical simulations to allow the formation of physically meaningful mechanisms that can provide insight into observations. At the same time, this constraint highlights a limitation on the testability of the KBR spacetime from an accretion-based perspective.

By contrast, in the theoretical part of the paper,  the matter content is modeled as test particles or linear perturbations that do not backreact on the spacetime. In this case, there is no requirement that matter be accreted, and larger values of $B$ can be consistently explored to isolate and understand how magnetic curvature modifies particle dynamics and wave propagation in the KBR geometry. The use of a wider range of B values in the theoretical analysis therefore serves a complementary purpose. It allows us to probe the intrinsic spacetime effects of magnetic curvature on orbital motion and oscillation spectra, independent of hydrodynamical constraints. This twofold strategy ensures that the parameter choices in both the numerical and analytical parts of the work are physically motivated, internally consistent, and tailored to the specific physical questions being addressed.

The accretion process around the KBR BH and the resulting QPO variability caused by the confinement of basic modes within the accretion flow have been statistically studied using the BHL mechanism. The findings demonstrate that the creation of unique shock-dominated morphologies and their corresponding timing signatures are jointly determined by BH spin and the magnetic curvature parameter of spacetime. Specifically, two shock-dominated topologies are quantitatively identified in the rapidly rotating regime: (i) a compact toroidal structure experiencing virtually quasi-periodic instabilities and (ii) a wide angle shock cone exhibiting intense flip-flop oscillations. Multi-state behavior in accreting BH sources can be naturally explained by these two physical mechanisms, which arise sequentially within the same system as shock-dominated morphologies and are not mutually exclusive.

PSD analysis shows two strong families of QPOs linked to different flow configurations. While the toroidal flow mainly stimulates HFQPOs through internal acoustic and epicyclic oscillations, LFQPOs are caused by large-scale shock oscillations and repetitive spiral shock generation. The fact that the same frequencies are present at various periods and radii suggests that the KBR geometry, not fleeting variations in the flow morphology, determines the dominant oscillations. In this way, rather than producing geometric frequencies, the accretion flow functions as a nonlinear filter that amplifies them. Our discussion of observed QPO ranges is intended as a phenomenological assessment of astrophysical relevance rather than as a direct observational fit. A quantitative test of the model will require future source-specific timing analyses, radiative modeling, and statistical inference.

The projected frequencies are within the reported LFQPO and HFQPO ranges of X-ray binaries when scaled from geometrized units to a stellar-mass BH with $M = 10M_\odot$. The model explains why variations in timing states do not cause the underlying frequency peaks to shift and easily explains the coexistence of numerous QPOs in a single source. These findings imply that magnetically curved BHs with fast rotation offer a consistent framework for investigating deviations from Kerr spacetime in the strong-field regime and for interpreting QPO occurrences.

The QNM analysis presented in this work reveals a clear and
systematic imprint of the magnetic field and rotation on the dynamical response
of the rotating KBR BH. Our results show that the
multipole number primarily governs the oscillation frequency of the ringdown
signal, exhibiting the expected eikonal scaling, while the magnetic field
parameter enhances the damping of scalar perturbations by modifying the
effective potential barrier and facilitating energy dissipation into the
horizon. In contrast, the spin parameter increases the real part of the QNM
frequencies through frame-dragging effects, with a comparatively milder
influence on the decay rates. The smooth evolution of the quasinormal spectrum
in the Argand plane and the absence of any growing modes in the time-domain
profiles confirm the stability of the spacetime under scalar perturbations
across the explored parameter range. Taken together, these findings indicate
that QNMs provide a sensitive probe of magnetic-field-induced
deformations and rotational effects in strong-gravity environments, and may
offer a potential observational window into magnetized rotating BHs
through future high-precision gravitational-wave measurements.

Finally, the combined impact of test-particle dynamics, QNMs, and fully relativistic accretion simulations yields a coherent and internally consistent picture. In each case, it is shown how the magnetic curvature parameter modifies the dynamics of the KBR spacetime. The shifts in characteristic orbital and epicyclic frequencies inferred from geodesic motion are mirrored by corresponding trends in the QNM spectrum, reflecting the common spacetime scales governing linear perturbations. These same scales manifest nonlinearly in the accretion flow through changes in shock stability, flip--flop instabilities, and toroidal-type accretion structures, leading to characteristic variability timescales in the numerical simulations. Although the connections between these regimes are qualitative, their mutual consistency supports the interpretation that the magnetic curvature parameter acts as a unifying control parameter linking particle motion, spacetime oscillations, and accretion dynamics within a single geometric framework.

\section*{Appendix}

\subsection*{Calculation of QNMs}
Owing to the presence of the magnetic-field–induced conformal factor and the angular function $P(\theta)$, the resulting angular equation is, in general, a magnetically deformed spheroidal equation with explicit dependence on $\omega$, $m_l$, and $B$. In the weak-field regime $(B^2\ll1)$, these deformations contribute only in subleading order, and the angular sector reduces to the standard scalar spheroidal harmonic equation used below:

\begin{equation}
\frac{1}{\sin\theta}\frac{d}{d\theta}\!\left(\sin\theta \frac{du_\theta}{d\theta}\right) + \left[a^2\omega^2\cos^2\theta - \frac{m_l^2}{\sin^2\theta} + \mathcal{A}_{lm_l}\right] u_\theta = 0,
\end{equation}

which generalizes the spheroidal harmonic equation. In the large limit $l$, the eigenvalue can be approximated using a WKB quantization rule as

\begin{equation}
\mathcal{A}_{lm_l}^{R} \simeq \Big(l+\tfrac{1}{2}\Big)^2 - \frac{a^2\omega^2}{2}\Bigg(1 - \frac{m_l^2}{(l+1/2)^2}\Bigg),
\end{equation}

and for damped modes ($\omega = \omega_R - i\omega_I$), the imaginary correction reads

\begin{equation}
\mathcal{A}_{lm_l}^{I} \simeq -2a^2\omega_R \omega_I \langle \cos^2\theta \rangle_{\text{WKB}}.
\end{equation}

The radial equation, after transformation to the tortoise coordinate $r_*$,

\begin{equation}
\frac{dr_*}{dr} = \frac{\Sigma(r)}{Q(r)}, \qquad \Sigma(r) = r^2+a^2,
\end{equation}
where
$
Q(r) = (1+B^2 r^2)\,\Delta(r).$

Oscillation frequency of ring-down gravitational waves, $\omega_R$ is given by,
\begin{widetext}
    \begin{equation}
\omega_R=  \frac{a m_l \left[2 A_1 M-r_0 \left\lbrace\left(a^2 B^2-2\right)^2+A_2\right\rbrace\right]}{r_0^3 \left\lbrace\left(a^2 B^2-2\right)^2+A_2\right\rbrace+a^2 r_0 \left\lbrace\left(a^2 B^2-2\right)^2-A_2\right\rbrace+2 a^2 A_1 M-6 A_1 M r_0^2},
\end{equation}
\end{widetext}
where we have used $A_1=a^4 B^4-3 a^2 B^2+2$ and $A_2=4 B^2 M^2 \left(a^2 B^2-1\right).$

The explicit form of $\omega_I$ is found to be
\begin{widetext}
    \begin{equation}
        \omega_I = -\frac{(2 l+1)^2 (2 n+1) \sqrt{-A_3^2 \left(A_4 \left(A_3 A_8 \left(a^2 B^2-2\right)^2+A_5 \left(a^2+r_0^2\right)\right)-2 a A_7 \left(a^2 B^2-2\right)^2 m_l\right)}}{\left(a^2 B^2-2\right) \left(a^2+r_0^2\right) \left(2 (2 l+1)^2 \left(a^2+r_0^2\right) \left(\left(a^2+r_0^2\right) \omega _R-a m_l\right)-a A_3 A_6 \left(B^2 r_0^2+1\right)\right)}
    \end{equation}
\end{widetext}
where $A_3=r_0^2 \left(\frac{A_2}{\left(a^2 B^2-2\right)^2}+1\right)-\frac{4 M r_0 \left(a^2 B^2-1\right)}{a^2 B^2-2}+a^2$, 

$A_4=\frac{1}{2} a^2 \left(\frac{4 m^2}{(2 l+1)^2}-3\right) \omega _R^2+2 a m \omega _R+l^2+l+\frac{1}{4}$, 

$A_5=r_0 \Big[r_0 \Big \lbrace B^2 r_0 \Big(8 A_1 M-3 r_0 (4 B^2 M^2 \left(a^2 B^2-1\right)+\left(a^2 B^2-2\right)^2)\Big)+(5 a^2 B^2-7) (4 B^2 M^2 \left(a^2 B^2-1\right)+\left(a^2 B^2-2\right)^2)\Big \rbrace-8 M (a^2 B^2-2)^2 (a^2 B^2-1)\Big]+a^2 (4 B^2 M^2 \left(a^2 B^2-1\right)+\left(a^2 B^2-2\right)^2)$, 

$A_6=2 (2 l+1)^2 m_l-a \left(-4 m_l^2+12 l (l+1)+3\right) \omega _R$, 

$A_7=-a^3 m_l+\left(a^2+r_0^2\right) \left(a^2-3 r_0^2\right) \omega _R+5 a r_0^2 m_l$ 

and 
$A_8=r_0^2 \left(10-8 a^2 B^2\right)+a^2 \left(a^2 B^2-2\right)+3 B^2 r_0^4.$ 

In the eikonal limit ($l\gg 1$), the QNMs can be related to the properties of unstable null geodesics (photon sphere). The real part of the frequency is approximately the orbital frequency of photons at the unstable circular orbit, whereas the imaginary part is governed by the Lyapunov exponent, which measures the instability timescale of these orbits. Explicitly,

\begin{equation}
\omega_{QNM} \simeq m_l\,\Omega_c - i\Big(n+\tfrac{1}{2}\Big)\,|\lambda_c|,
\end{equation}

Here, $\Omega_c$ is the angular velocity of the circular photon orbit and $\lambda_c$ is the Lyapunov exponent.   For KBR BH, both $\Omega_c$ and $\lambda_c$ change when the magnetic parameter $B$ is nonzero. The radial potential for photon orbits acquires a factor $(1 + B^2 r^2)$, which shifts the photon sphere radius compared to Kerr. The eikonal QNM frequencies also shift, with larger deviations for higher $B$.  This provides a simple geometric picture of QNMs. Photon sphere properties affect BH shadows, while QNMs appear in gravitational-wave ringdowns.

\section*{Acknowledgments}
This research was funded by the National Natural Science Foundation of China (NSFC) under Grant No. U2541210. All numerical simulations were performed using the Phoenix High Performance Computing facility at the American University of the Middle East (AUM), Kuwait. We thank the anonymous referee for constructive feedback and valuable suggestions that improved the quality of this manuscript.\\\\

\section*{Data Availability Statement}
The data sets generated and analyzed during the current study were produced using high-performance computing resources. These data are not publicly available due to their large size and computational nature, but are available from the corresponding author upon reasonable request.

\def\prc{Phys. Rev. C}
\def\pre{Phys. Rev. E}
\def\prd{Phys. Rev. D}
\def\jcap{Journal of Cosmology and Astroparticle Physics}
\def\apss{Astrophysics and Space Science}
\def\mnras{Monthly Notices of the Royal Astronomical Society}
\def\apj{The Astrophysical Journal}
\def\aap{Astronomy and Astrophysics}
\def\actaa{Acta Astronomica}
\def\pasj{Publications of the Astronomical Society of Japan}
\def\apjl{Astrophysical Journal Letters}
\def\pasa{Publications Astronomical Society of Australia}
\def\nat{Nature}
\def\physrep{Physics Reports}
\def\araa{Annual Review of Astronomy and Astrophysics}
\def\apjs{The Astrophysical Journal Supplement}
\def\aapr{The Astronomy and Astrophysics Review}
\def\procspie{Proceedings of the SPIE}


\bibliographystyle{JHEP}
\bibliography{reference}

\end{document}